\documentclass[twocolumns]{aastex631}

\usepackage[english]{babel}

\usepackage{soul}  
\setstcolor{red}
\usepackage{wasysym}
\usepackage{rotating}
\usepackage{multirow}
\usepackage{chemfig}
\usepackage{enumitem}
\usepackage{ulem,xcolor}

\usepackage{amsmath} 
\usepackage{graphicx} 
\usepackage{array} 

\begin{document}

\title{Efficient searches of small signals across two-channel noisy data: a  challenge in Big Data observations}

\author[0000-0002-1518-1946]{Maryam Aghaei Abchouyeh}
\affiliation{Department of Physics and Astronomy, Sejong University, 98 Gunja-Dong, Gwangjin-gu, Seoul 143-747, Republic of Korea}

\author[0000-0002-9212-411X]{Maurice H.P.M. van Putten}
\altaffiliation{Corresponding author, mvp@sejong.ac.kr}
\affiliation{Department of Physics and Astronomy, Sejong University, 98 Gunja-Dong, Gwangjin-gu, Seoul 143-747, Republic of Korea}
\affiliation{INAF-OAS Bologna, via P. Gobetti, 101, I-40129 Bologna Italy, Italy}

\author[0000-0002-2102-7398]{Seyong Kim}
\affiliation{Department of Physics and Astronomy, Sejong University, 98 Gunja-Dong, Gwangjin-gu, Seoul 143-747, Republic of Korea}



\begin{abstract}

{{Searching for novel small signals in noisy data is preferably pursued by correlation in two or more independently operating channels.}
Signals of potential interest exist in tails beyond $\kappa\sigma$, where $\kappa$ denotes a multiple of the standard deviation $\sigma$ of the data.  
Since moving data is a major cost factor in Big Data analysis and heterogeneous computing more generally, efficiency may be  optimized by restricting the {correlations computation to the} tails
of two-channel data exceeding $\kappa\sigma$.
Already, a moderate value $\kappa\gtrsim2$ realizes a data-reduction by at least an order of magnitude.}
{Here, we study this approach using a novel {\it Excess Probability Ratio} (EPR), correlating Boolean data resulting from tails beyond a cut-off $\kappa\sigma$. We compare and rank EPR performance against conventional direct cross-correlation (DCC) and Pearson coefficient (PC), applicable {to the original} data with no cut-off.
This benchmark is performed over different combinations of background noise (Gaussian, Poisson and Uniform) and signals (Gaussian, Poisson, Uniform, Chirps and Sine waves).
Results show performance of EPR to be comparable to {that of} PC, providing a new approach for significant improvements in efficiency {with} essentially no loss of {sensitivity, } relevant to {the} present era of Big Data observatories.
}

\end{abstract}



\section{Introduction}
\label{Sec:Intro}
{The discovery potential of modern astronomical observatories critically relies on signal processing in search for new or un-modeled signals in noisy channels. 
The mathematical formulation of {distinguishing signals from noise} {was} originally formulated for single-channel data, based on a minimum sampling rate by H. Nyquist \citep{nyqu1928}.} 
{This line of work was subsequently extended by Shannon with mathematical foundations} of information theory, applied to {data transmission} by binary encoding across noisy channels \citep{shan1948,shan1949}.
Further developments over the past century have focused on efficient implementations of signal processing in the modern world of digital devices \citep{wien1949, kay1993, kay1998,proa2006, oppe2009}. 
This is driven by {ever-increasing} applications in industry and science. Today, this extends to modern detectors and observatories \citep{amar2012,putten2014,brau2015,abbo2016,abbo2017,igan2022,agaz2023,abbo2023,cao2024}, whose operation and discovery power increasingly {tilt} towards Big Data processing rather than hardware.
As such, Big Data requires optimization for efficiency -  {\it green} computing which is more broadly recognized in the growth of the Internet, cloud computing, and high-performance computing.

When detectors produce massive amounts of data, observations may be limited by bandwidth in moving data and/or data-storage with accompanying costs in time, hardware and energy. Ideally, the infrastructure for data-processing is compute rather than bandwidth limited. Efficiency in the processing of Big Data observations may hereby become a critical factor in optimizing discovery power. {For instance,} the {\it Square Kilometer Array} (SKA) is expected to produce on the order of zettabytes per year - roughly equal to the total global Internet traffic in 2022 \citep{Ahar2013,Farnes2018}.

{A candidate detection for small new or un-modeled signals in noisy channels must satisfy the ground rule of a correlated response in two or more detectors, operating at similar sensitivity \citep[e.g.][]{MAA2023,putten2024}.
 Additionally, while background statistics may vary, the signal can be transient or continuous, and the data maybe be integer, real or even Boolean/binary valued.}
 
\begin{figure}
\centering
\includegraphics[width=0.85\linewidth]{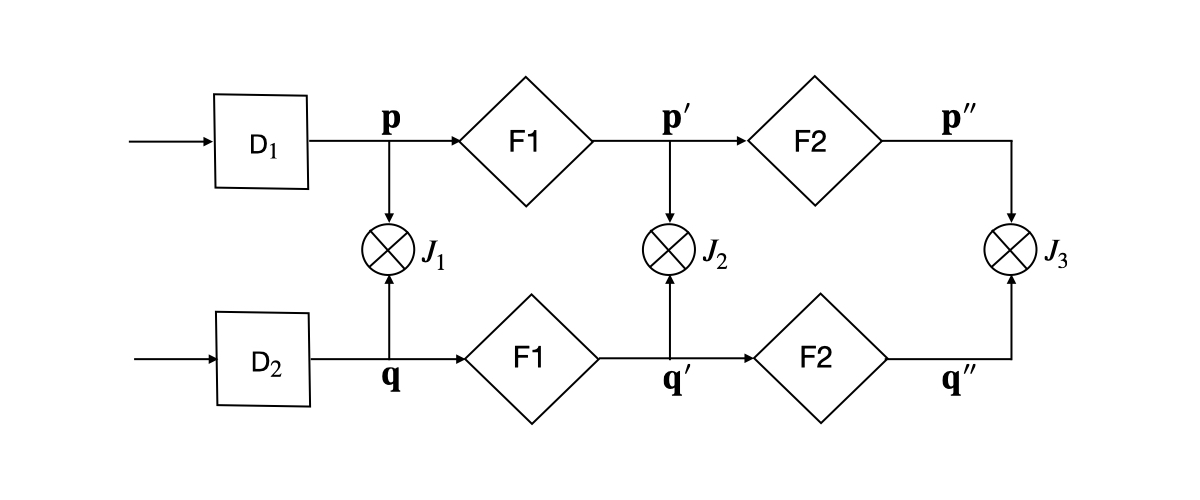}
\caption{Schematic of a two-channel pipeline processing data $\left({\bf p},{\bf q}\right)$ of detectors $\left(D_1,D_2\right)$, across different junctures $J_i$ ($i=1,2,3$) for cross-correlation.
Cross-correlations can be performed by DCC, PC and EPR, discussed in the present work, applied to raw data for DCC or PC (\S\ref{Sec:DCCandPC}) at $J_1$, PC and EPR at $J_2$ following  a filter (F1),  or further down the pipeline by EPR following a second filter (F2). For EPR applied to real or complex-valued data $\left({\bf p, q}\right)$, these filters may include a threshold (\ref{EQN_kappa}) for conversion into Boolean data, {or filters for signal amplification prior to cut-off application.}}
\label{fig:Process}
\end{figure}

In this paper, we study sensitivity to picking up small signals in two-channel observations according to a detection threshold for different methods of correlation {for} a variety of combinations of background statistics and signal types {motivated by astronomical sources (\ref{Sec:combinations}), following the process indicated in Fig. \ref{fig:Process}}. 

In seeking to improve efficiency, we consider the possibility of data-reduction by performing correlations {on} {\it tails} of two-channel data, exceeding some cut-off. We consider tails according to a multiple $\kappa$ of the standard deviation $\sigma$ of data. Analysis of tails rather than the full data of observation {promises} improvement in efficiency whenever the size of tails is significantly small relative to the original detector data. 
{Following the general form of tails defined in \cite{Putten2017}, here for data ${\bf A}=\{A_0, A_1,\cdots\}$, the tail is defined to be a binary string ${\bf A}^\prime=\{A_0^\prime,A_1^\prime,\cdots\}$ according to} 

\begin{equation}
    \label{EQN_kappa}
    A_i^\prime = 1 \,\,{\rm if}\,\,A_i > \kappa \sigma\,\,{\rm and}\,A_i^\prime=0\,\,{\rm otherwise},
\end{equation}

\noindent {where} $\kappa$ controls data-reduction by ignoring the bulk comprising mainly background {containing} essentially unrecoverable information.

\begin{figure}
\centering
\includegraphics[width=0.47\linewidth]{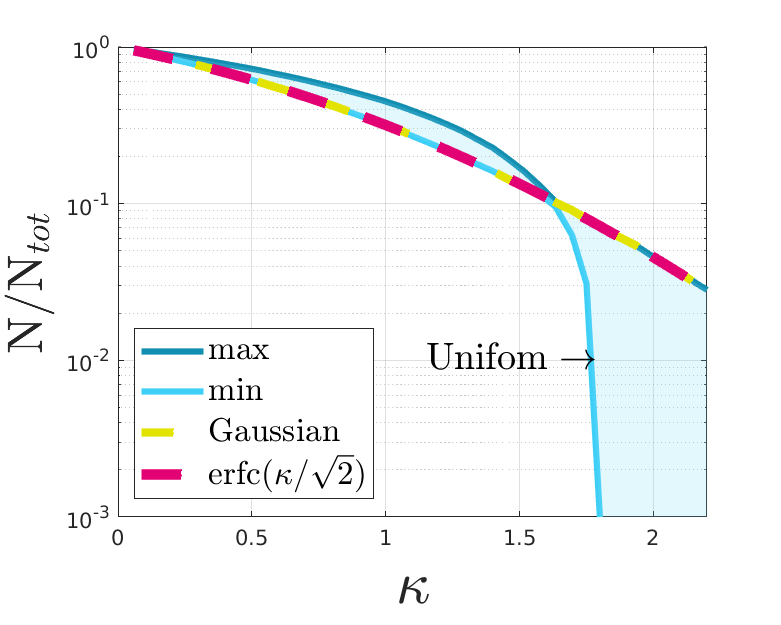}
\includegraphics[width=0.47\linewidth]{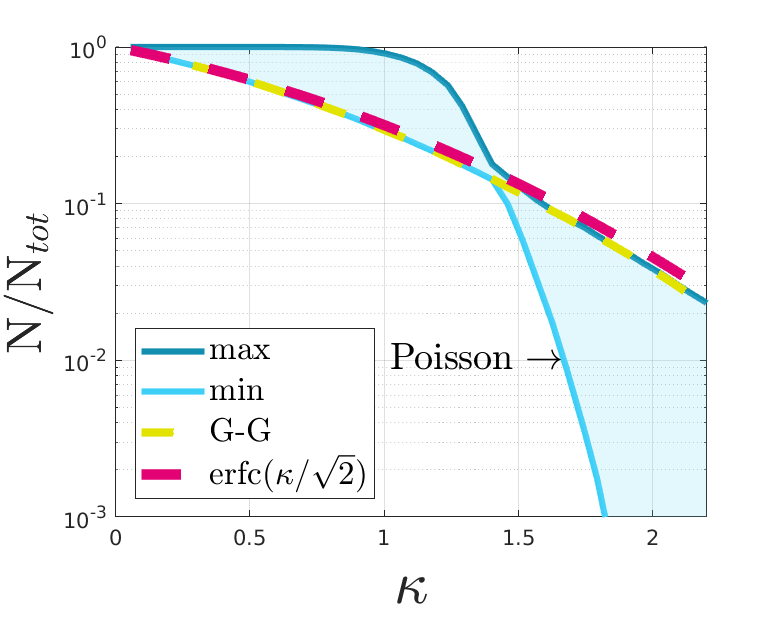}
\caption{
{(Left panel.) Data-reduction shown by fraction of data remaining after the cut-off (\ref{EQN_kappa}) as a function of $\kappa$ for a given background noise. 
Results shown cover the three backgrounds of Gaussian, Poisson and Uniform distributions. 
 The yellow dashed curve shows the result for Gaussian background, which nicely matches the theoretical complementary error function (red dashed line, left). 
(Right panel.) Essentially the same data reductions appear in the presence of signals, here studied across 15 different combinations of backgrounds and signals at strength $\alpha=0.5$ (\S \ref{Sec:combinations}). The yellow dashed line shows the result for a Gaussian signal in a Gaussian background (G-G). Notable is the exchange of leading curve between Uniform to Poisson background between the left and right panels.}} 
\label{fig:N-kappa}
\end{figure}

{By probability of detection, in contrast, tails possess the greatest density of information.}
Fig. \ref{fig:N-kappa} shows representative examples of data-reduction in tails versus $\kappa$. The results show reductions to $10\%$ of original detector total data already at moderate values $\kappa\simeq 2$. 
{A data-reduction by an order of magnitude offers a significant avenue to improve efficiency in processing a given amount of detector data or, alternatively, permits a larger batch of data to be processed with a similar amount of effort.}

{To study the feasibility of analyzing tails (\ref{EQN_kappa}) rather than full detector data}, we introduce a novel method of {\it Excess Probability Ratio} {(EPR) \citep{putten2019x,putten2019}}. 
We benchmark EPR  by ranking its performance against the performance of two commonly used correlations namely direct cross-correlations (DCC) and Pearson coefficients (PC), applicable to the {\it full} data-array $A_i$. 
DCC and PC naturally apply to continuous data, notably real or complex valued time-series. Unlike the former, PC is normalized, famously interpreted as a cosine. Additionally, {despite} PC,  DCC is a map of two arrays to another array and the maximum value is considered to be an indicator of similarity between the two arrays. 
Time slides or Monte Carlo simulations can be used for a direct comparison of PC with DCC on equal footing.
In contrast, EPR applies to Boolean-valued or more generally non-negative data, real or integer valued. 
A notable example is counting noise in photon detectors in {high-energy} satellite observations \citep[e.g.][]{Frontera2008}. Such tends to produce a background with uniform statistics in angular coverage of the sky and Poisson in the time-series of light curves. 
For data of this kind a natural first step is the extraction of tails by application of the cut-off (\ref{EQN_kappa}), creating binary strings of the Boolean data.
 
We define EPR in the same spirit of Shannon's binary encoding of information \citep{shan1948}, now applied to the problem of correlating two noisy data {channels,} with binary output following (\ref{EQN_kappa}).
Here also, time slides or Monte Carlo simulations can be used for a direct comparison with DCC and {EPR}. 
We evaluate EPR against DCC and PC {by sensitivity of detection} identified in response curves. To this end, we perform an extensive benchmark in MATLAB \citep{MATLAB}, covering various signal injections into different backgrounds of noise - Gaussian, Poisson or Uniform distributions. Results are ranked by response curves based on detection thresholds in each such combination as follows.

\begin{figure}[ht!]
\centering
\includegraphics[width=0.6\linewidth]{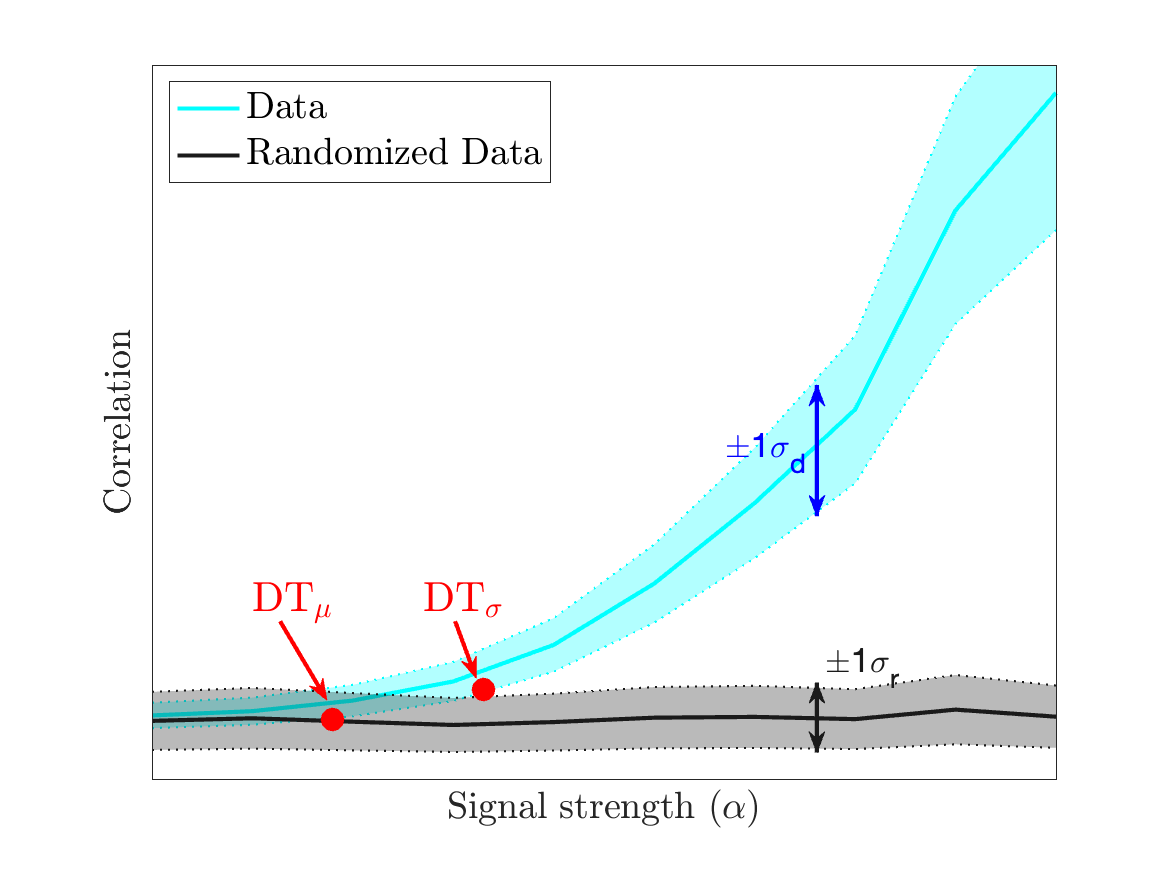}
\caption{Classification of sensitivity in response curves (cyan) in signal injection experiments by detection thresholds (DT, red), beyond which output exceeds expectation values from background (grey). 
Background is defined following time-randomization of data, effectively equivalent to noise without signal (when noise is white).
Faded cyan and gray bands indicate $\pm1\sigma_d$ and $\pm1\sigma_r$ {, respectively}. $DT_\mu$ and $DT_\sigma$ indicate the smallest signal amplitudes $\alpha$ at which the cyan band departs from the black line and gray band, respectively, here highlighted by red dots. DT's facilitate the ranking of different methods of analysis. }
\label{fig:departure}
\end{figure}

{For ranking}, we define detection threshold by $DT_\mu$ and $DT_\sigma$ with respect to the mean value $\mu$ and $1\sigma$ away from the mean value of background for each response curve. Looking for small signals, $DT_\mu$ is {\it the signal amplitude at which the mean (expectation value of the) response curve departs from that of the time-randomized data by $1\sigma_{d}$}, where $\sigma_{d}$ is the standard {deviation in the} former (Fig. \ref{fig:departure}). {Time-randomized data is shown to be equivalent to the background noise as any trend related to a potential small signal is lost in randomization.}
Accordingly, $DT_\sigma$ is the signal amplitude at which the mean response curve departs from that of the time-randomized version by $\sigma_{r}+\sigma_{d}$, where $\sigma_{r}$ is the standard deviation in the time-randomized data.  Fig. \ref{fig:Process} shows the application of these methods at different junctures in a two-channel pipeline. 

We organize this study as follows. First, we review the properties and definitions of PC and DCC \S\ref{Sec:DCCandPC}.
EPR is introduced in \S\ref{Sec:EPR_def} as a new method following \cite{putten2019} and \cite{putten2019x}. 
The setup for benchmarking and ranking of these three methods is in \S\ref{comparison} and \S\ref{Sec:results}. 
We conclude the discussion in \S\ref{Sec:outlook} 
on the performance of EPR: its potential to improve efficiency without significant loss of sensitivity. Included are some additional considerations and application for these methods.

\section{Definitions and notations}
\label{Sec:DCCandPC}

 {To study and compare methods of correlations, we produce two-channel arrays ${\bf D}_i$ with independent background noise ${\bf n}_i$ $(i=1,2)$ and a common signal $S$ by injections.} Noise introduces a background whose texture depends on its statistical distribution. With ${\bf n}_i$ independent in two channels $D_i$ and no common signal, any residual correlation defines a fiducial background for a given method of correlation. For DCC and PC applied to Gaussian noise, for instance, the expectation value of this residual is zero. In what follows, {the presence of a common signal} $S$ {in both} channels is simulated by injection, adding $S$ to both ${\bf n}_i$ over a range of amplitudes $\alpha$. For two detectors D$_1$ and D$_2$, their respective data arrays ${\bf p}={\bf D}_1$ and ${\bf q}={\bf D}_2$ satisfy

\begin{equation}
{\bf D}_i={\bf n}_i+\alpha {\bf S}
\hspace{0.2cm} 
\left(i=1,2\right).
    \label{data}
\end{equation}

\noindent We consider the response curve (Fig. \ref{fig:departure}) of a two-channel analysis by correlations between the two as a function of the amplitude $\alpha$ of the injected signal. This amplitude parametrizes the {strength} of the signal relative to background noise according to
\begin{equation}
    \label{EQN_alpha}
    \alpha \equiv \frac{\sigma_{sg}}{\sigma_{bg}}, 
\end{equation}
where $\sigma_{sg}$ and $\sigma_{bg}$ are the standard deviations of signal and background, respectively.

For reference, we next review DCC and PC, and their {domain of application}.

\subsection{Direct cross-correlation}

\label{Sec:dcc}
DCC computes the correlation {between two channels as a function} of a slide between the two. {For} real-valued arrays of size $N$, the discrete form of DCC maps ${\bf p}$ and ${\bf q}$ to a new array ${\bf C}$ of the same dimension \citep{brac1999,ifea2001,zito2003} by 

\begin{equation}
    \label{EQN_dcc}
    C_i=\sum_{j=1}^{N}p_jq_{j+i}.
\end{equation}

\noindent A peak value $\hat{C}$ in $\bf{C}$ defines a (maximal) correlation between ${\bf p}$ and ${\bf q}$, 
the location $i$ of which in ${\bf C}$ identifies the associated time-slide. With no normalization in \eqref{EQN_dcc}, $\hat{C}$ is unbounded.

The significance of $\hat{C}$ will be relative to a baseline, nominally defined by DCC in the absence of a common signal between two channels. Particularly for small signals, this baseline is equivalently derived following time-randomization of the data, i.e., by random index permutations of the arrays, removing any information in (time-ordered) signals. This baseline applies in the ideal case of white noise, when the autocorrelation function is trivial.
Accordingly, we arrive at the following

\begin{equation}
\mathcal{R}=\frac{\hat{C}}{\hat{C}_R},
    \label{EQN_dcc_max}
\end{equation}

\noindent where $\hat{C}_R$ is the maximum of DCC following time-randomization of the original data.

\subsection{Correlation by Pearson coefficient}

\label{Sec:pc}
For real-valued arrays ${\bf p}$ and ${\bf q}$ with mean values $\mu$ and, respectively, $\nu$, the Pearson coefficient is defined by \citep{pear1895,stig1989} 

\begin{equation}
    \label{EQN_Pearson}
    P=\dfrac{({\bf p}-\mu)\cdot({\bf q}-\nu)}{|{\bf p}-\mu|\ |{\bf q}-\nu|}.
\end{equation}

\noindent {With $\theta$ being the angle between the two arrays ${\bf p}$ and ${\bf q}$,} PC is the $\cos\theta$ between two arrays by virtue of the normalization in (\ref{EQN_Pearson}), satisfying $-1<P<1$.

While PC maps ${\bf p}$ and ${\bf q}$ to a scalar, we are at liberty to apply time-slides to the input arrays. This creates a time-series of PC over {the} said time-slide, facilitating a direct comparison with DCC. Alternatively, Monte Carlo simulations can be used for {the same} comparison. 

\section{Excess Probability Ratio}

\label{Sec:EPR_def}

For Boolean-valued data, we next consider a novel {\it Excess Probability Ratio} (EPR), initially introduced in \cite{putten2019} to explore correlations across pairs of spectrograms.
EPR has a natural application to cut-offs that set a floor to a candidate signal or feature in the data.
Following (\ref{EQN_kappa}), we here refer to data values above cut-off as ``hits''. 
{Focused on two-channel data ${\bf p'}$ and ${\bf q'}$ passed through the filter (\ref{EQN_kappa}), we shall drop the prime for ease of read.}

\begin{figure}
\centering{
\includegraphics[scale=0.35]{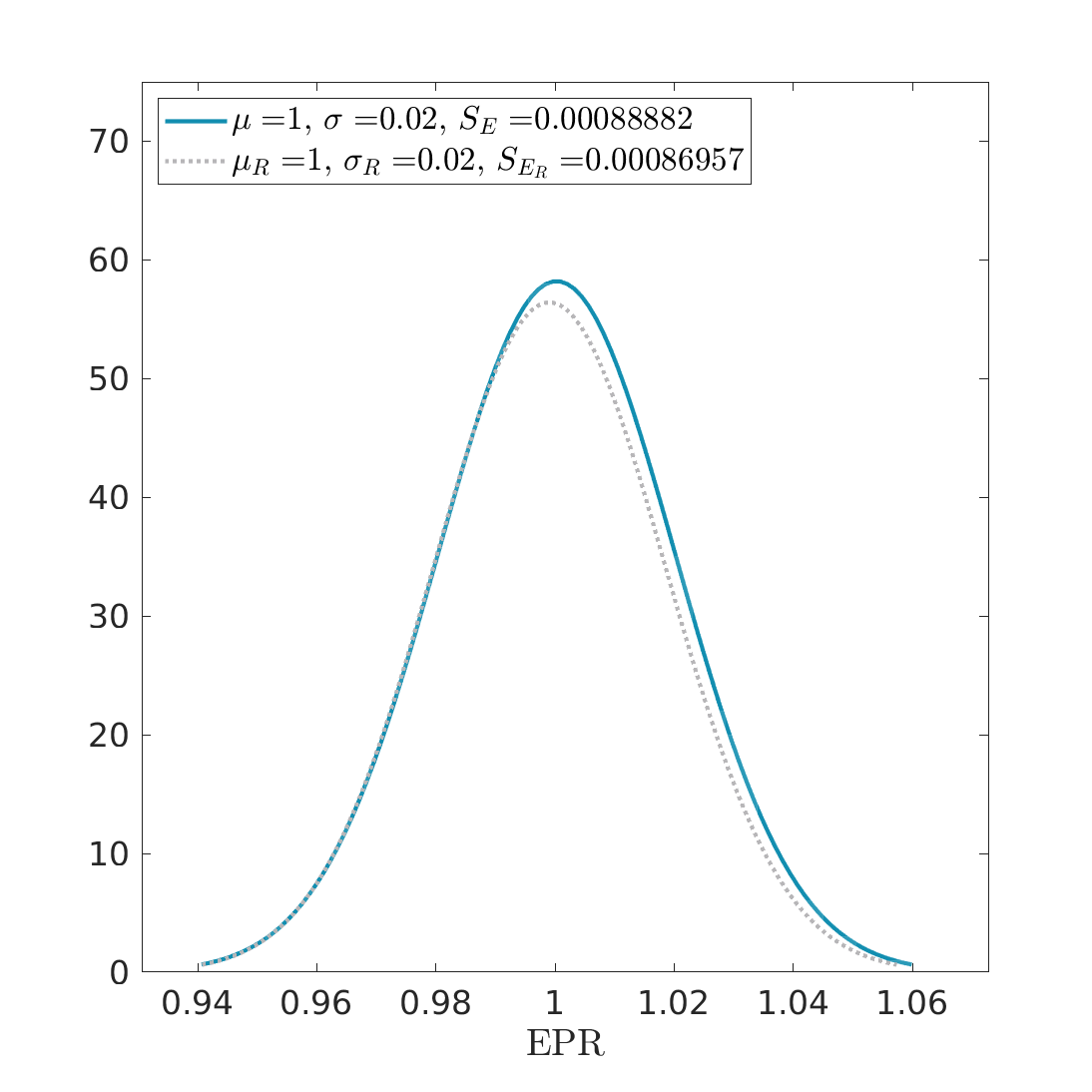}
\includegraphics[scale=0.35]{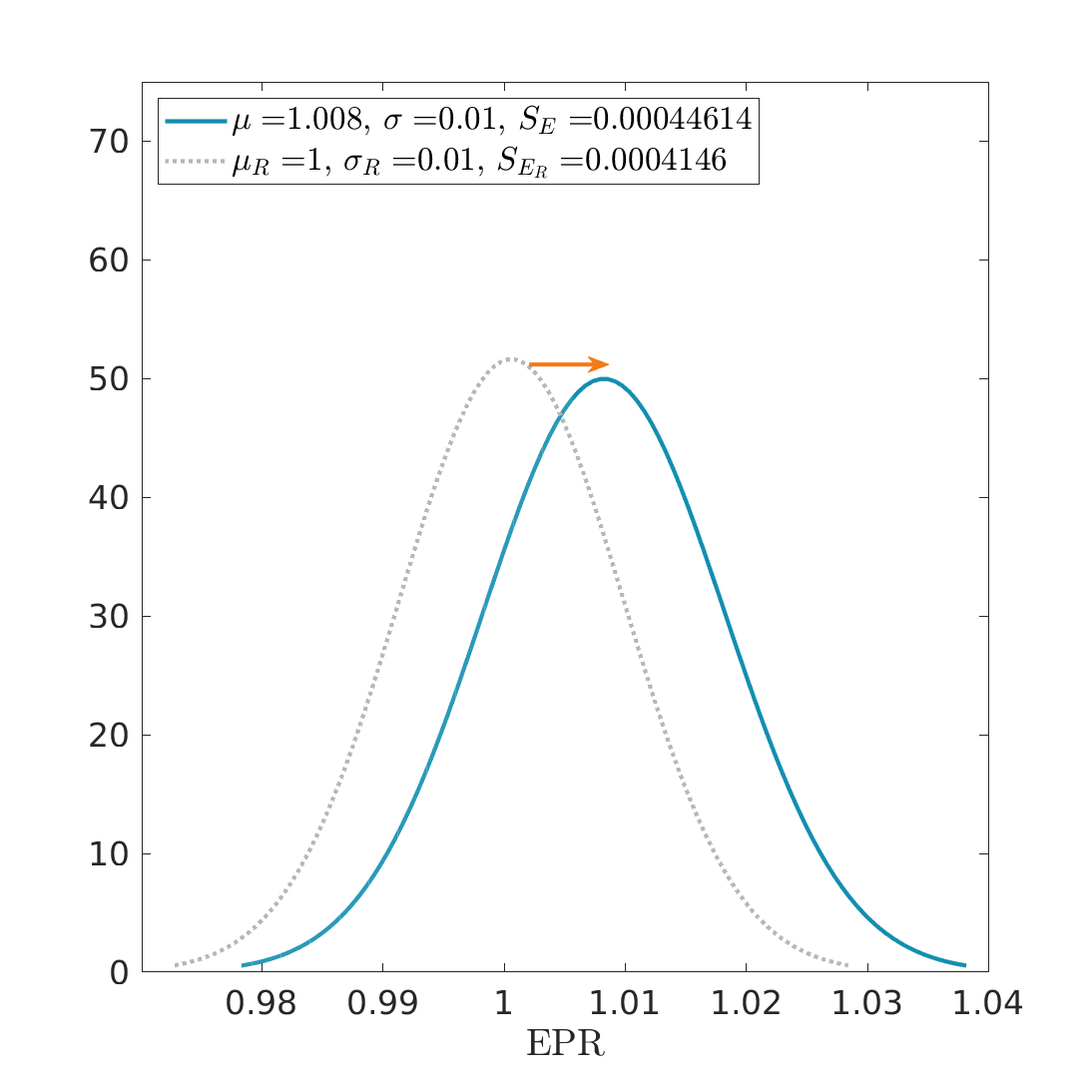}
}
\caption{(Left panel.) PDF of EPR {for background noise} without signal injection. For random numbers (white noise), EPR has expectation value 1 (green curve). The same outcome is produced following time-randomization (gray dotted line). (Right panel.) With signal injection, the expectation value of EPR shows a positive shift correlated to signal amplitude (green curve), while time randomization preserves the baseline.} 
\label{fig:EPR1}
\end{figure}

EPR measures the {\it excess probability} in finding joint hits across two channels passing (\ref{EQN_kappa}). According to \cite{Putten2017},

\begin{equation}
\mbox{EPR}=\frac{P_{12}}{P_1P_2}.
    \label{EQN_EPR}
\end{equation}

\noindent Here, $P_{12}=N^{-1}{\bf p\cdot q}$ is the fraction of the data ${\bf p\cdot q}$ {jointly} passing (\ref{EQN_kappa}), while the mean values
$P_{1}= N^{-1} \sum_{i=1}^{N}p_i$ and $P_{2}= N^{-1} \sum_{i=1}^{N}q_i$ represent the fractional hits in ${\bf p}$ and, respectively, (${\bf q}$. 
{With the number of hits} $n_p=NP_1$ in ${\bf p}$ and  $n_q=NP_2$ in ${\bf q}$, 
and $n=NP_{12}$ in ${\bf p\cdot q}$,
the joint probability
\begin{equation}
  \mbox{EPR}_{p,q}=\frac{\bf p\cdot q}{NP_1P_2} = \frac{nN}{n_pn_q},
\label{EQN_EPR_A}
\end{equation}
{represents a ratio of simultaneous hits to the product of individual hits}.

Combined with (\ref{EQN_kappa}), EPR quite generally {applies} to non-negative real data. In a two-channel observation 
(Fig. \ref{fig:Process}),
this may include conversion to the time-frequency domain first {(following F1 or F2 in Fig. \ref{fig:Process})} before applying a {cut-off (\ref{EQN_kappa}) to pre-amplify a potential signal.} 
Such is opportune in searches for transient signals - distinct features in the time-frequency domain, i.e., image-based searches in spectrograms. 

For real-valued data $({\bf p, q})$, we apply EPR following (\ref{EQN_kappa}) after subtracting their respective mean values to ensure focus on fluctuations similar in spirit to PC (\ref{EQN_Pearson}). 
Similar to PC also, we apply time slides to (\ref{EQN_EPR_A}) or Monte Carlo simulations to facilitate a direct comparison with DCC.

\subsection{The expectation value of EPR}\label{EPR_range}

{By (\ref{EQN_EPR_A}), EPR produces a non-negative output $0\leq {\rm EPR} \leq N$. 
Its expectation value satisfies $\mathbb{E}\left[{\rm EPR}\right]=1$
when the two arrays are statistically uncorrelated, while 
$\mathbb{E}\left[{\rm EPR}\right]>1$
indicates a correlation, signaling a common signal. {This means the EPR can be used as a test statistic for detecting a correlation or deviation from statistical independence.}

{To see this, consider a pair of arrays $\left({\bf p},{\bf q}\right)$ of statistically independent random sequences of Boolean-valued numbers of length $N$.
By statistical independence of the two arrays, the expectation value of the numerator of EPR in (\ref{EQN_EPR_A}), factored independently over ${\bf p}$ and ${\bf q}$, satisfies
\begin{eqnarray}
\mathbb{E}\left[{\bf p}\cdot{\bf q}\right]=\mathbb{E}\left[\sum_{i=1}^Np_iq_i\right]=\sum_{i=1}^N \mathbb{E}\left[p_iq_i\right]
= \sum_{i=1}^N P_2 \mathbb{E}\left[p_i\right] = P_2 \sum_{i=1}^N\mathbb{E}\left[p_i\right] = NP_2P_1.
\label{EQN_E1}
\end{eqnarray}
The expectation value of EPR hereby reduces to
$\mathbb{E}\left[ {\rm EPR}_{p,q} \right] = 1$.
}
{Illustrative is also the limiting case ${\bf q}={\bf p}$ for which} 
\begin{eqnarray}
{\rm EPR}_{p,p}= N 
\frac{\sum_{i=1}^N p_i^2}
{\left(\sum_i^N p_i\right)^{2}}.
\end{eqnarray}
{Since ${\bf p}$ is Boolean valued, 
$ p_i \in \{0,1\} $, we have $\sum p_i=\sum p_i^2\leq N$. Consequently, $\left ( \sum p_i\right)^2 \leq N \sum p_i^2$, and hence
${\rm EPR}_{\bf p,p}\geq 1$. Accordingly, $\mathbb{E}\left[{\rm EPR}\right]$ shifts when $({\bf p,q})$ are correlated. }
This can be seen by orthogonal decomposition, using a correlation coefficient $\lambda$ in ${\bf q} = {\bf q}_\perp + {\bf q}_{\parallel}$, ${\bf q}_{\parallel} = \lambda {\bf \hat{p}}$, ${\bf \hat{p}}={\bf p}/\left|{\bf p}\right|$.

\begin{figure}
\centering{     \includegraphics[scale=0.35]{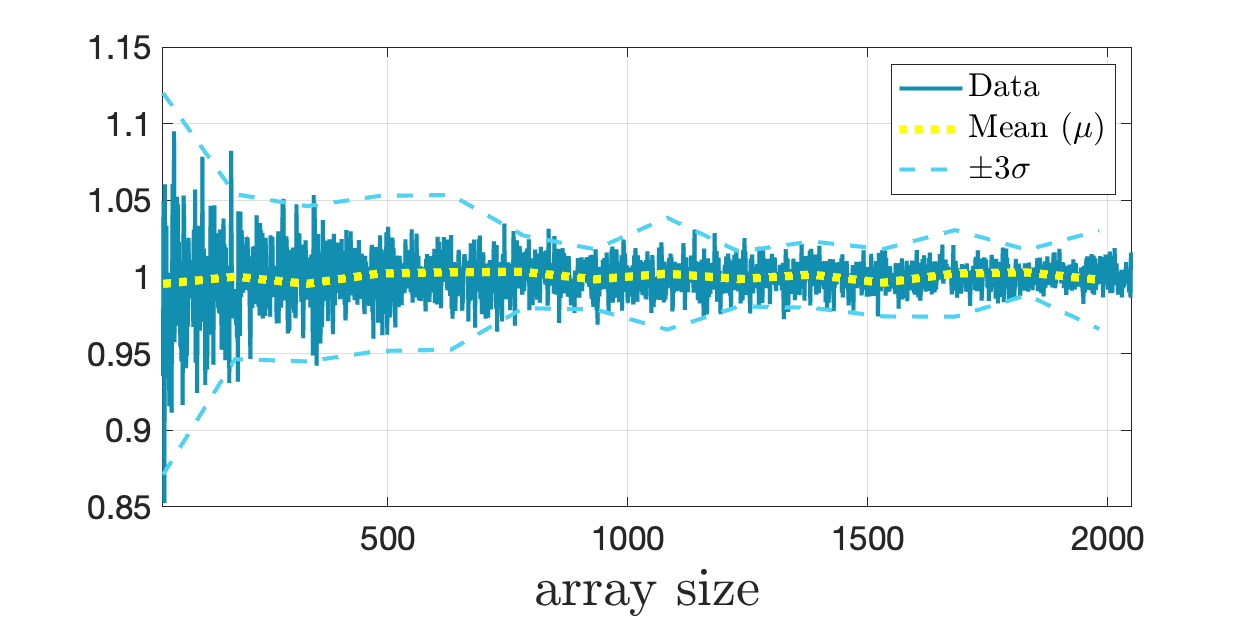}         
\includegraphics[scale=0.35]{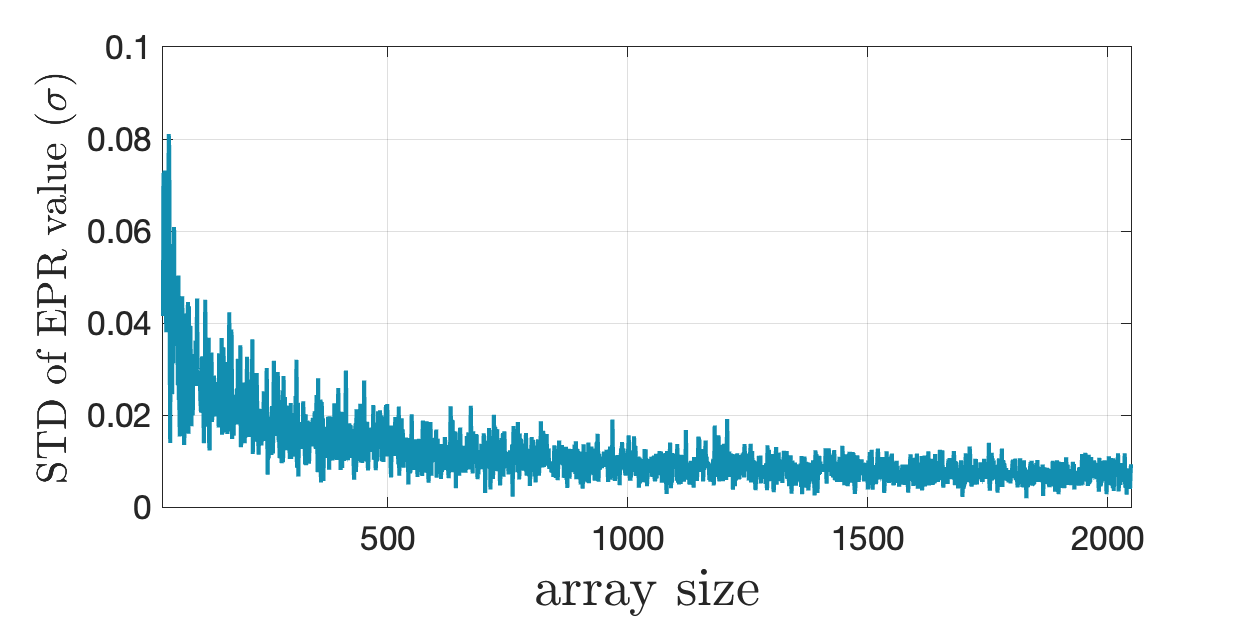}
}
\caption{EPR applied to a pair of uncorrelated random arrays drawn from Poisson distribution. (Left panel.) $\overline{\rm EPR} =1$ and its standard deviation $\sigma$ decreases with array size $N$. (Right panel.)
The standard deviation $\sigma\propto 1/\sqrt{N}$ follows the canonical result of random walks. 
}
\label{fig:EPR2}
\end{figure}

\subsection{Example of EPR}
\label{Sec:EPR_prop}

Illustrative for EPR (\ref{EQN_EPR_A}) is the application to two independent data arrays. 
EPR measures correlation by shift above baseline as shown in Figs. \ref{fig:EPR1}-\ref{fig:EPR2}. In this response, the width of its distribution remains effectively the same.

 \section{Experimental setup}
  \label{comparison}
In (\ref{EQN_EPR_A}), EPR is defined for Boolean-valued strings of data, that may derive from real-valued data following a cut-off (\ref{EQN_kappa}) in a two-channel pipeline (Fig. \ref{fig:Process}). 
To benchmark sensitivity, we consider EPR against conventional cross-correlations DCC and PC.

{Below, we compare and rank the sensitivity of these three methods in searches for small signals} according to their detection thresholds DT {(Fig. 2)}. In light of their broad range of applications, we include various cases of background noise and signal injections \eqref{data}.
This study is implemented using MATLAB simulated data, allowing full control over the relevant parameters.

\subsection{Selection of background and signals}
\label{Sec:combinations}

To cover some generality in our comparison study, we select various combinations of background and signal injections in \eqref{data} motivated by astronomical observations.

\subsubsection{Samples of astronomical statistics}

Observed in some astronomical phenomena are some key statistical distributions. These include but not limited to Gaussian, Poisson and Uniform distributions observed for instance in the following:

\begin{figure}[ht!]
    \centering
    \begin{longtable}{cc}  
        \includegraphics[width=0.45\textwidth]{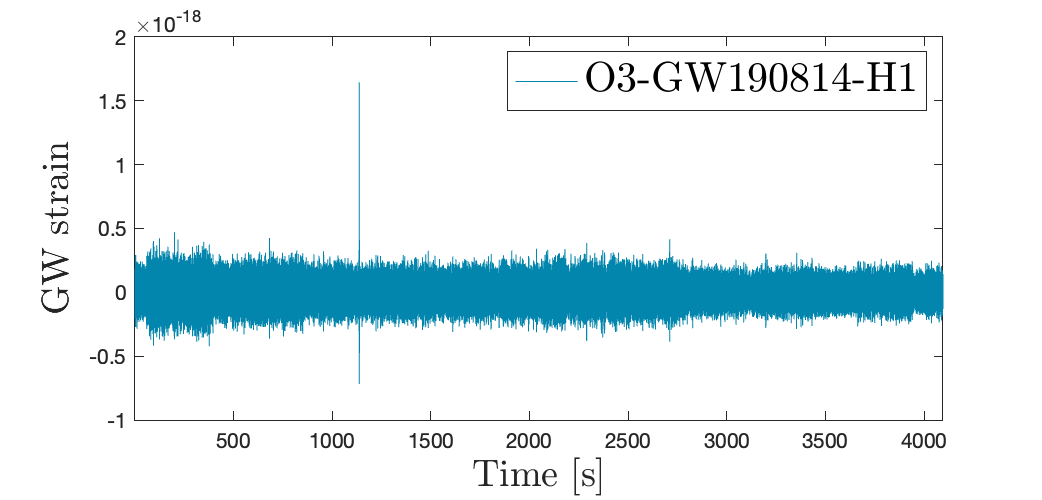} & 
        \includegraphics[width=0.45\textwidth]{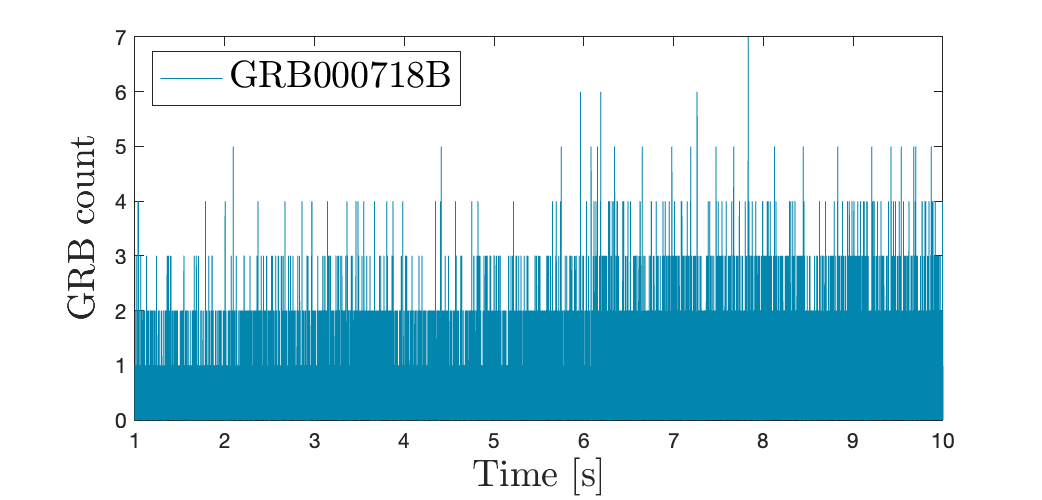} \\
        (a) & (d) \\

        \includegraphics[width=0.45\textwidth]{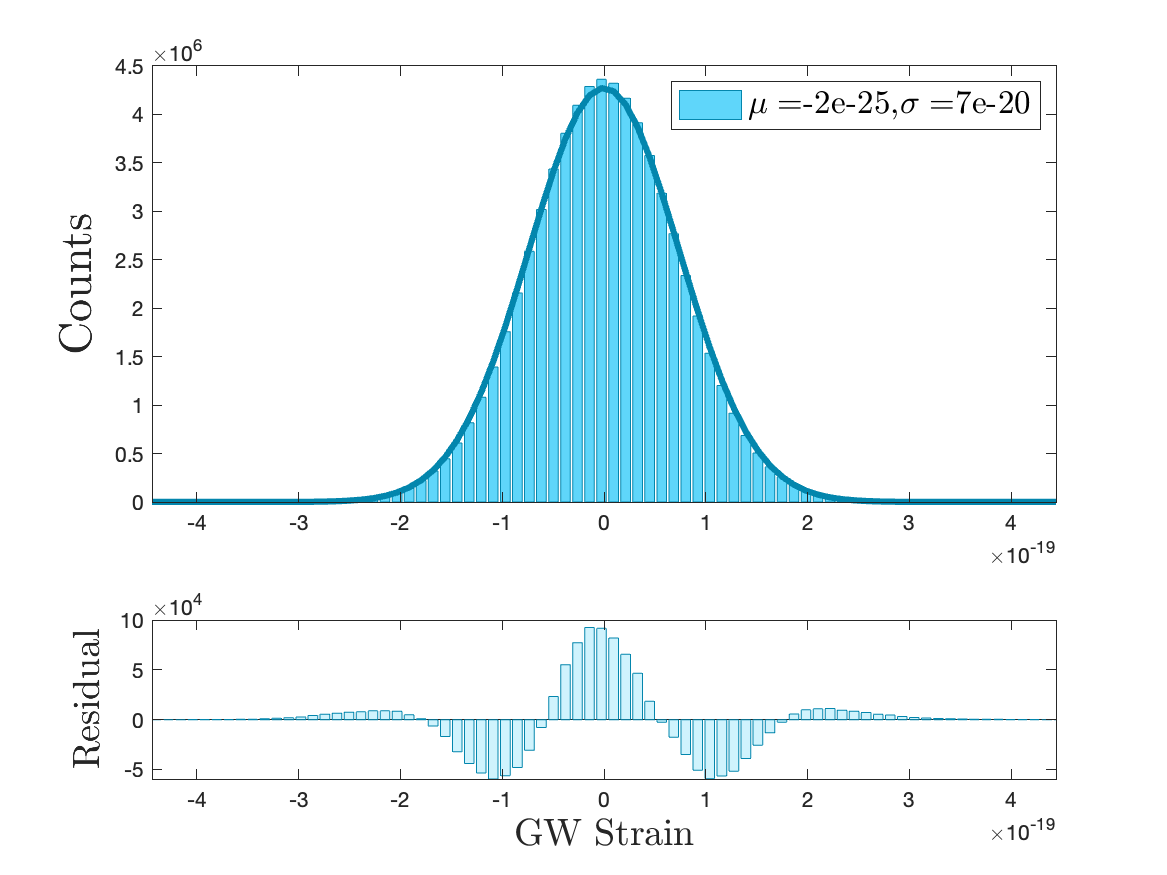} & 
        \includegraphics[width=0.45\textwidth]{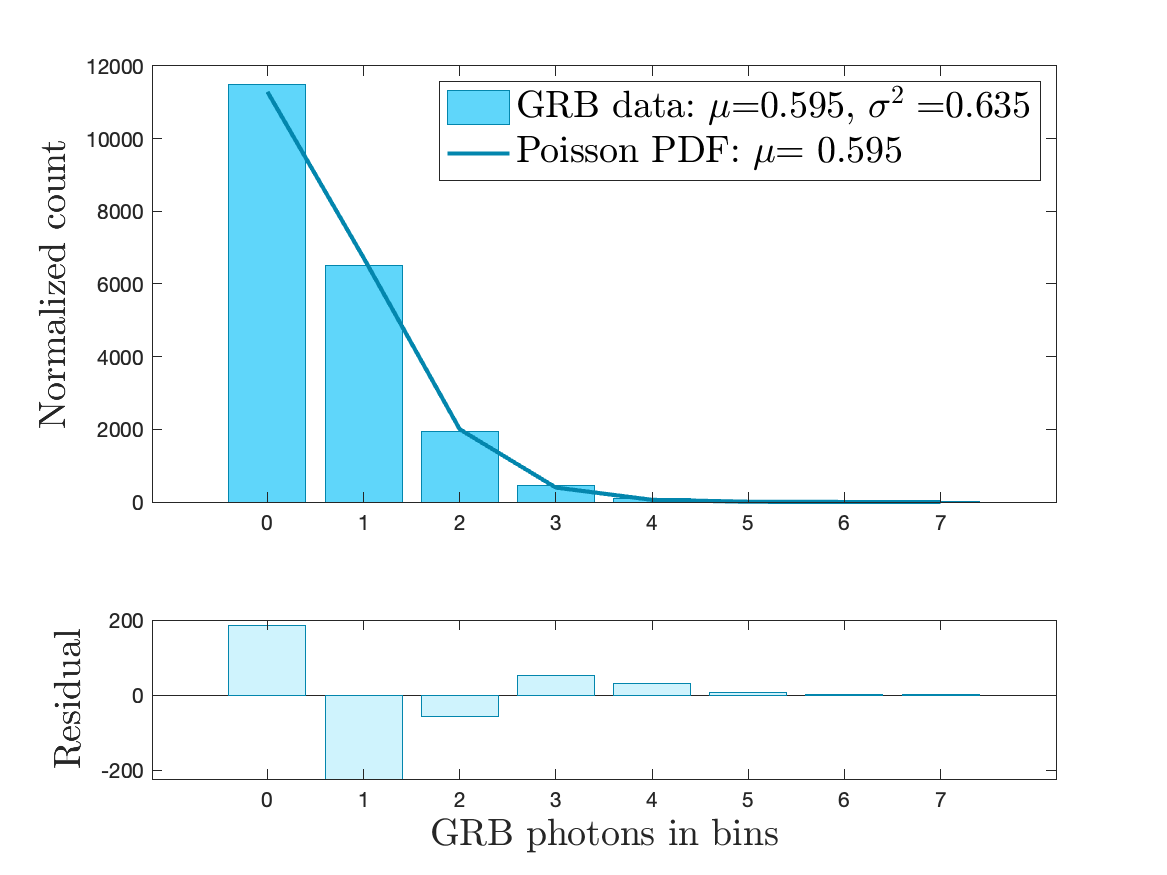} \\
        (b) & (e) \\

        \includegraphics[width=0.45\textwidth]{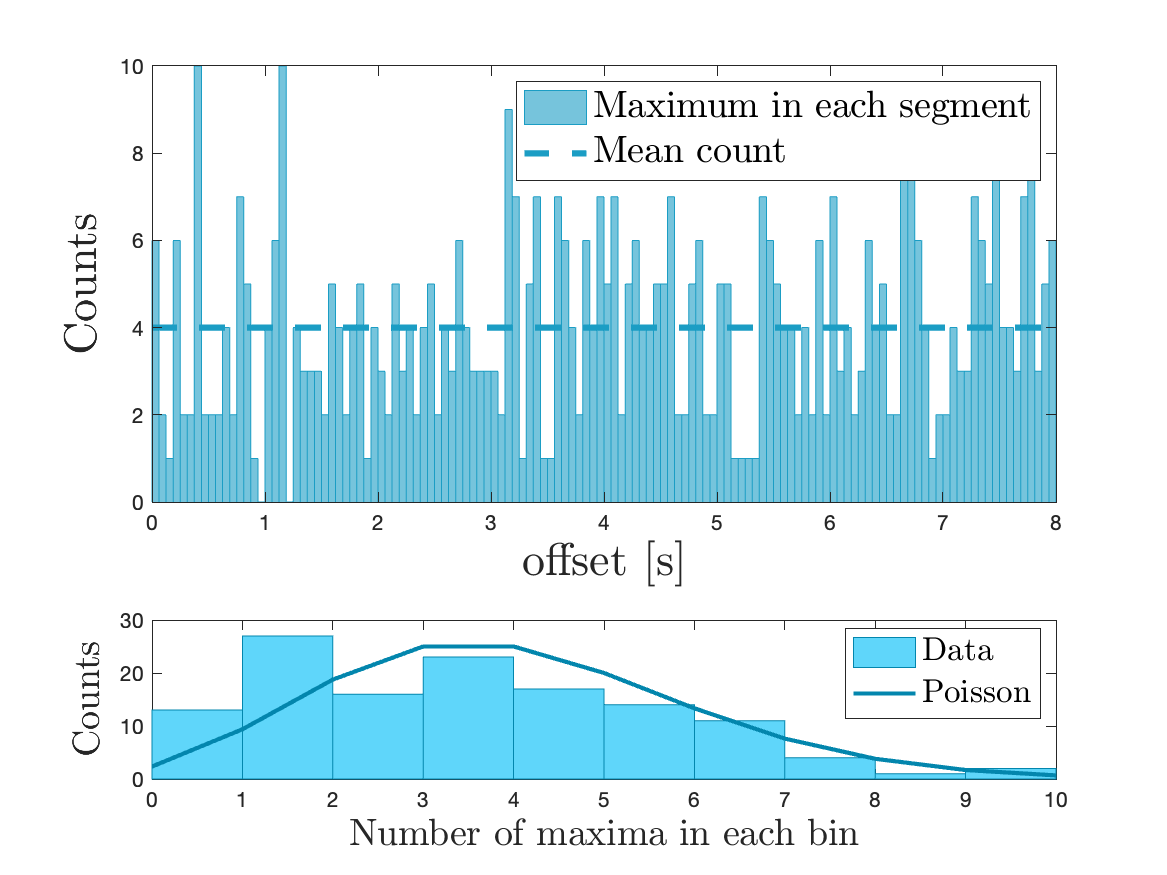} & 
        \includegraphics[width=0.41\textwidth]{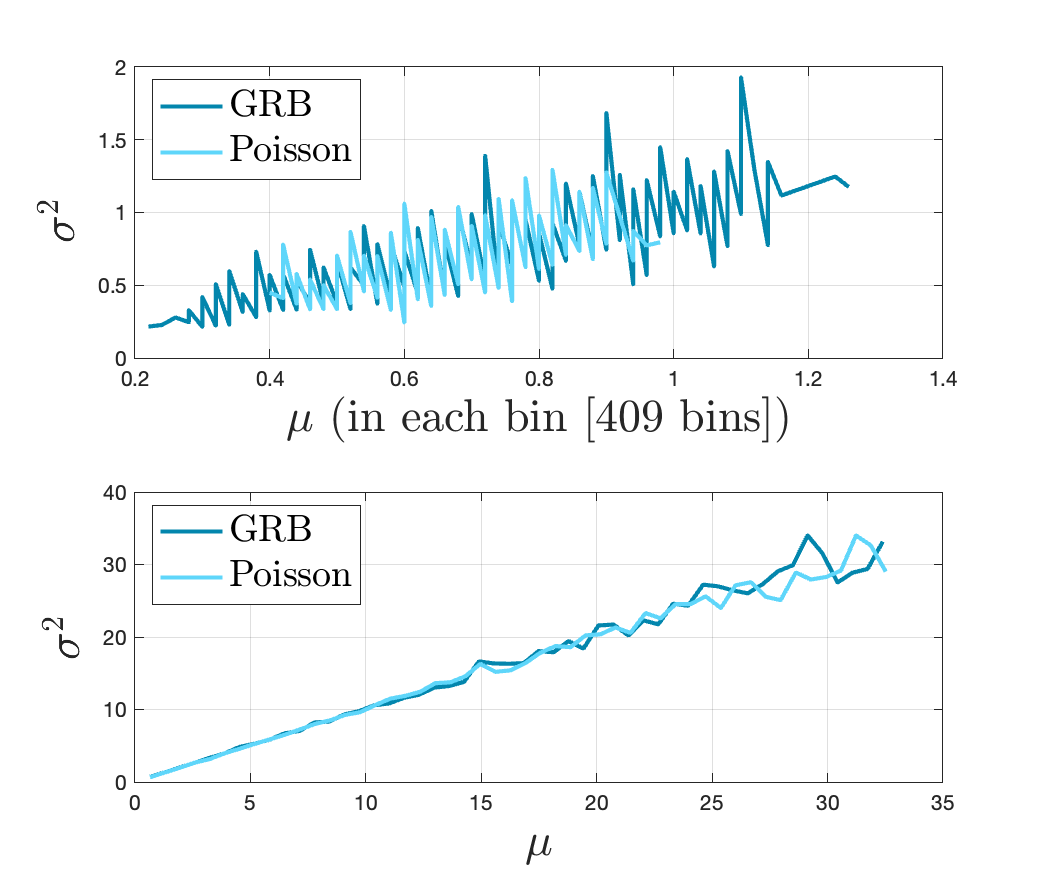} \\
        (c) & (f) \\
        
    \end{longtable}
    \caption{
    (a) Strain data for 4096 seconds around GW190814 from H1 observed during O3. 
    (b) PDF of a frame of GW strain data (4096 seconds) around GW190814 showing a typical Gaussian distribution. 
    (c) Partitioning this frame in 8 second segments, shown are the indices of peak magnitude in strain in each of {these 512 segments}. Cumulatively, they satisfy a uniform distribution in time {with mean count 4 in segments partitioned over 128 bins}. {Bars show the counts in each bin, that follow a Poisson distribution.} 
    (d) 
    Shown is the first 10 seconds of the {\it BeppoSax} event GRB000718B light curve sampled at 2KHz. 
    (e) Histogram of light curve of GRB000718B, satisfying a nearly perfect Poisson distribution. (f) Dark green shows the behavior of GRB000718B alongside a corresponding ideal Poisson distribution (light green) with the same mean. The data nicely follow the Poisson trend {with standard deviation} $\sigma$ and mean $\mu$, {shown here} for interlaced sums of 1 to 50 bins. The upper panel shows the distribution in each bin, while the lower panel shows the behavior of superposition of bins.}
    \label{fig:example}
\end{figure}

\begin{itemize}
\item 
LIGO detector strain noise. This generally is colored {\it Gaussian}, illustrated in Fig. \ref{fig:example} showing the PDF of {4096 seconds} data around the event GW190814.

\item 
{GRB light curves. These are typically represented by a time-series of binned photon counts. This generally satisfies a Poisson distribution with the property that their standard deviation is the square root of mean counts.} 
It results in a trend in standard deviation with brightness of the GRB emission. Fig. \ref{fig:example} shows the statistics of GRB000718B by a histogram of photon counts together with said trend by summation over neighboring bins \citep{Frontera2008}.

\item {
Uniform distributions are common in daily life and astronomy. For instance, in event timing such as the time index of maxima in data segments of given size \citep{Putten2023}. An example of this case is included in Fig. \ref{fig:example}.}

\end{itemize}

Against this background, we consider transient signals that are common in astronomy. 
{Specifically, we include frequency chirps of either long or short duration, that may be ascending or descending in, e.g., compact binary merger and, respectively, spin-down of their compact remnants which may have their bearing on long and short GRBs. Additionally, we consider transients in the form of noise, of the same type as background.} 
{Our focus is on small signals to benchmark sensitivity of a method of correlation according to a threshold of detection shown in Fig. \ref{fig:departure}}. Our default array size is $N=2^{14}$ data points. Significantly smaller arrays may encounter small number statistics, which is outside the scope of this work. 

Specifically, we consider the following combinations {of background and test signals}:
\begin{itemize}
\item 
    {\it Background}: Gaussian, Poisson, Uniform
\item 
    {\it Transient signals}: Gaussian, Poisson, Uniform, Chirp, Sine 
\end{itemize}
For each of these $15 = 3\times 5$ combinations of background and signal, we compute the response curve to  signal {with} amplitude $\alpha$, and determine the thresholds $DT_{\mu}$ and $DT_{\sigma}$ {defined by the smallest signal amplitude at which the mean and respectively, $1\sigma$ of the response curve departs from that of the randomized curve (\S \ref{Sec:Intro}; and Fig. \ref{fig:departure}).}

\begin{figure}[ht!] 
\hspace{-2cm}
 \begin{tabular}
      {wc{20mm}|wl{45mm} wl{45mm} wl{45mm}} \hline  & \hspace{1cm} Gaussian  & \hspace{1cm} Poisson & \hspace{1cm} Uniform  \\ 
      
      \hline Gaussian & \parbox[c]{0em}{
      \includegraphics[width=2in]{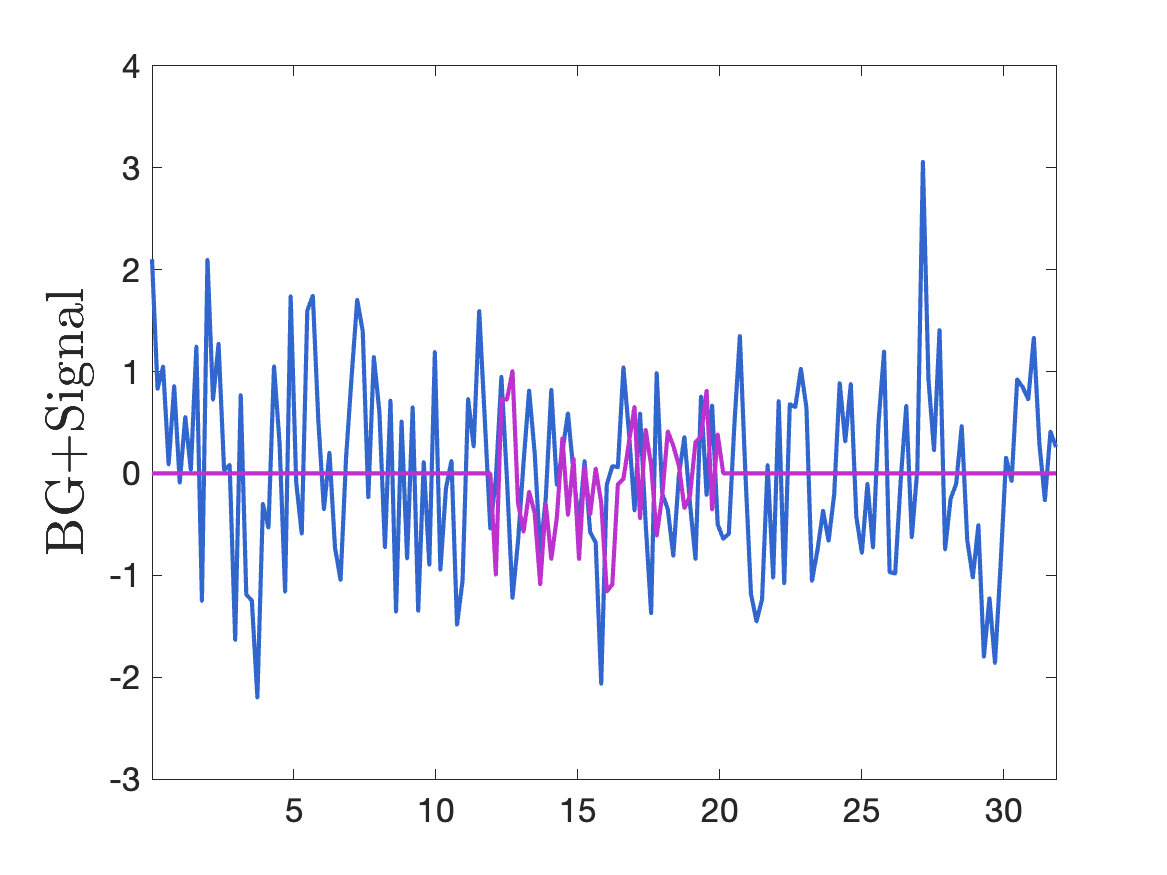}} & \parbox[c]{0em}{
      \includegraphics[width=2in]{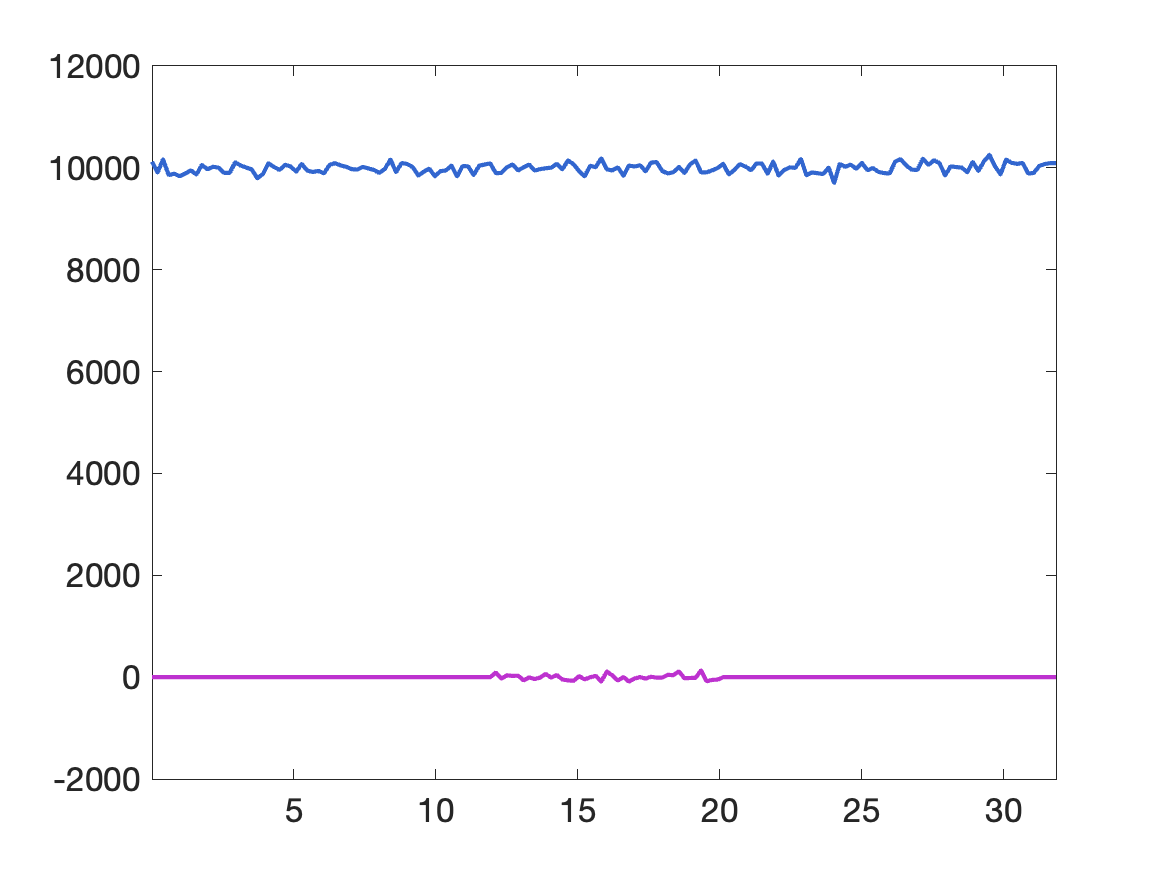}} & \parbox[c]{0em}{
      \includegraphics[width=2in]{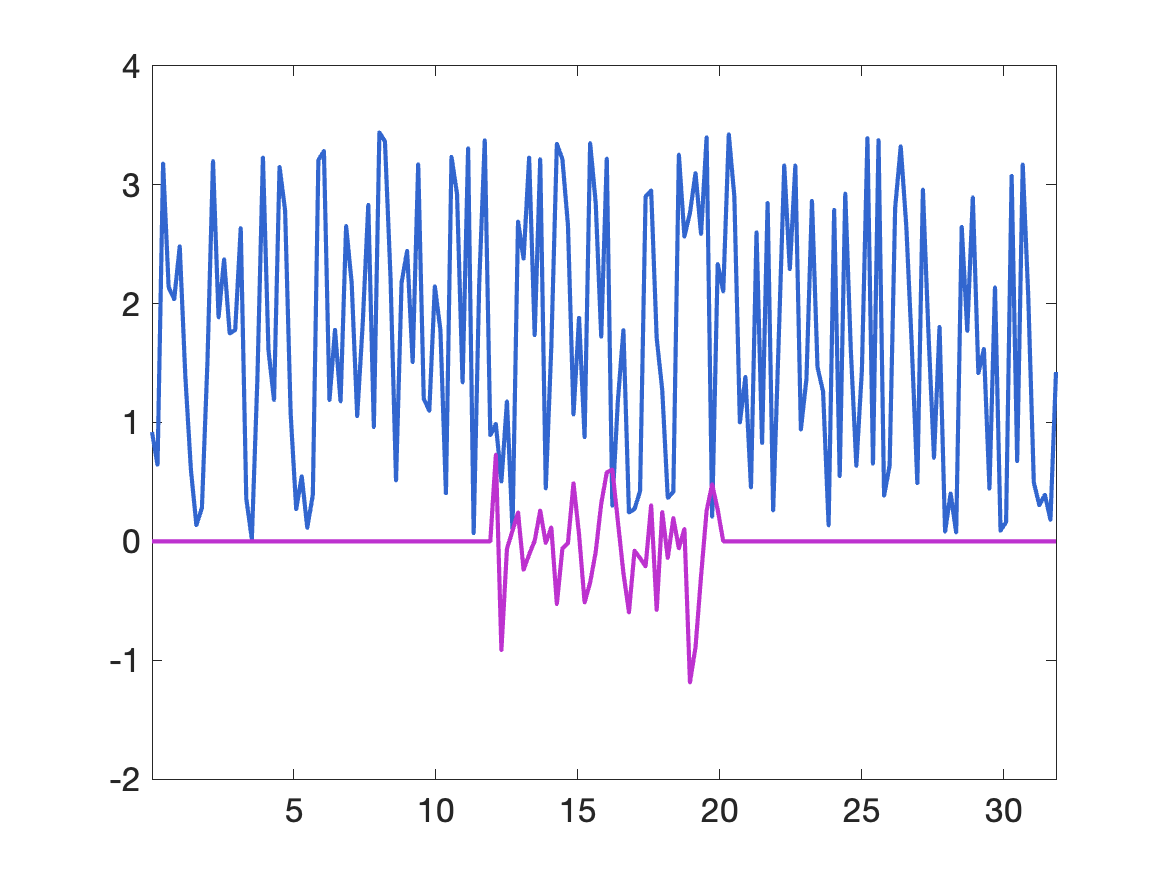}}\\
      \hline
      Poisson &\parbox[c]{0em}{
      \includegraphics[width=2in]{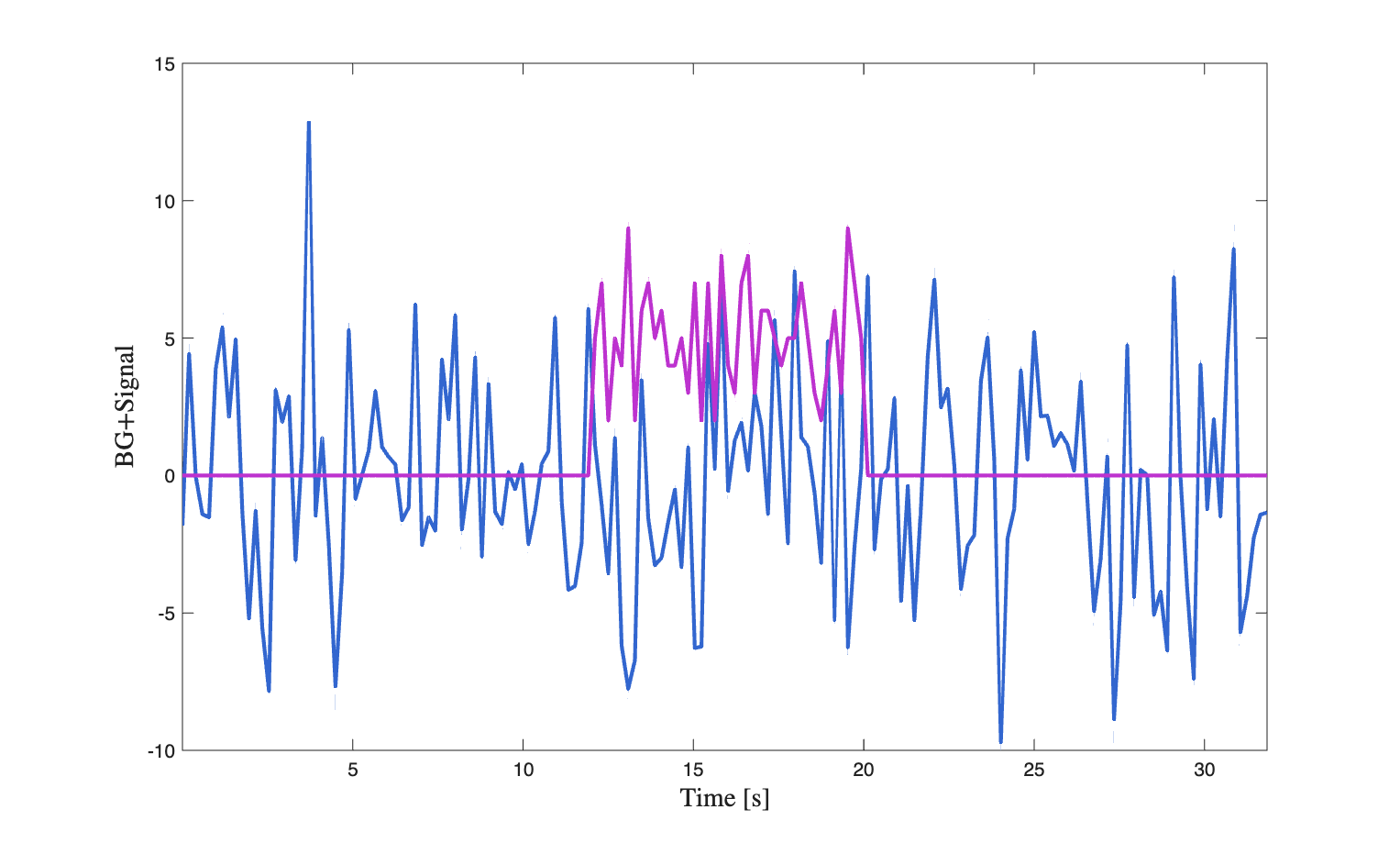}} & \parbox[c]{0em}{
      \includegraphics[width=2in]{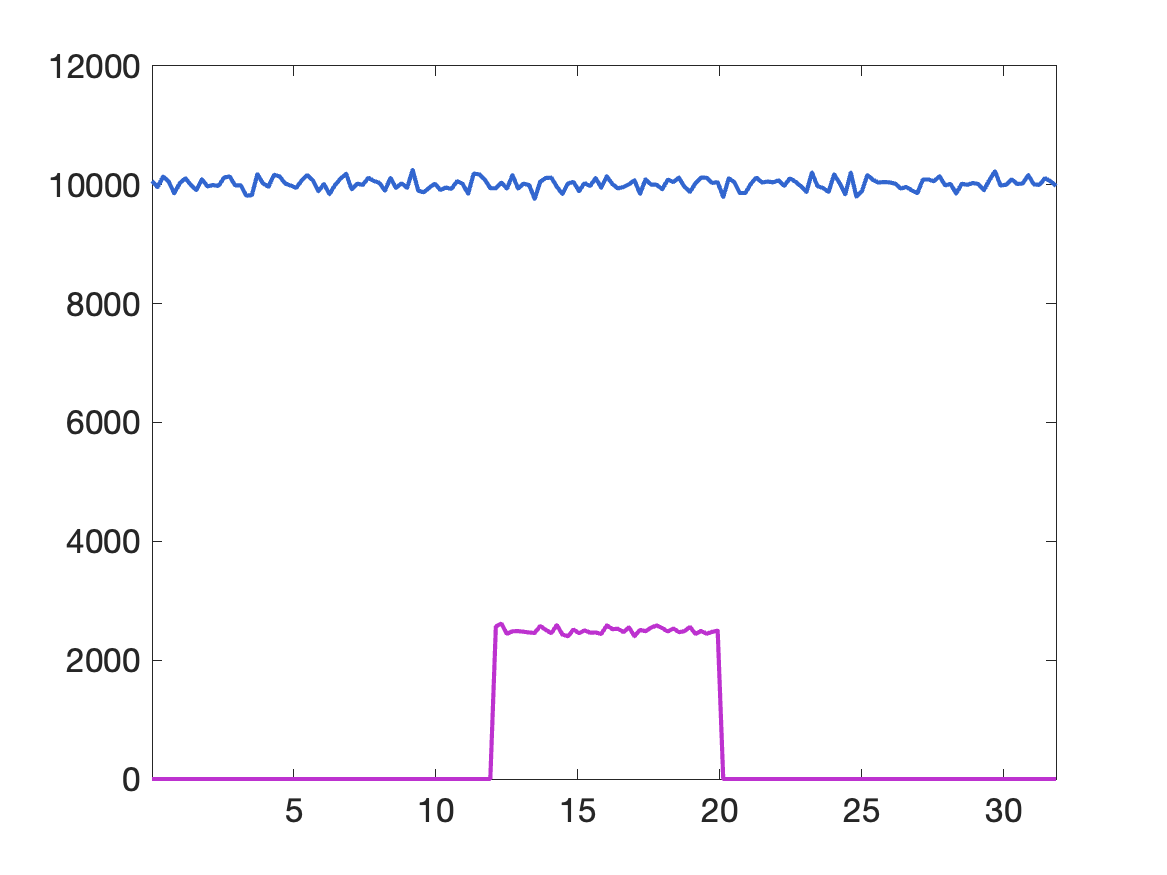}} & \parbox[c]{0em}{
      \includegraphics[width=2in]{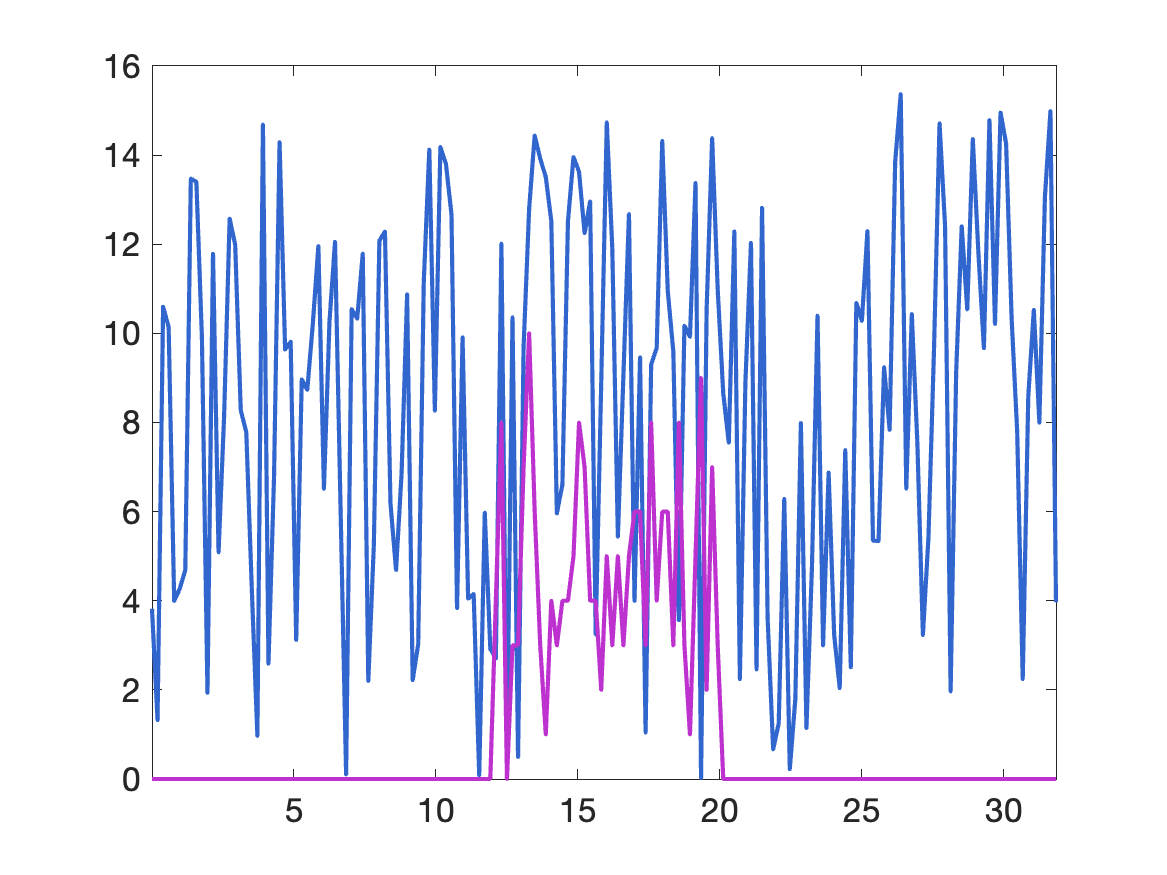}}\\
      \hline
      Uniform &  \parbox[c]{0em}{
      \includegraphics[width=2in]{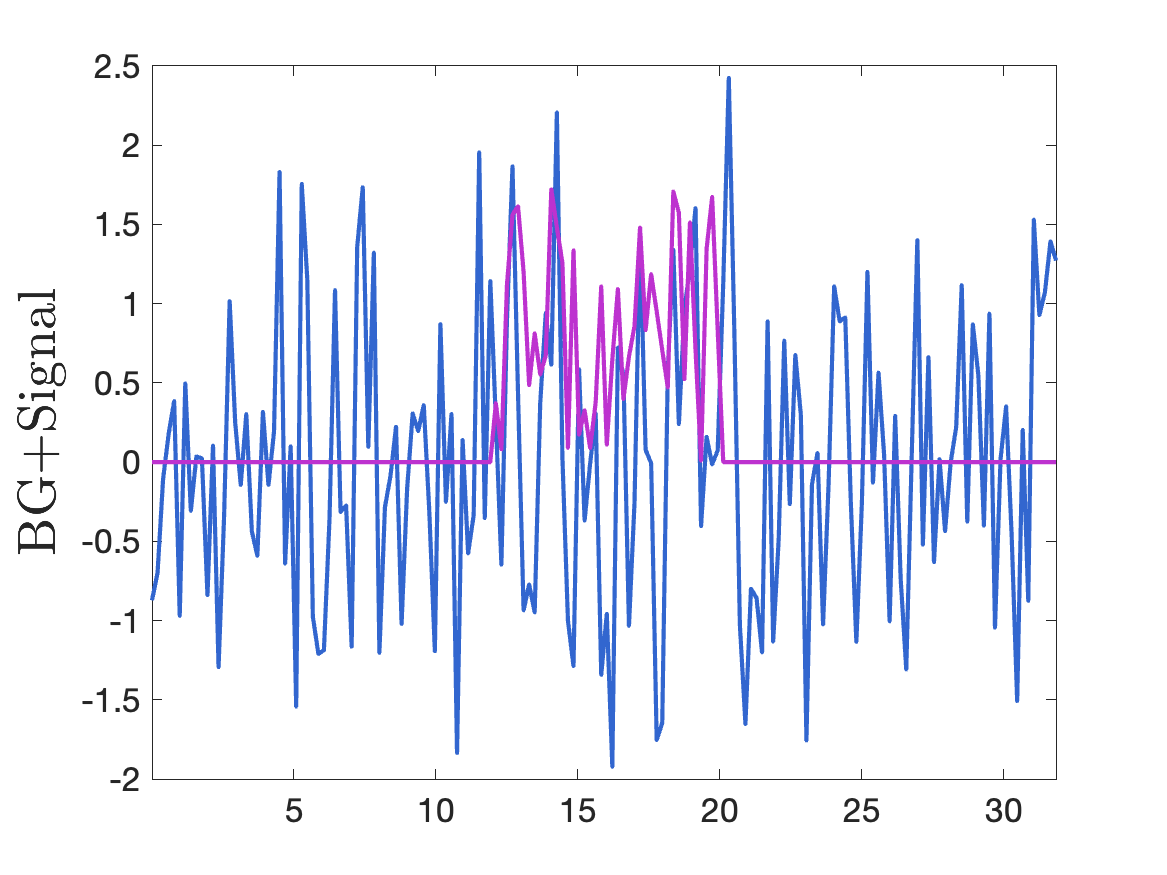}} & \parbox[c]{0em}{
      \includegraphics[width=2in]{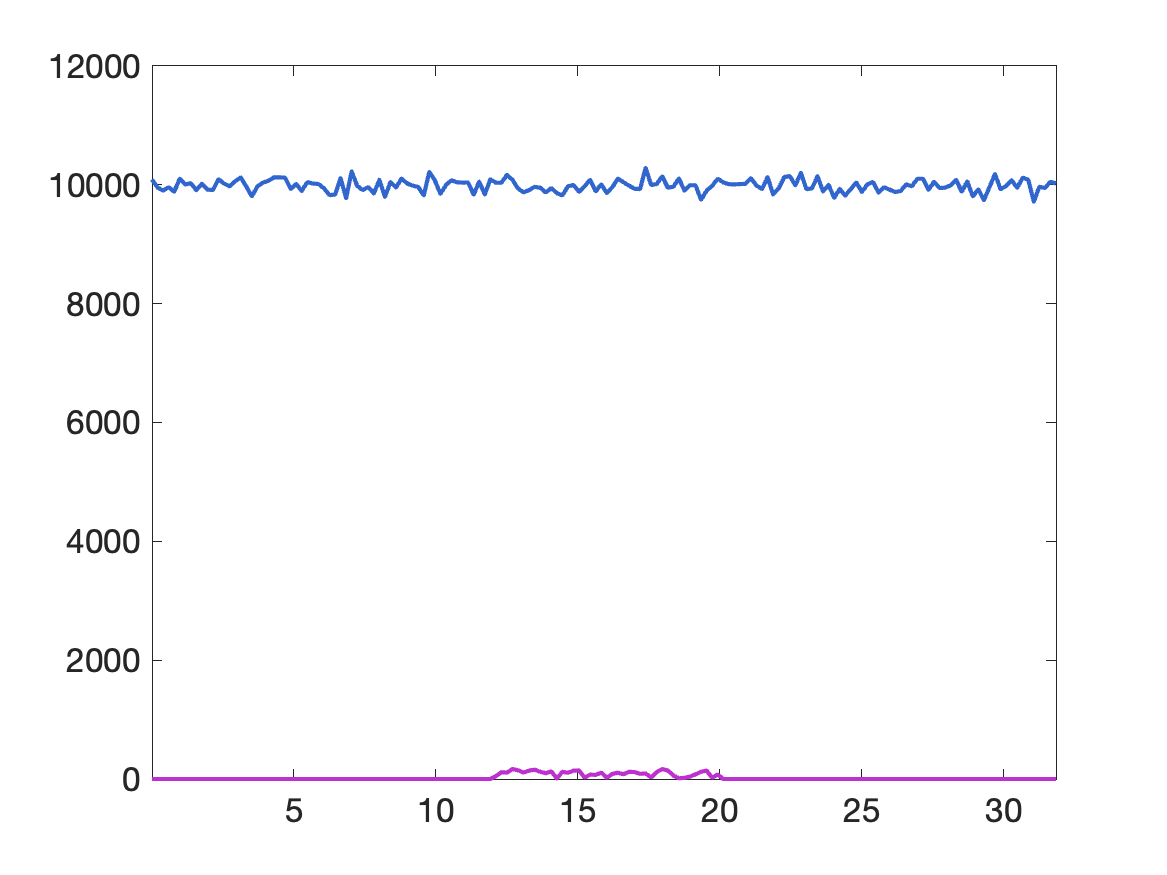}}& \parbox[c]{0em}{
      \includegraphics[width=2in]{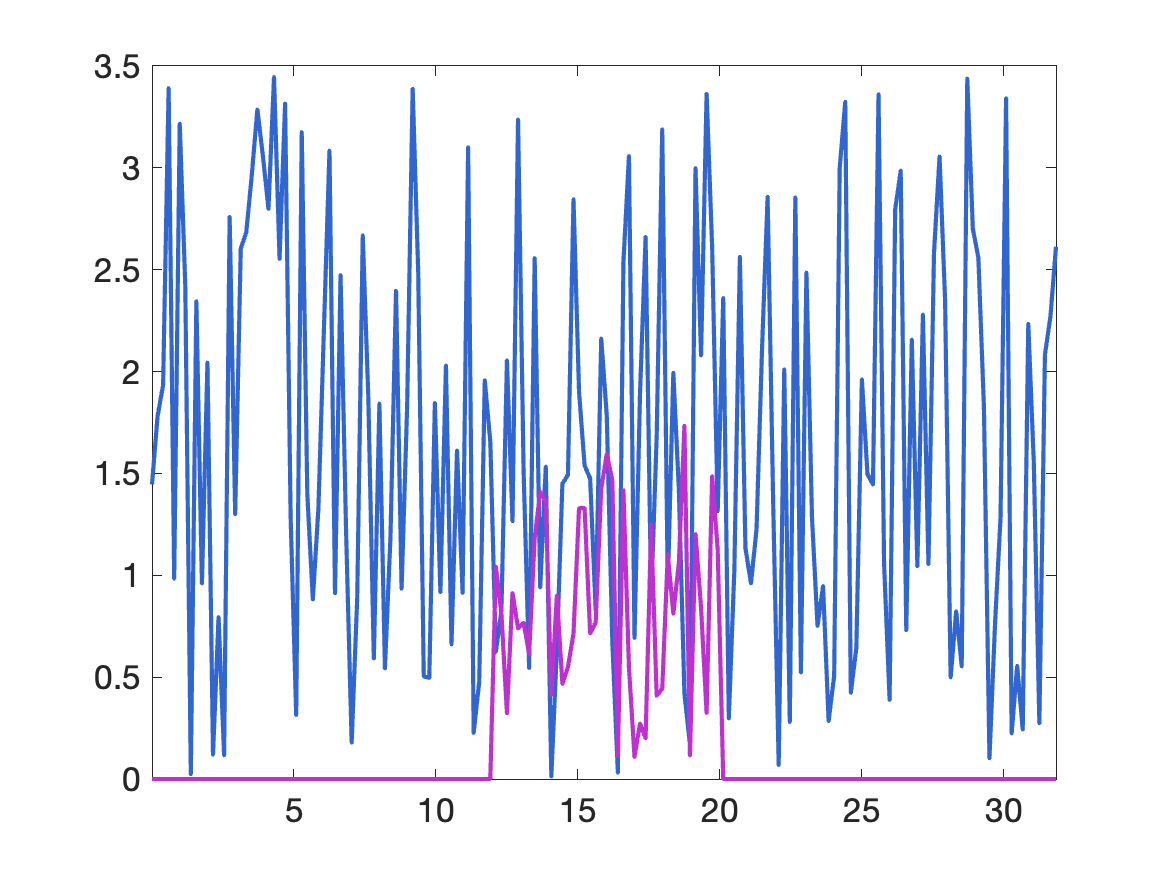}}\\
      \hline
      Chirp & \parbox[c]{0em}{
      \includegraphics[width=2in]{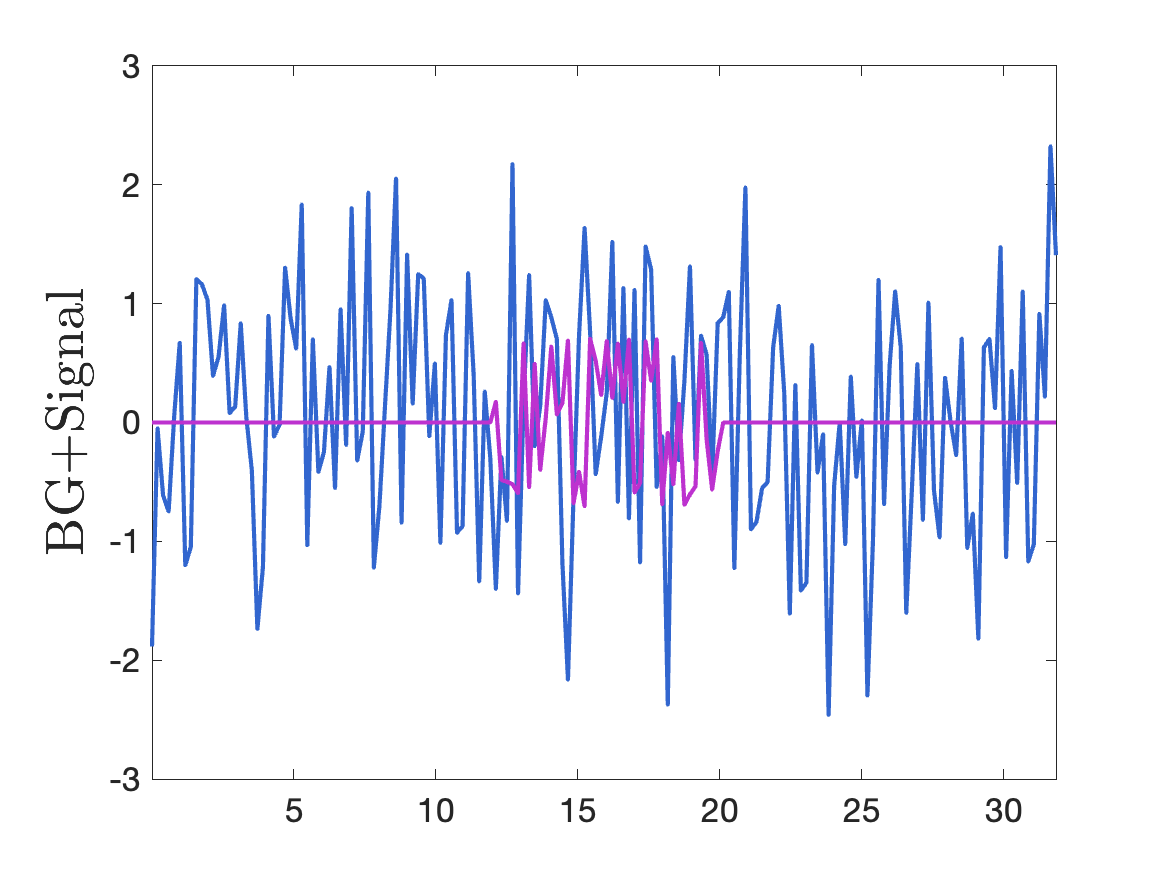}} & \parbox[c]{0cm}{
      \includegraphics[width=2in]{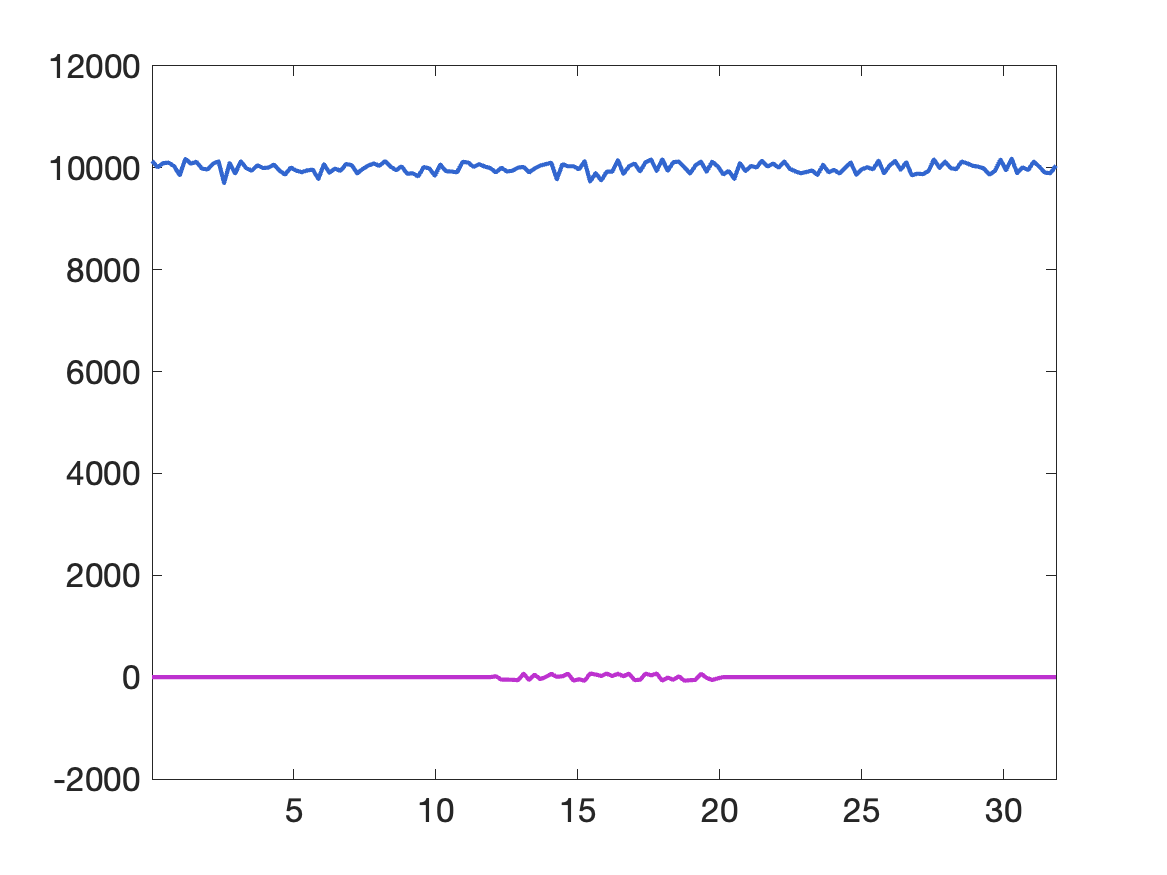}} & \parbox[c]{0cm}{
      \includegraphics[width=2in]{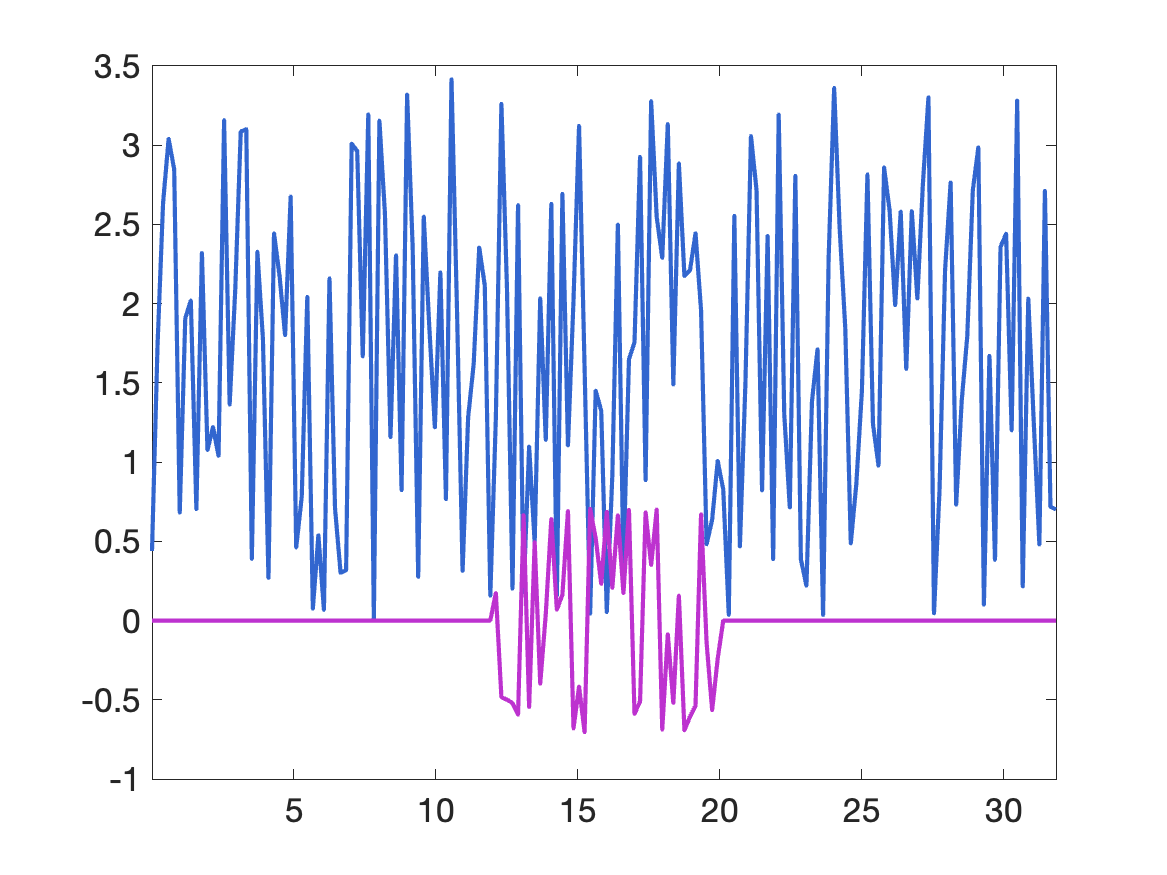}} \\
      \hline
      Sine&  \parbox[c]{0em}{
      \includegraphics[width=2in]{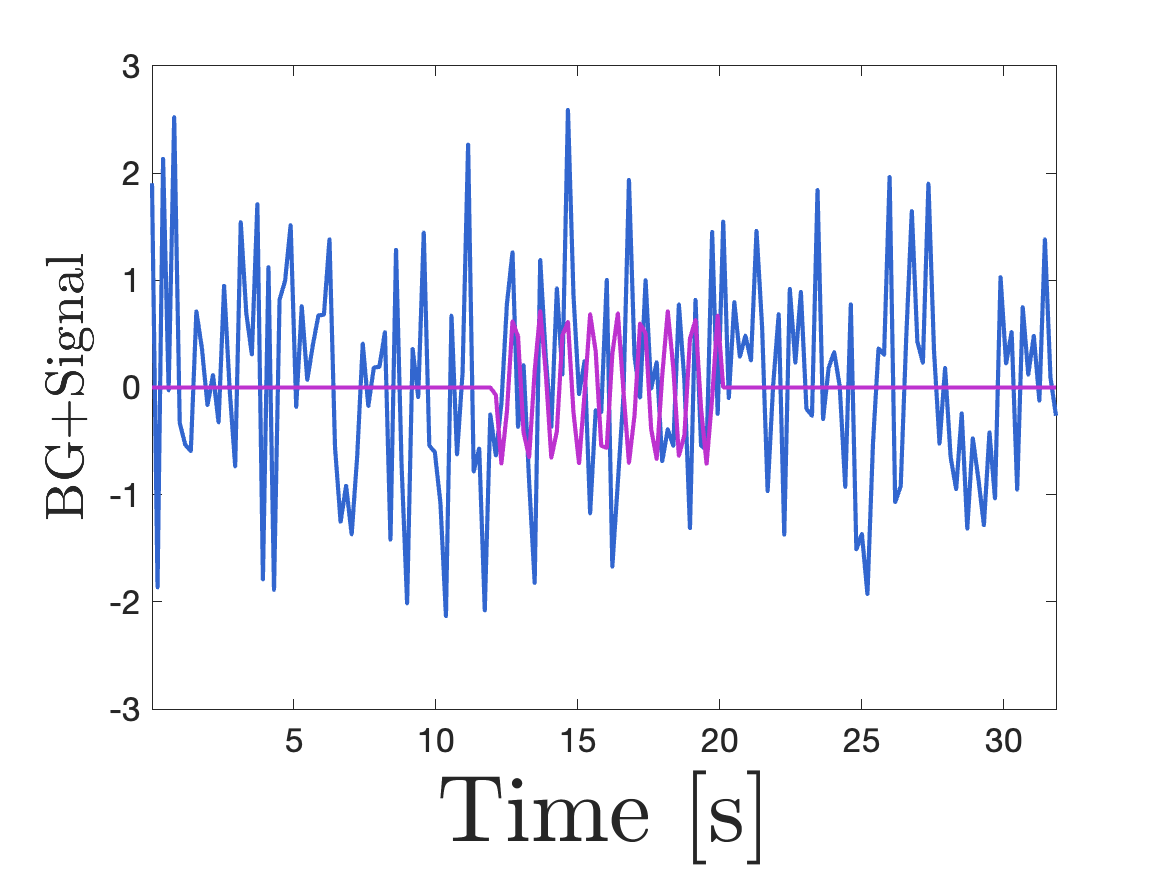}} & \parbox[c]{0em}{
      \includegraphics[width=2in]{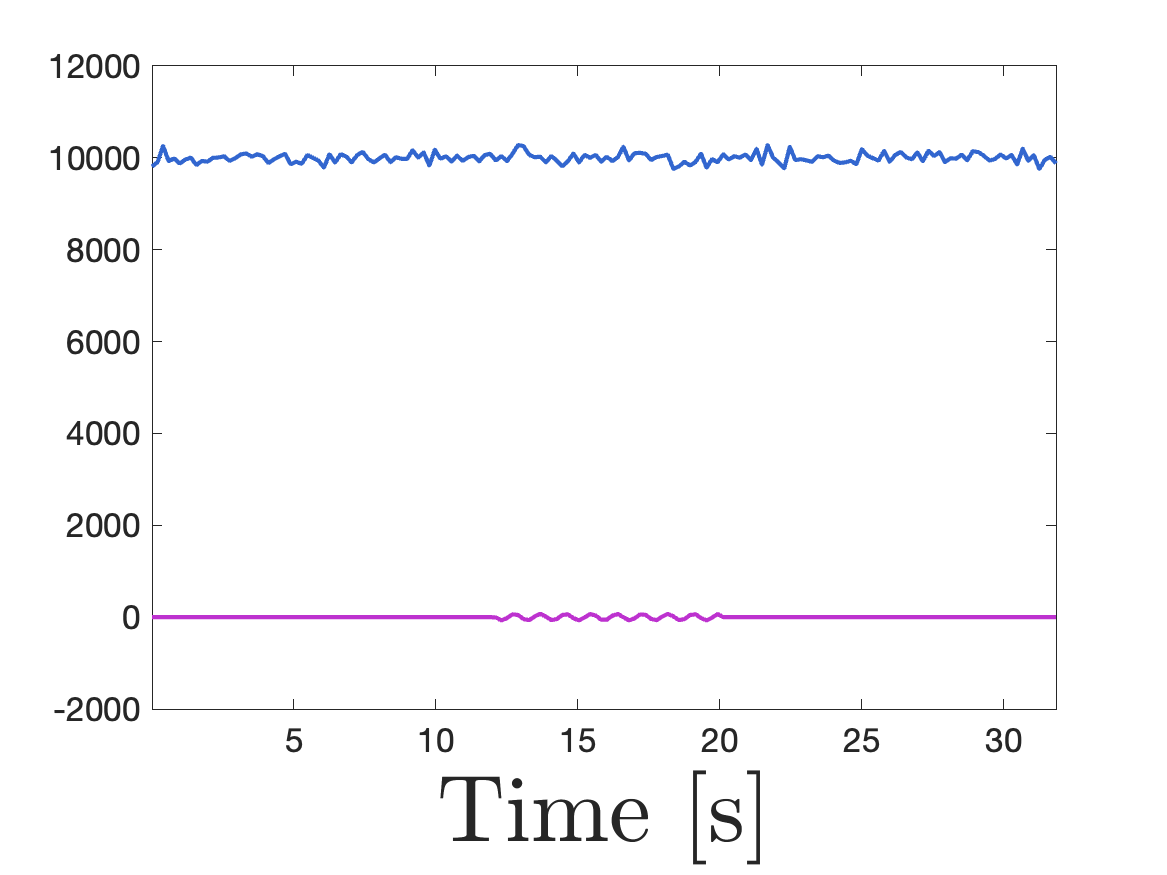}} & \parbox[c]{0em}{
      \includegraphics[width=2in]{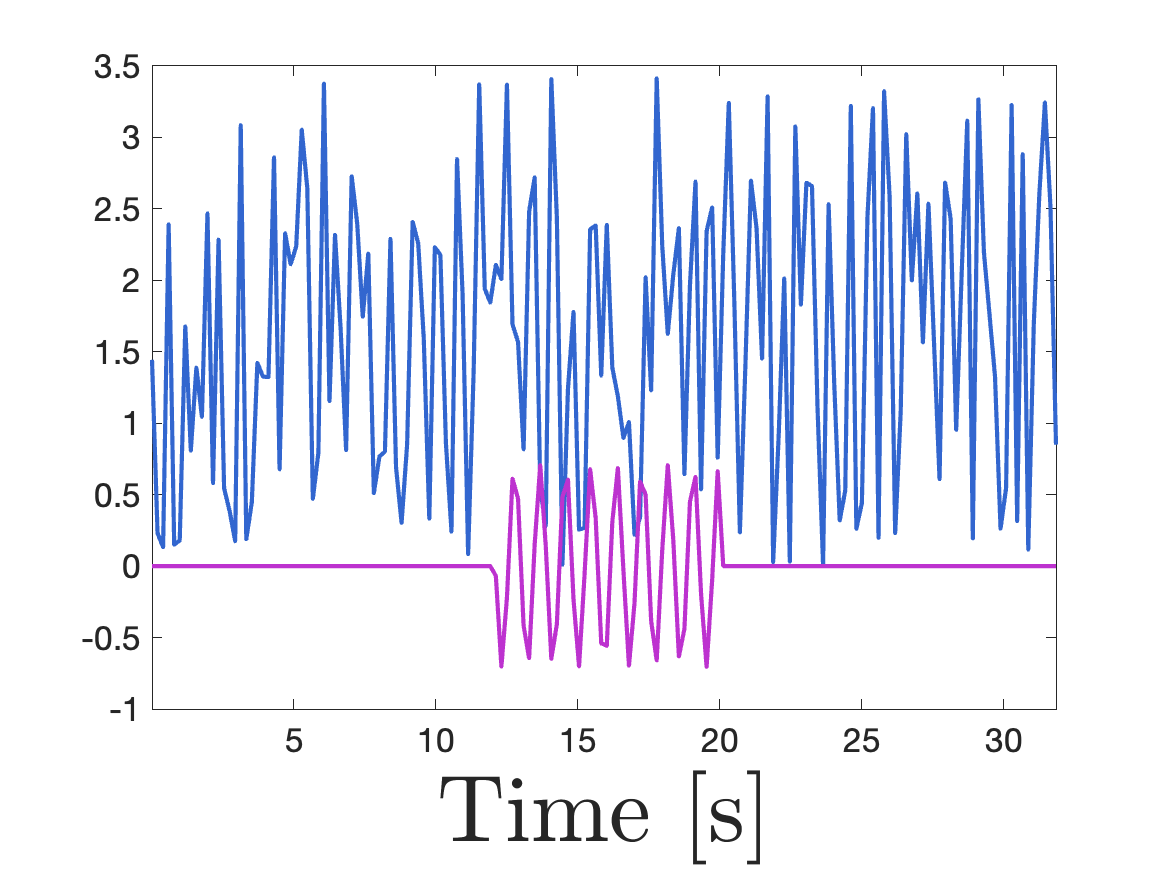}} \\
      \hline
  \end{tabular}
  \caption{Background (horizontal) and signal (vertical) combinations based on \S \ref{Sec:combinations} with Poisson parameters $\lambda=10000$ and $\lambda_s$=5 and signal strength \eqref{EQN_alpha} $\alpha=0.5$.}
    \label{T7}
\end{figure}

In the set-up of this benchmark {with control} \eqref{EQN_alpha}, we first prepare background and signal to have equal standard deviations. Their ratio is next adjusted by a choice of $\alpha$.}
This is done by a normalization to their standard deviations followed by a multiplication of signal to arrive at the desired $\alpha$. This procedure readily applies to all cases except those involving Poisson background or signal. {When Poisson statistics is involved, we turn to the following.}

\subsection{The case of Poisson background or signal}

The Poisson distribution is defined by integer valued data in $\mathbb{N}$ with PDF 
\begin{equation}
    \label{EQN_Poisson}
    f(k)=\dfrac{\lambda ^k e^{-\lambda}}{k!},
\end{equation}
where $k$ is the count and $\lambda$ is a real positive number, equal to the mean value of the distribution. It has the distinct property of having a standard deviation $\sigma=\sqrt{\lambda}$. This property prevents normalization of $\sigma$ independent of mean, otherwise available for the other distributions.
Poisson distribution approaches a Gaussian distribution for $\lambda\gg1$, where the skewness of Poisson distribution is negligible. 

\subsubsection{Poisson background and signal}

A Poisson background with a fixed $\lambda=\lambda_{bg}$ has a standard deviation $\sqrt{\lambda_{bg}}$, that cannot be independently normalized.
For a given signal in \eqref{data}, we  control signal strength according to \eqref{EQN_alpha} in the following steps keeping background fixed. 
For signals other than Poisson, the signal is first prepared to have the same standard deviations as background. 
Subsequently, we multiply by $\alpha$ {according to} its definition in \eqref{EQN_alpha}:
  
\begin{equation}
\label{EQN_mf1}
{\rm Raw} ({\rm signal, background}) \rightarrow (\alpha {\rm S}, {\rm BG}),
\end{equation}
where S and BG stand for {signal and background, respectively.}

When the signal {is} also Poisson, however, normalization is prohibited for both signal and background. To consider a range of signal strength (\ref{EQN_alpha}), we create signals over a range of $\lambda_s$ using the  relation $\sigma_s = \sqrt{\lambda_s}$, and consider

\begin{equation}
    \label{EQN_mf2}
    \alpha=\dfrac{\sqrt{\lambda_s}}{\sqrt{\lambda_{bg}}}.
\end{equation}
This makes sure that signal strength follows our definition \eqref{EQN_alpha}.

\subsubsection{Poisson signal and non-Poisson background}

When a Poisson signal is added to a background different from Poisson, similar considerations as above apply.
In this case, we are at liberty to normalize background to arrive at a desired (\ref{EQN_alpha}). 
As above, we first equate standard deviations of background and signal by normalizing background data, followed by {\it dividing} it by $\alpha$ to satisfy \eqref{EQN_alpha}:

\begin{equation}
\label{EQN_mf3}
{\rm Raw} ({\rm  signal,\  background}) \rightarrow ({\rm S}, \alpha^{-1} {\rm BG}).
\end{equation}

{Fig. \ref{T7} shows} the background and signal combinations mentioned in \S \ref{Sec:combinations}.
While large $\lambda$ reduces the case of Poisson to Gaussian, we include the case of $\lambda_s=5$ to cover a  
a genuine case of Poisson noise.
For illustrative purposes, spectrograms of these combinations are included in Appendix \ref{Sec:AppA}. 

\subsection{Benchmark procedure}\label{Sec:procedure}

Our benchmark comparison of DCC, PC and EPR (\S2-3) in two-channel signal processing (Fig. \ref{fig:Process}) starts with \eqref{data}: simulated background with signal injections accompanied by time-randomized data by random {index permutation. The detection} threshold $DT_\mu$ (Fig. \ref{fig:departure}) is determined by response curves of correlations as a function of signal strength $\alpha$. As mentioned in \S \ref{Sec:dcc}, DCC outputs an array ${\bf R}$, whose peak value is representative for the correlation strength between the two input arrays. This output carries a baseline in the mean and standard deviation $\sigma_d$.
Alternatively, one may define a  baseline by {maximum} of the output $R$ following time-randomization of the input data $({\bf p,q})$. In this way, \eqref{EQN_dcc_max} and $\sigma_d$ give the cyan curve in Fig. \ref{fig:departure} with $\pm1\sigma_d$ band for DCC. 
The same for randomized {\bf p} and {\bf q} produces the gray band and curve. 

However, the procedure to calculate the $1\sigma_d$ band for PC and EPR is different, yet similar for both according to \eqref{EQN_Pearson} and \eqref{EQN_EPR}.
In both cases, the output is a scalar rather than a vector. This suggests the need for time slides or, equivalently, iterations by Monte Carlo simulations. 
For the latter, we perform a large number of iterations for each $\alpha$, thus obtaining standard deviations in the response for each $\alpha$ and the associated mean value. 
The resulting response curve - in mean and width - is {equivalent to } the cyan band and cyan curve shown in Fig. \ref{fig:departure}, for both PC and EPR.
Following randomized index permutations of {\bf p} and {\bf q}, the same procedure produces the gray band and gray curve as the baseline.
{The same results can be obtained using}, instead, time slides applied to either one of these two input arrays over all strides of size $N_{ts}<N$.
To be effective, however, $N$ needs to be relatively large to mimic time randomization.

\begin{figure}
\centering{
\includegraphics[scale=0.24]{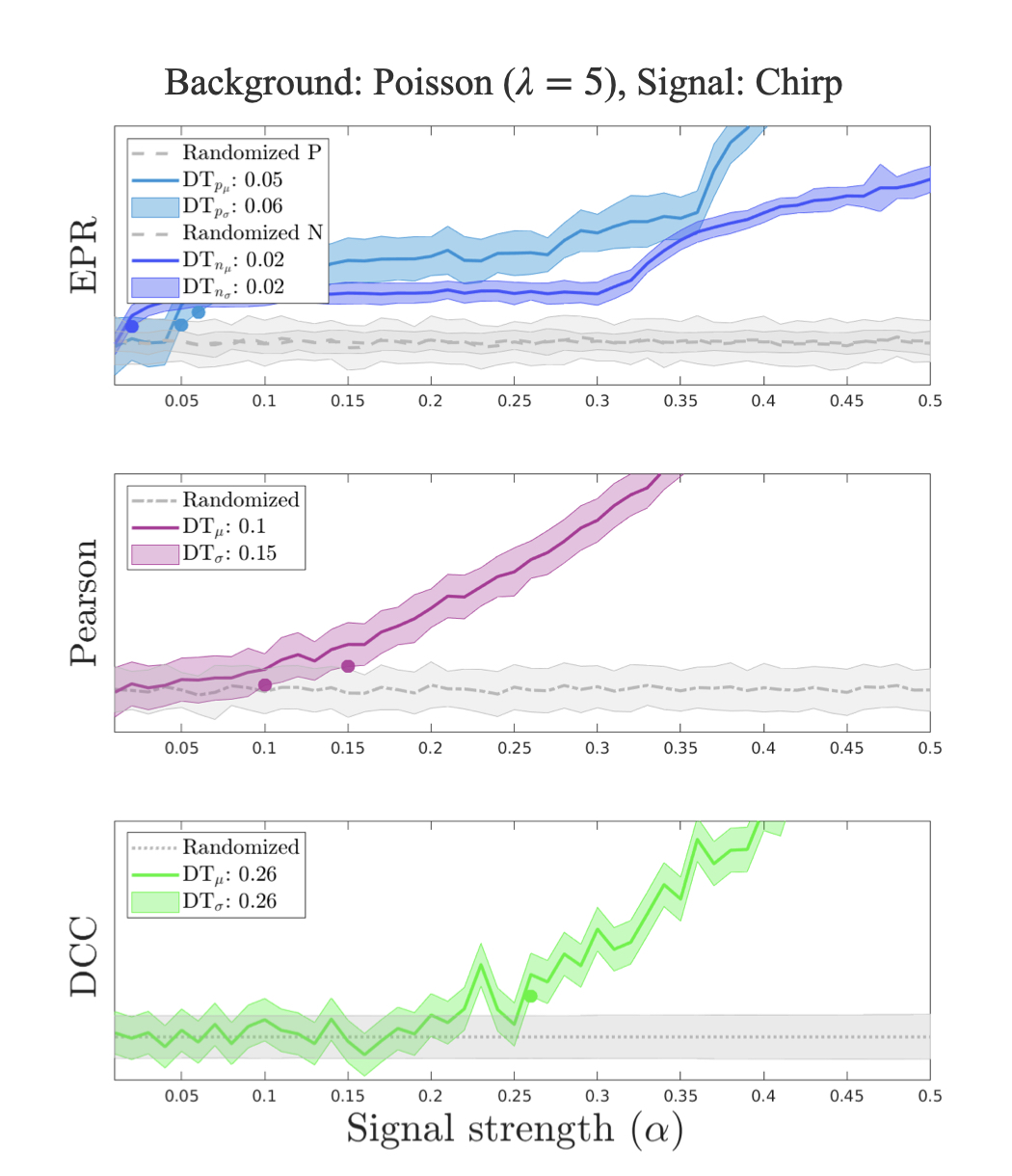}
\includegraphics[scale=0.24]{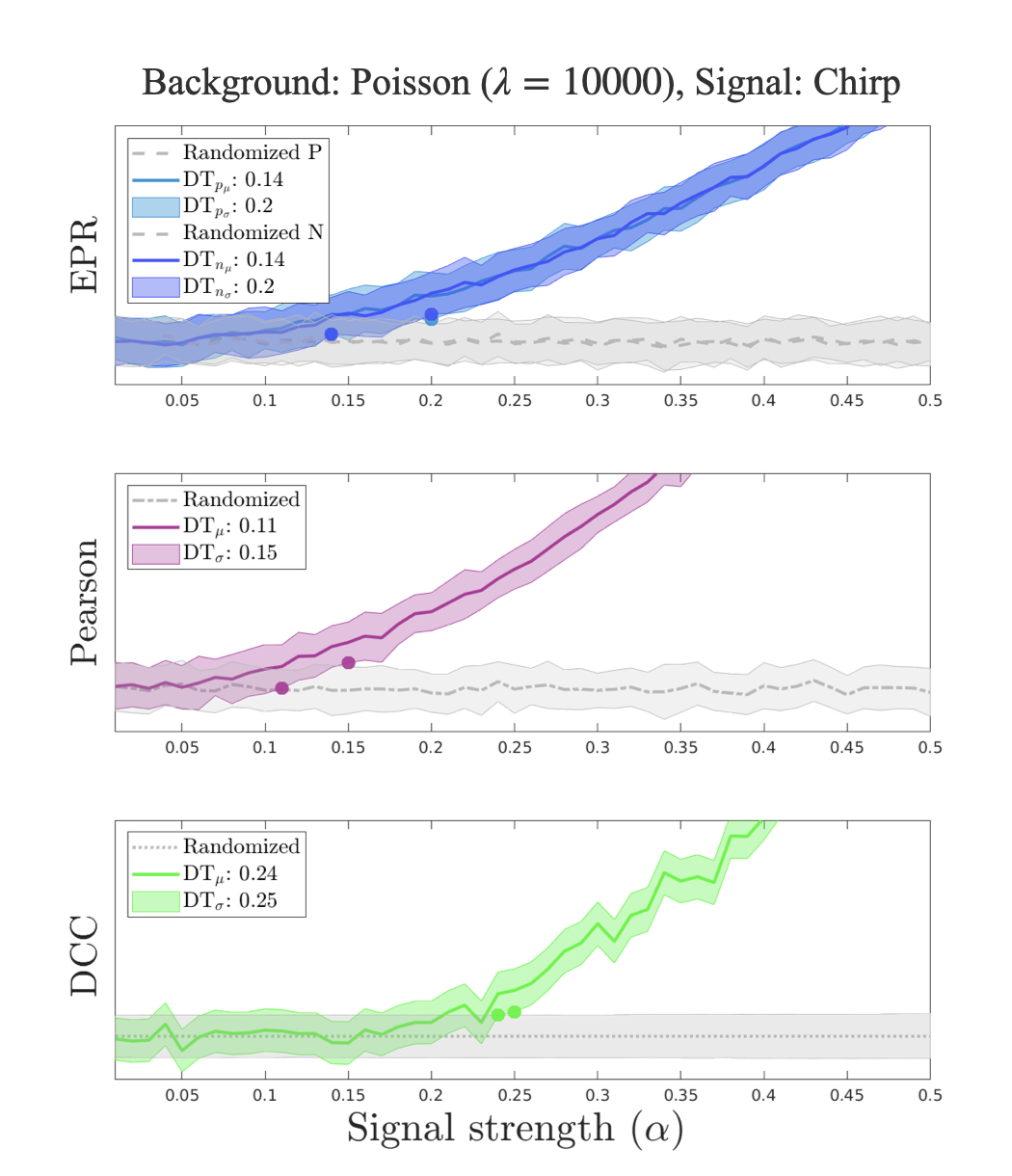}}
\centering{
\includegraphics[scale=0.24]{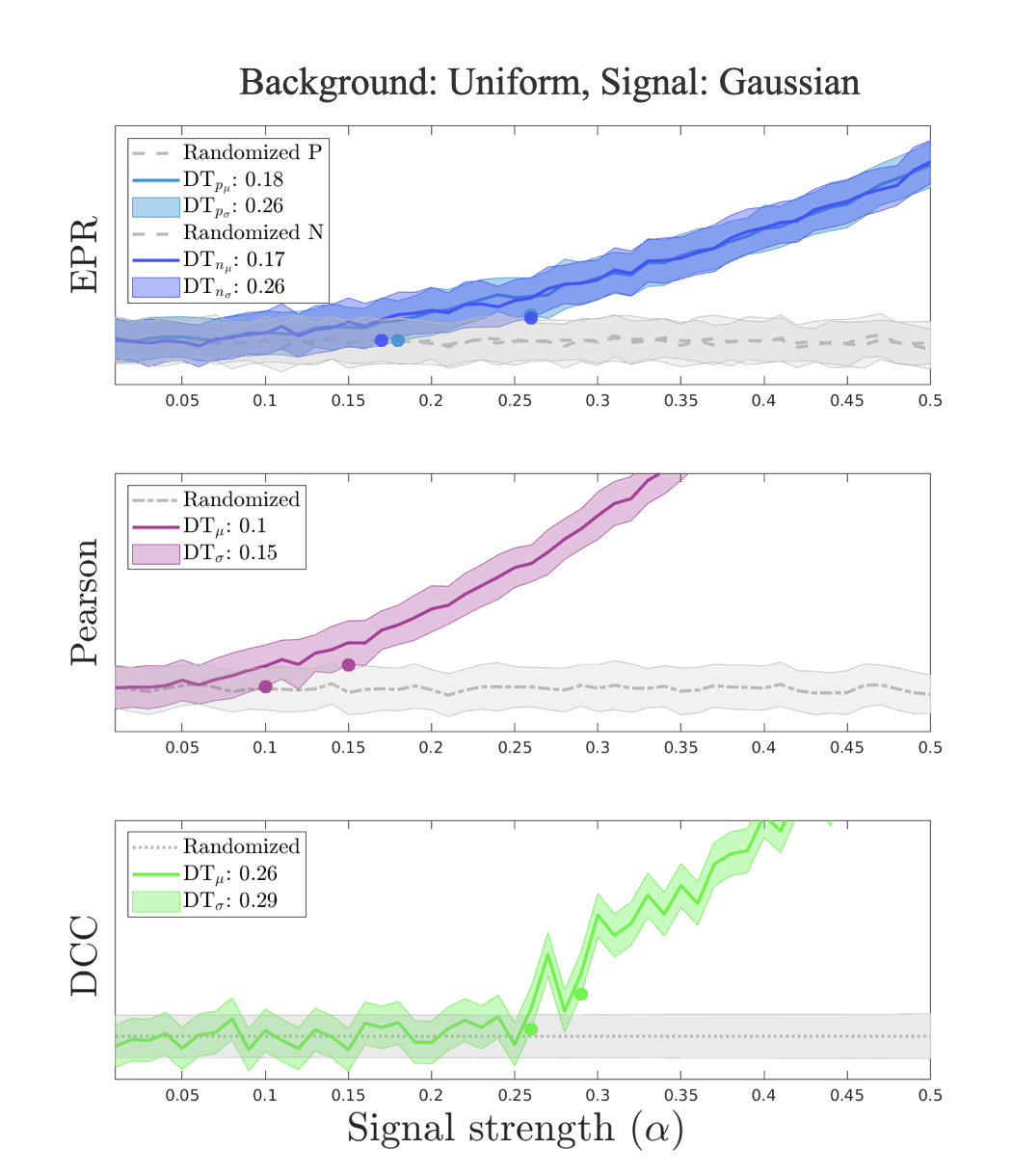}
\includegraphics[scale=0.24]{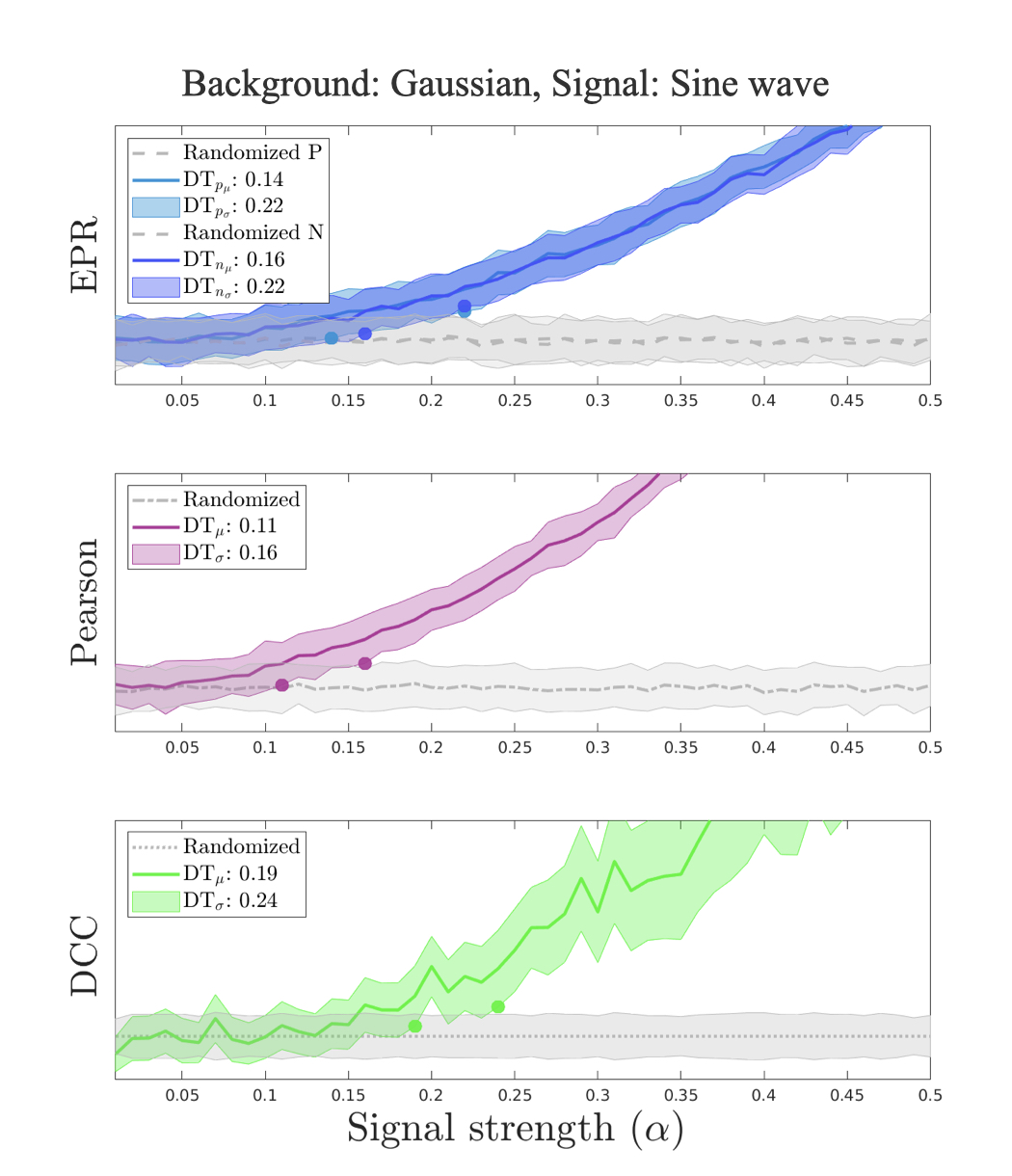}}
\caption{
     Sample of response curves {and detection thresholds} for EPR \eqref{EQN_EPR}, PC \eqref{EQN_Pearson} and DCC \eqref{EQN_dcc_max}. (Left top panel.) {Poisson background with $\lambda=5$ and a chirp like signal injected to the two channels. EPR is applied with $\kappa =0.5$ in (\ref{EQN_kappa})}. 
     For EPR, shown are two response curves following a split of real-valued data {into positive (``P") and negative (``N") branches } for conversion to Boolean data, together with their time-randomized version. (Right top panel.) The near-Gaussian case of Poisson $\lambda=10000$ with otherwise the same chirp signal. (Bottom left.) Background noise and signal are now respectively drawn from uniform and Gaussian distributions preserving the same $\kappa$ as in the upper panels. (Bottom right.) The same now for Gaussian and Sine wave respectively as background and signal. }
    \label{fig:DT}
\end{figure}

EPR (\S \ref{Sec:EPR_def}) following (\ref{EQN_kappa})
may require an additional step when detector output data are signed, extending over positive and negative values. 
This may be approached as follows.
    
\begin{enumerate} 
\item Convert detector output data $({\bf p,q})$ to absolute values before applying (\ref{EQN_kappa}).
This may apply to data that are symmetrically distributed about zero. In case of an offset, however, the response will be a superposition of unequal responses to positive and negative data, unless this offset is filtered out first. 
In this approach also there is a small chance of having excess correlation or de-correlation.

\item Split detector output into positive (P) and negative (N) values,  before applying (\ref{EQN_kappa}). The response now bifurcates into two branches, P and N, that are generally distinct unless the detector data are distributed symmetrically about zero. 
\end{enumerate}

In this work, we shall use the second alternative, that appears preferred for Poisson noise combined with time-harmonic or Gaussian signals.
Fig. \ref{fig:DT} presents a representative sample of response curves in these benchmarks of DCC, PC and EPR.

\begin{figure}
\centering{
\includegraphics[scale=0.24]{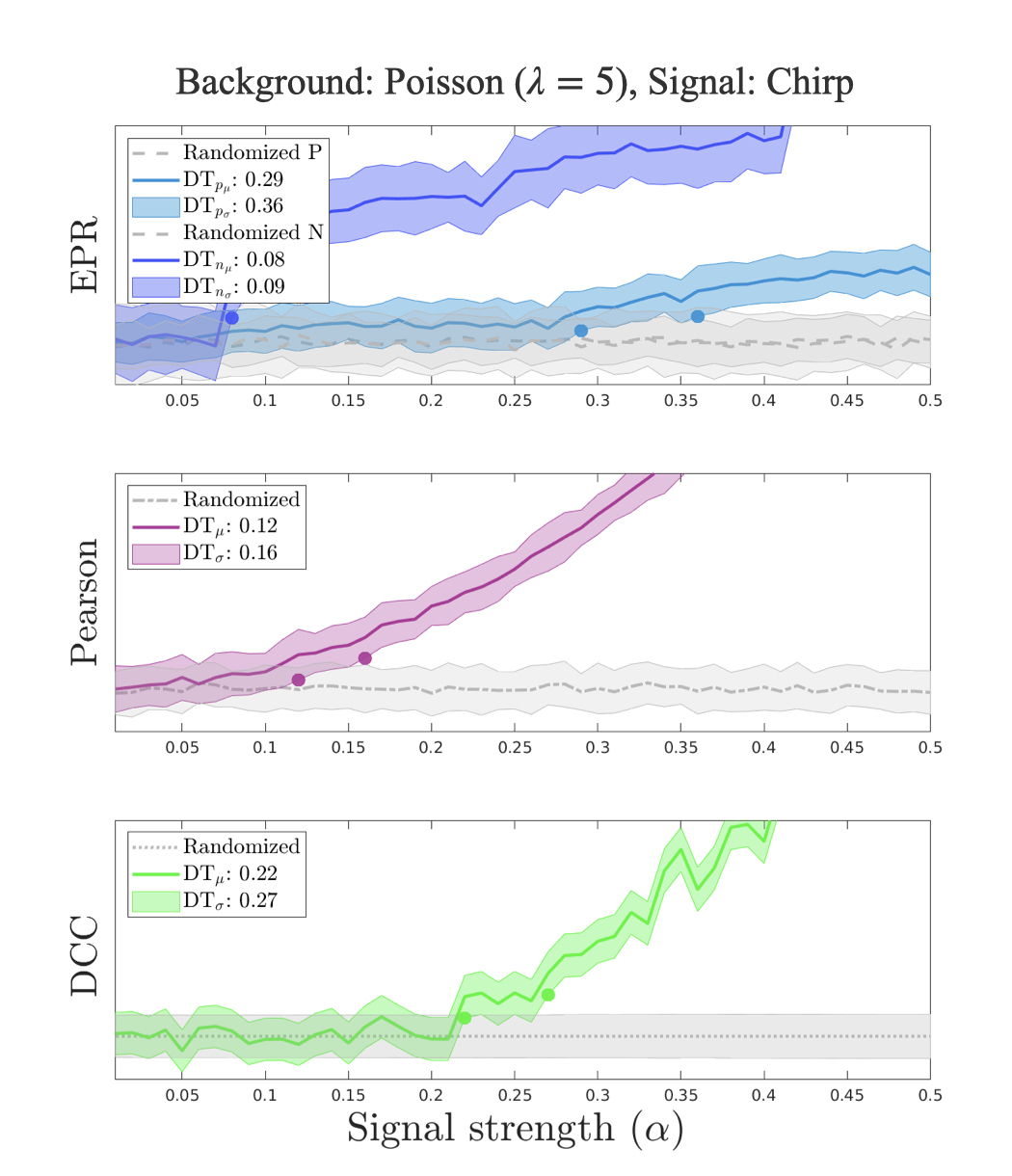}
\includegraphics[scale=0.24]{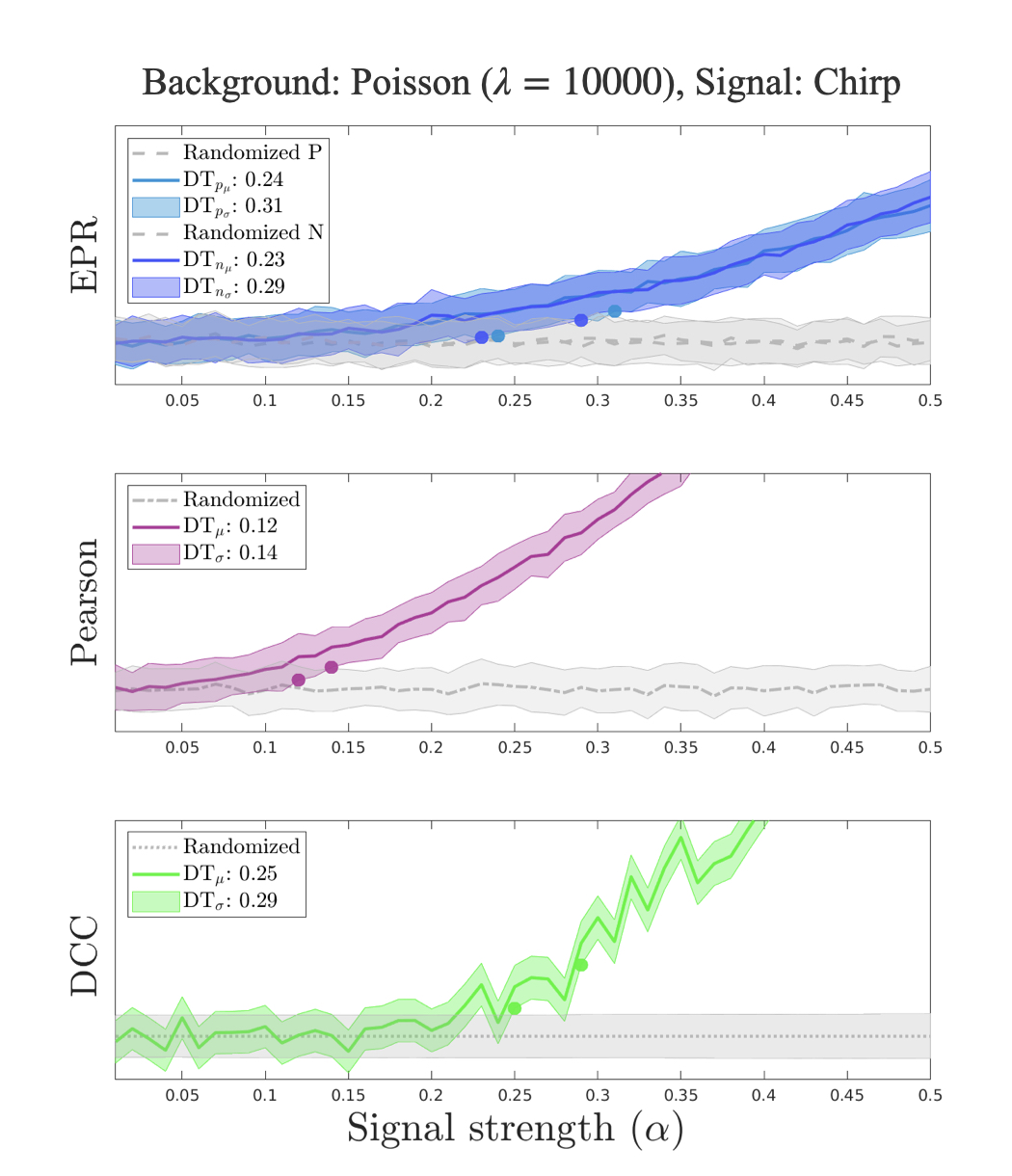}}
\centering{
\includegraphics[scale=0.24]{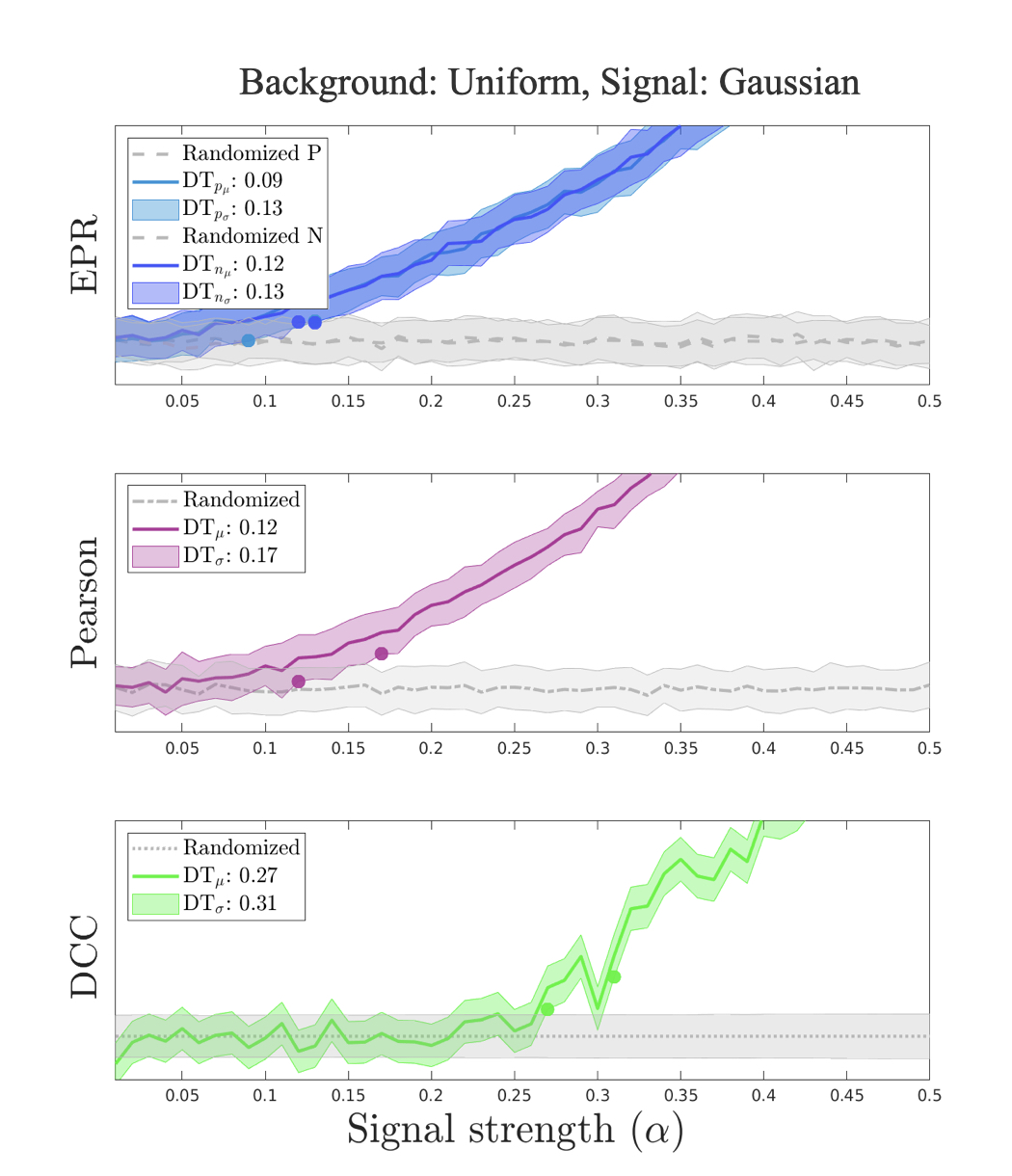}
\includegraphics[scale=0.24]{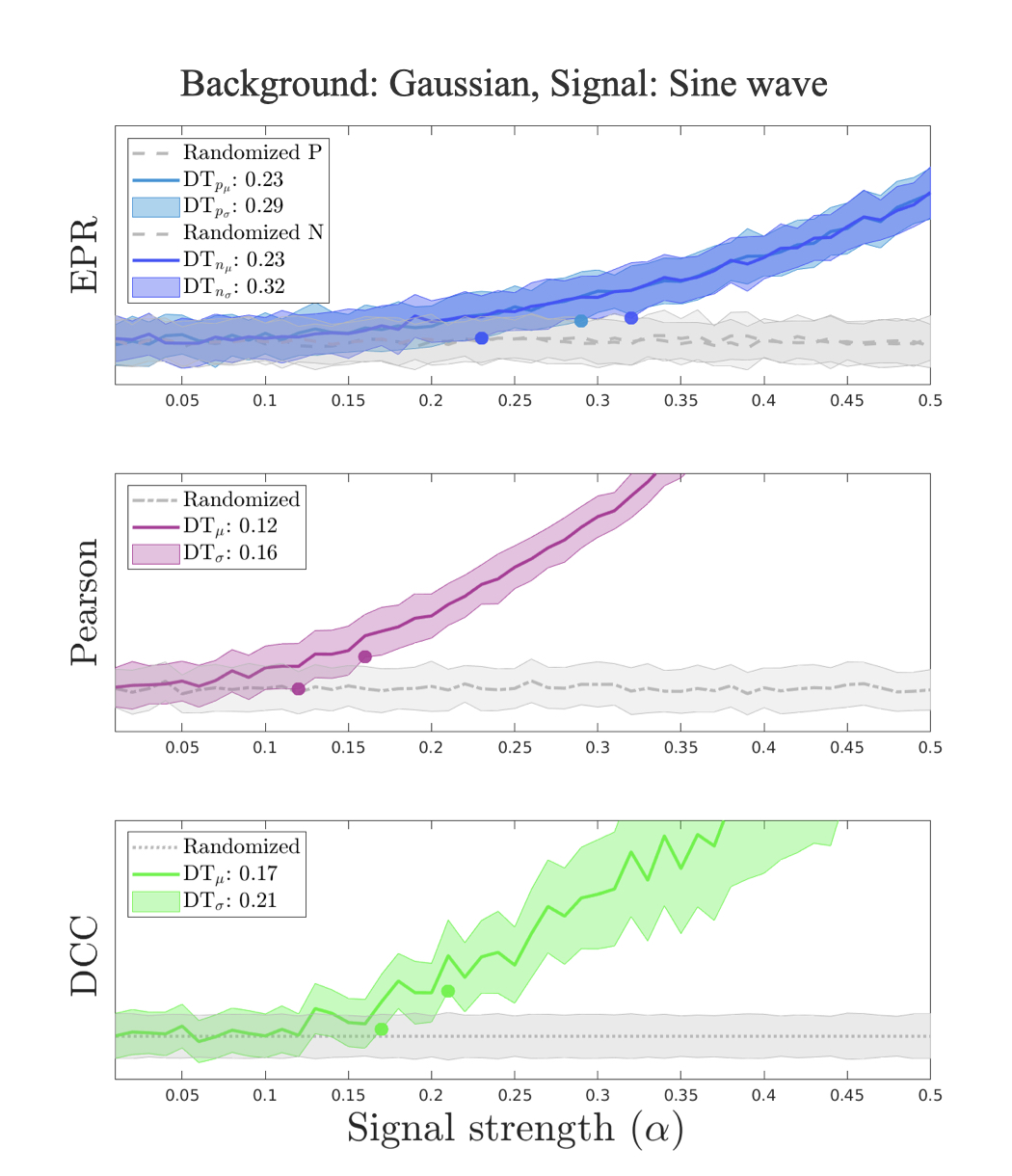}}
\caption{Detection thresholds as in Fig. \ref{fig:DT} now with $\kappa=1.6$ for EPR.
While $\kappa$ may change the ranking of PC and EPR based on $DT_\mu$ and $DT_\sigma$, {Fig. \ref{T1}} shows this is limited to a few specific cases. }
    \label{fig:DT2}
\end{figure}

\section{Benchmarks and ranking}\label{Sec:results}

Following \S\ref{Sec:procedure}, we prepare a {database of} response curves for the 15 combinations of background and signals analyzed by DCC, PC and EPR, quantified for sensitivity by {their detection thresholds} schematically indicated in Fig. \ref{fig:departure}. Fig. \ref{fig:DT} presents some representative samples of this benchmark study.

Fig. \ref{T1} summarizes the ranking of the three methods, where EPR is included for two choices of $\kappa = 0.5$ and $\kappa = 1.6$. It also includes two mean values for Poisson background $\lambda=5,\ 10000$.
The ``winner" is indicated by color in each entry of the lower part.
We observe that DCC \eqref{EQN_dcc_max} falls behind PC \eqref{EQN_Pearson} and EPR \eqref{EQN_EPR} in our ranking by $DT_\mu$ and $DT_\sigma$.
In searches of small signals in noisy data, the competition is therefore between PC and EPR. 

For PC and EPR, {Fig. \ref{T1} shows} varying performance with the type of noise, more so than the type of signal. 
For Gaussian and Gaussian-like noise (Poisson with large $\lambda$) PC wins out over EPR. 
For Uniform noise, EPR typically outperforms PC for larger $\kappa$. 
Specific cases exist when PC and EPR show very similar performance, 
e.g., for a Poisson background and signal and $\kappa=0.5$ for EPR.

\begin{figure}
  [ht] 
  \hspace{-2cm}
   \begin{tabular}
      {wc{15mm}|wl{35mm} wl{35mm} wl{35mm} wl{35mm}} \hline  & \hspace{1cm} Gaussian  & \hspace{0.5cm} Poisson ($\lambda=10000$)& \hspace{0.7cm} Poisson ($\lambda=5$) & \hspace{1cm} Uniform  \\ 
      \hline Gaussian & \parbox[c]{0em}{
      \includegraphics[width=1.7in]{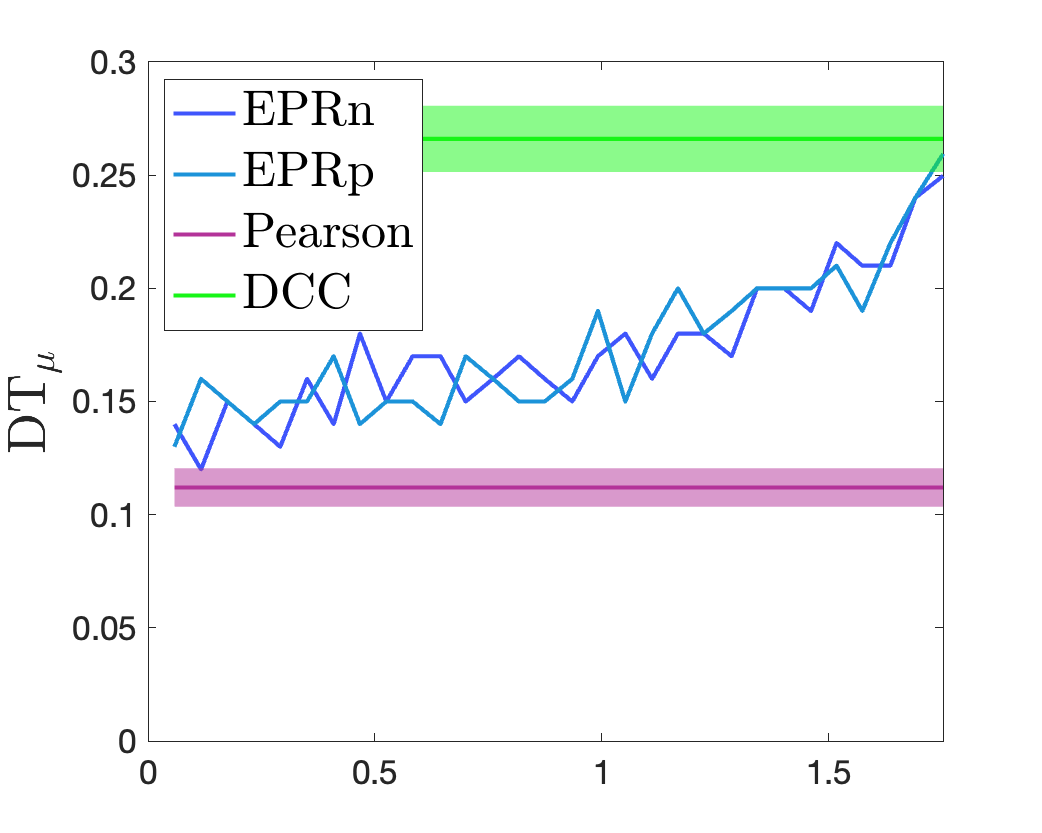}} & \parbox[c]{0em}{
      \includegraphics[width=1.7in]{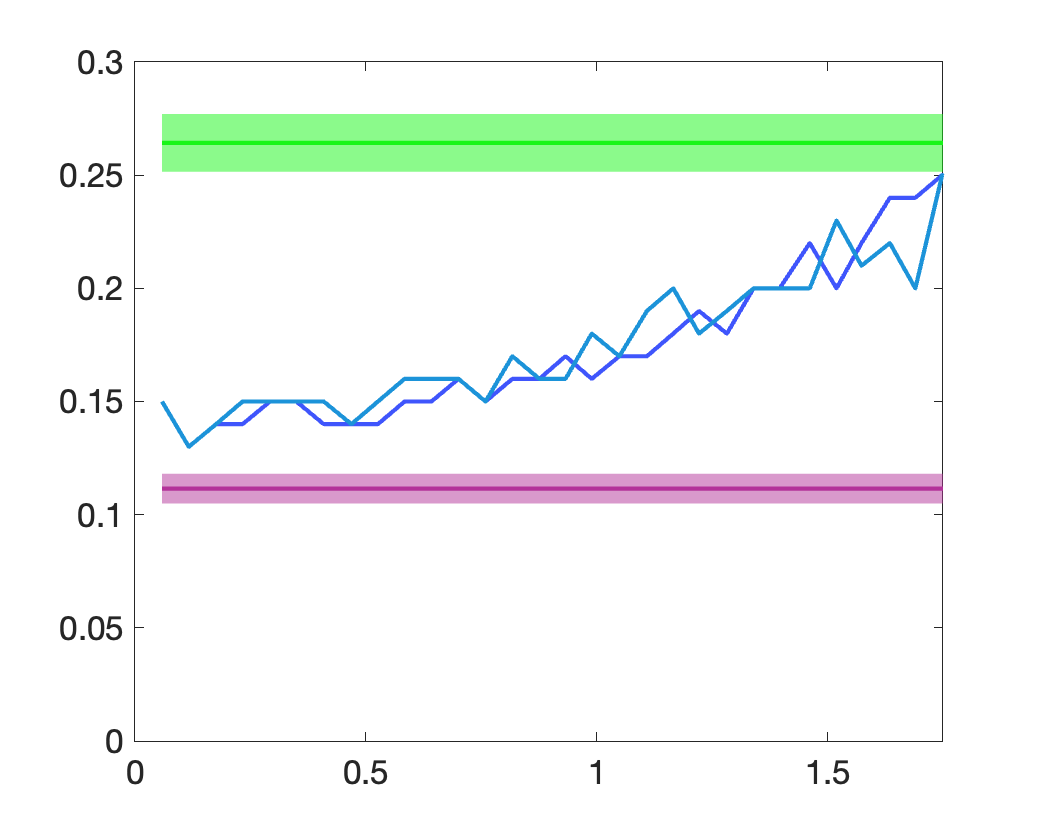}} &\parbox[c]{0em}{
      \includegraphics[width=1.7in]{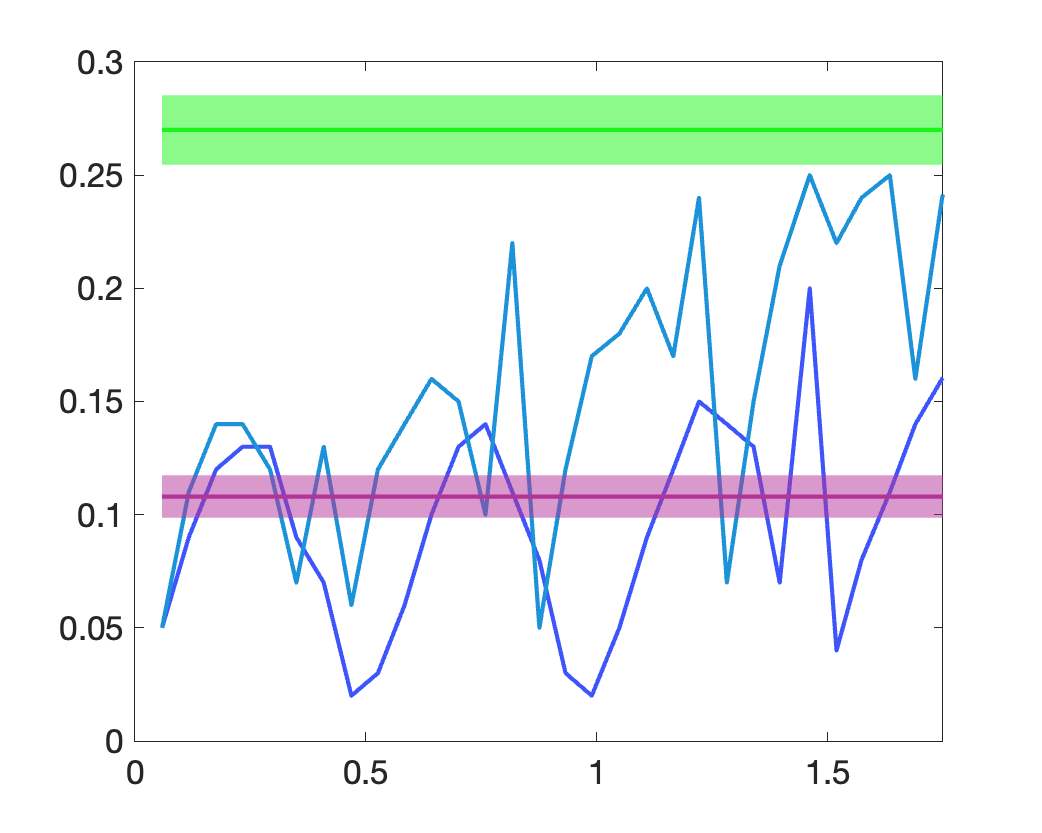}} & \parbox[c]{0em}{
      \includegraphics[width=1.7in]{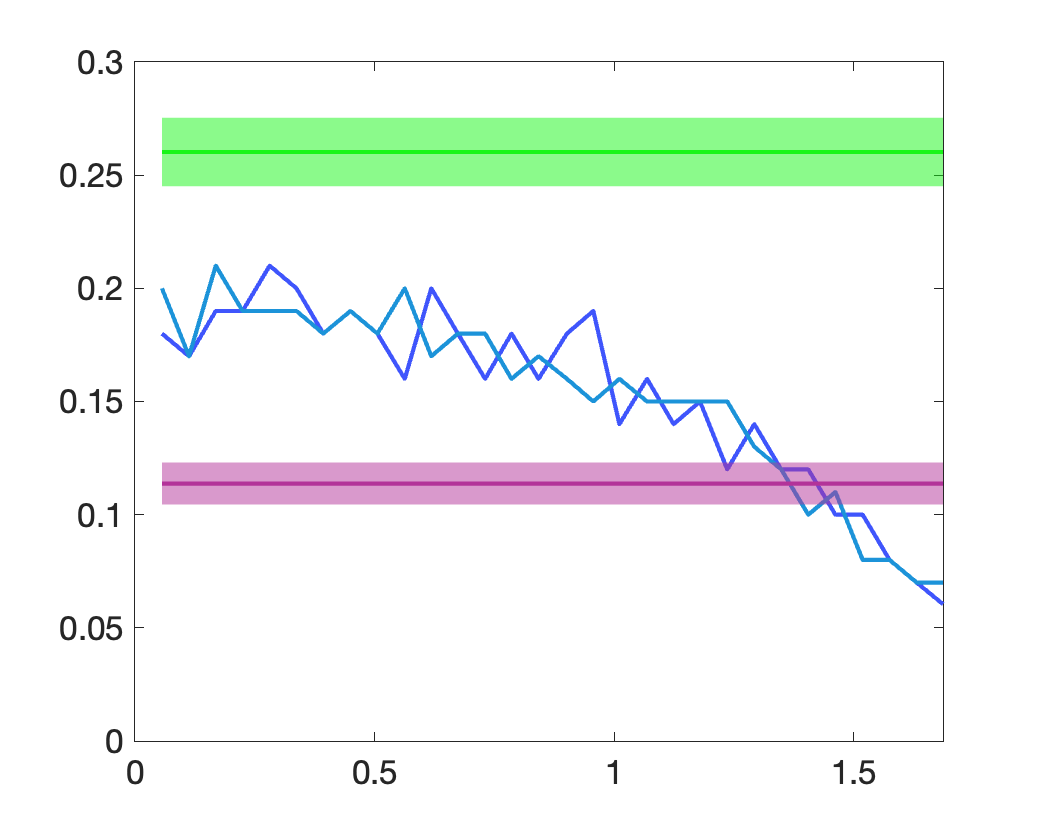}}\\
      \hline
      Poisson &\parbox[c]{0em}{
      \includegraphics[width=1.7in]{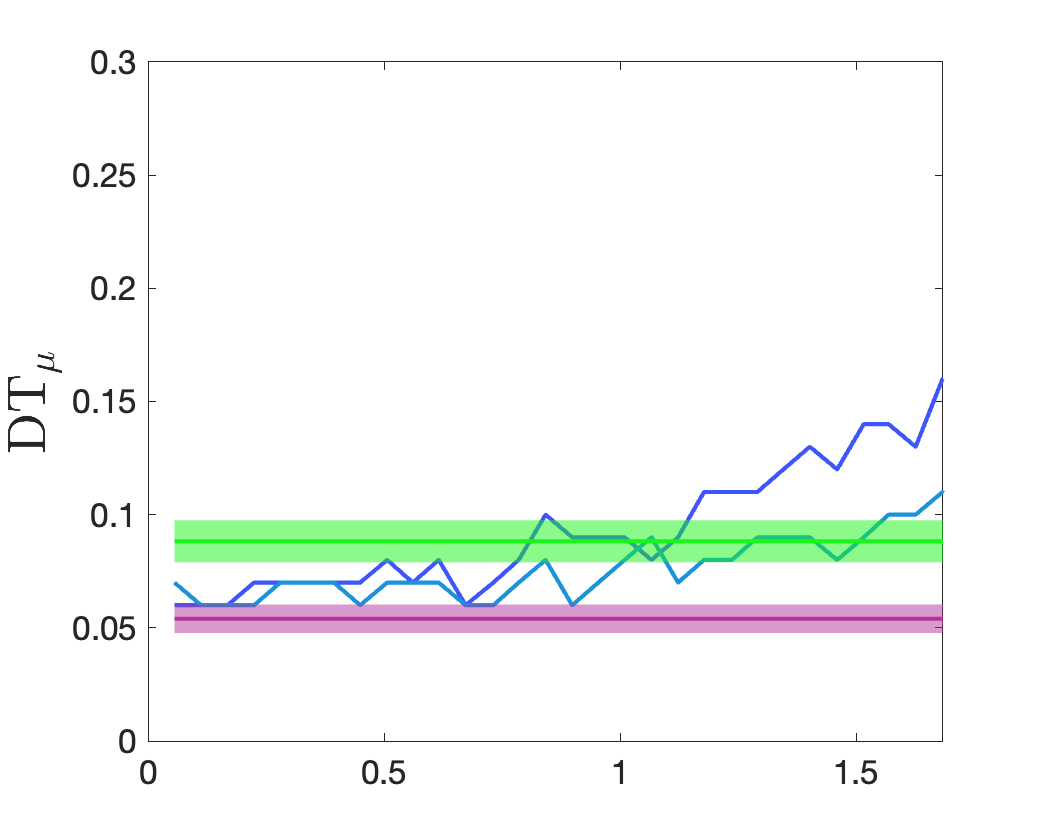}} & \parbox[c]{0em}{
      \includegraphics[width=1.7in]{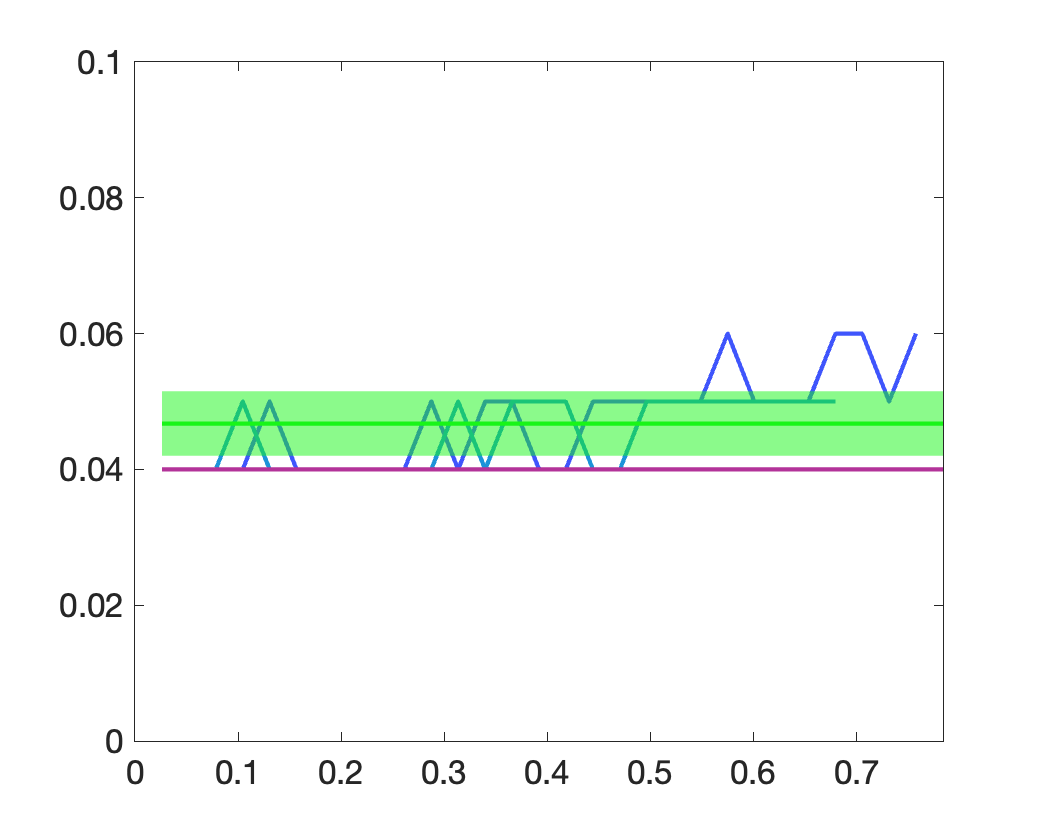}} &\parbox[c]{0em}{
      \includegraphics[width=1.7in]{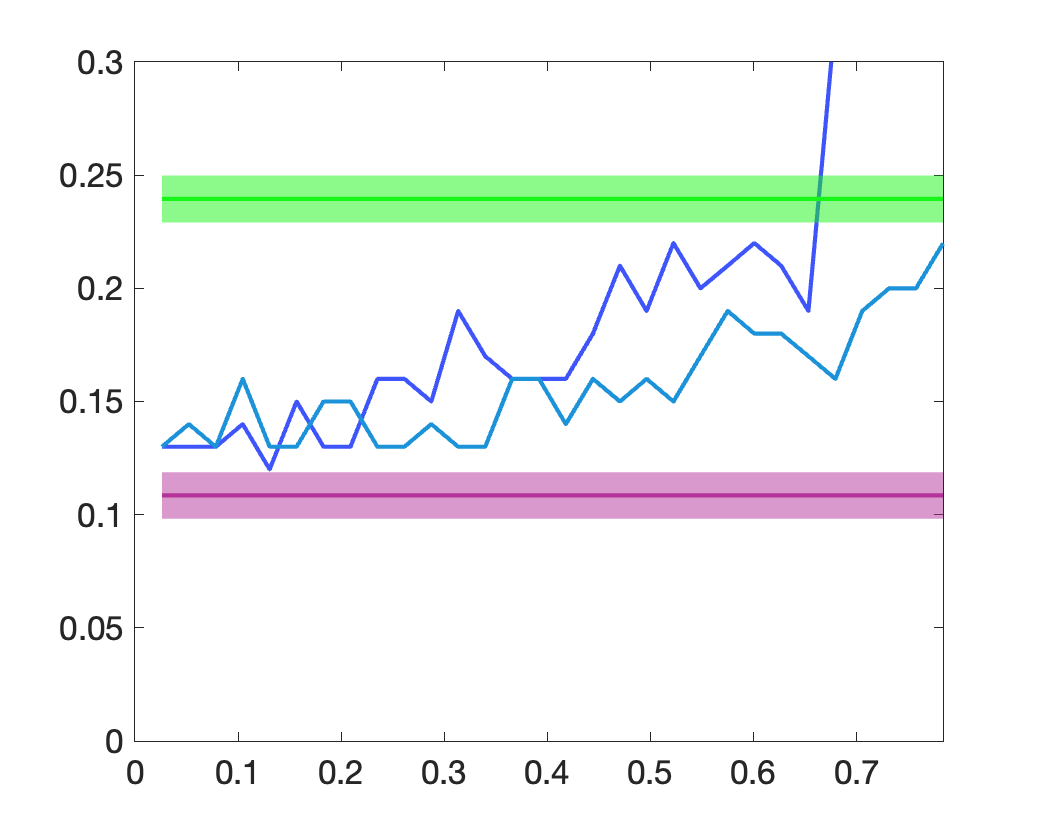}} & \parbox[c]{0em}{
      \includegraphics[width=1.7in]{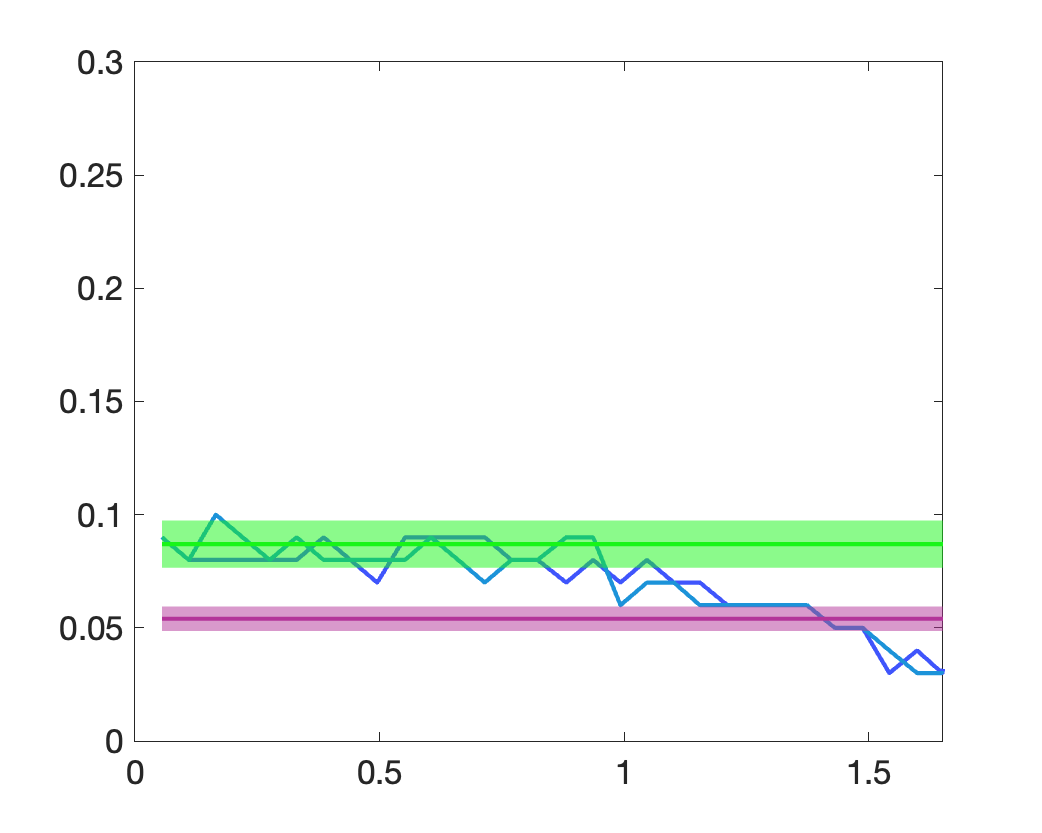}}\\
      \hline
      Uniform &  \parbox[c]{0em}{
      \includegraphics[width=1.7in]{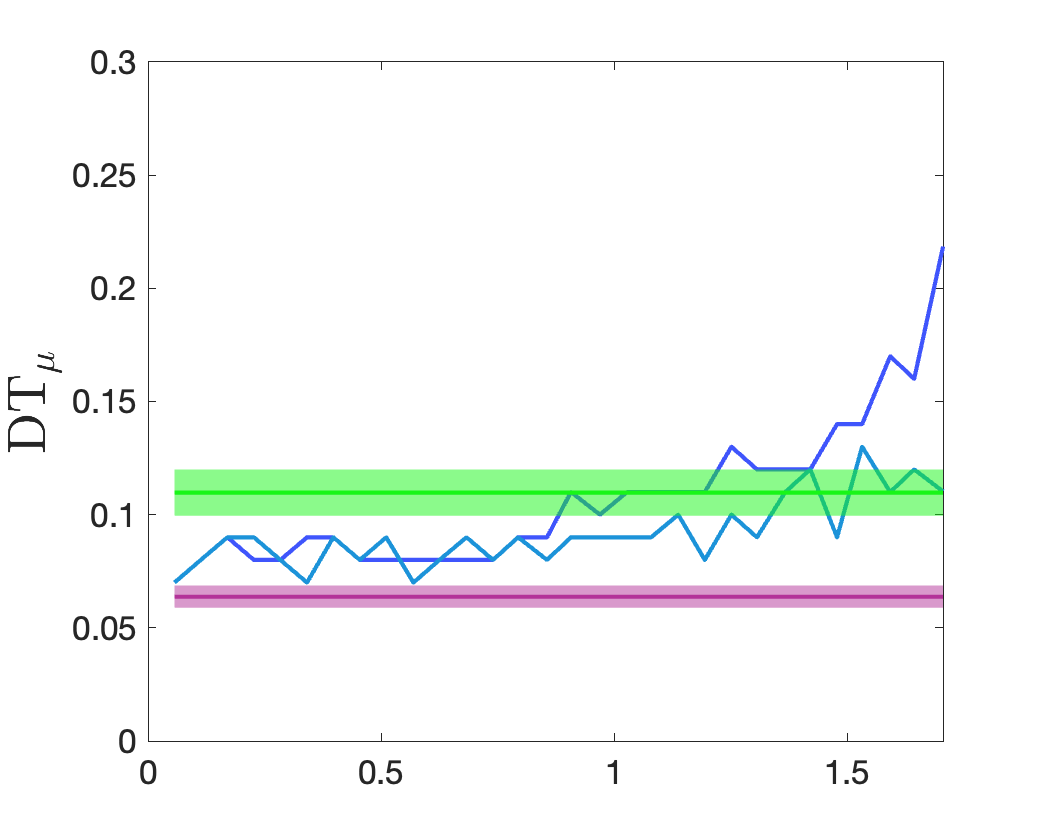}} & \parbox[c]{0em}{
      \includegraphics[width=1.7in]{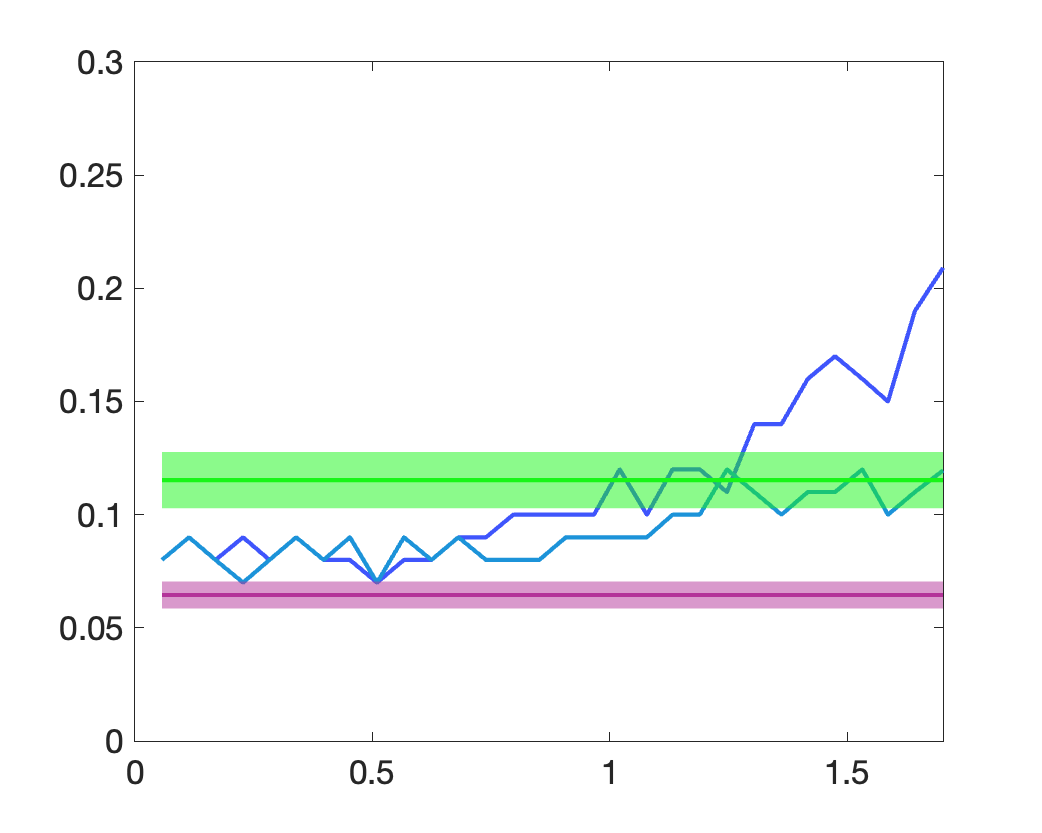}}&\parbox[c]{0em}{
      \includegraphics[width=1.7in]{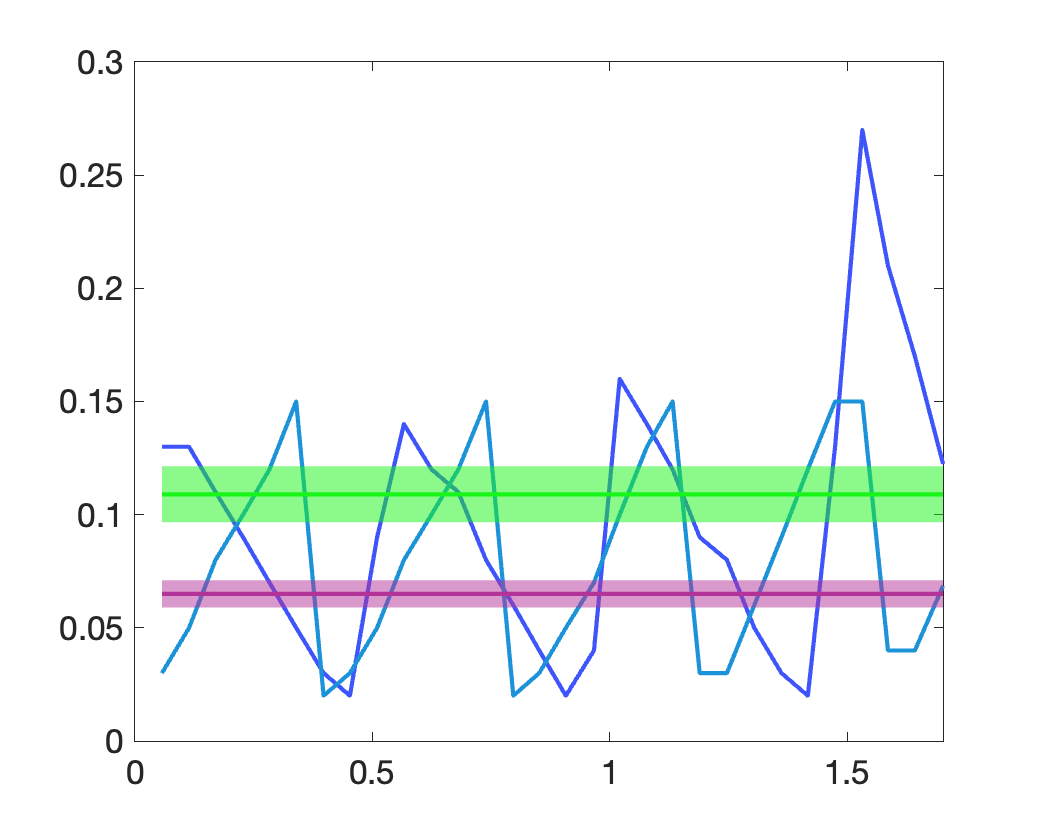}} & \parbox[c]{0em}{
      \includegraphics[width=1.7in]{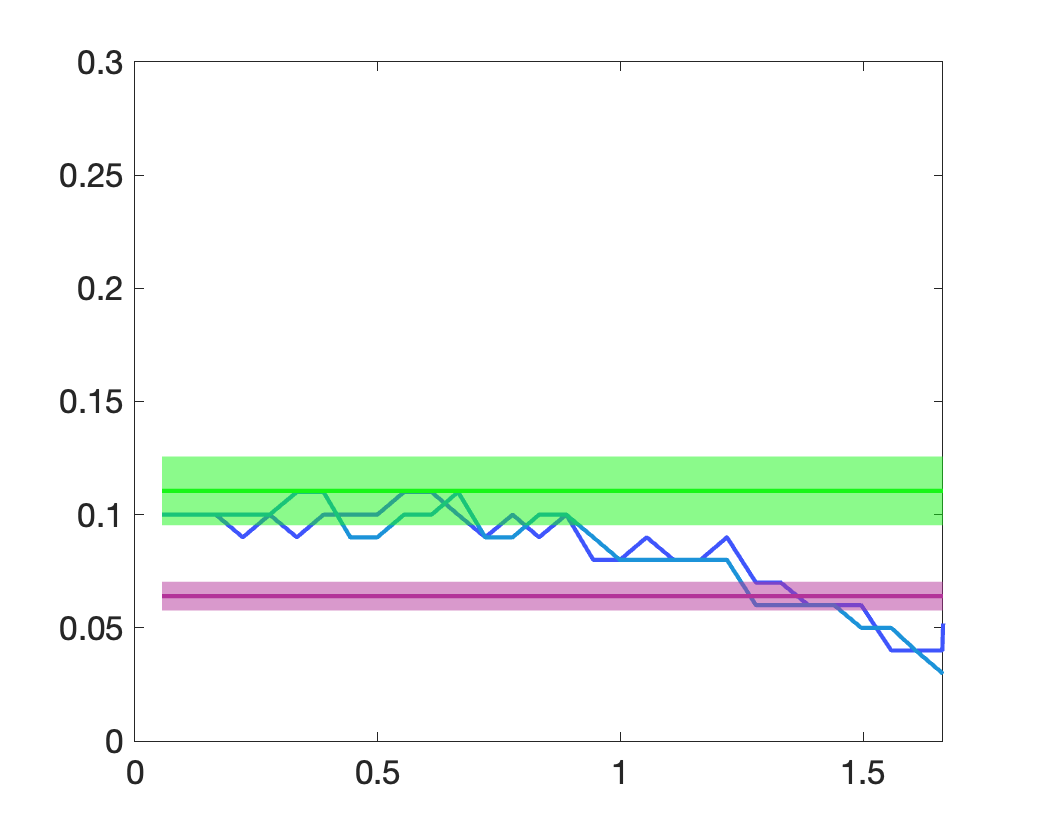}}\\
      \hline
      Chirp & \parbox[c]{0em}{
      \includegraphics[width=1.7in]{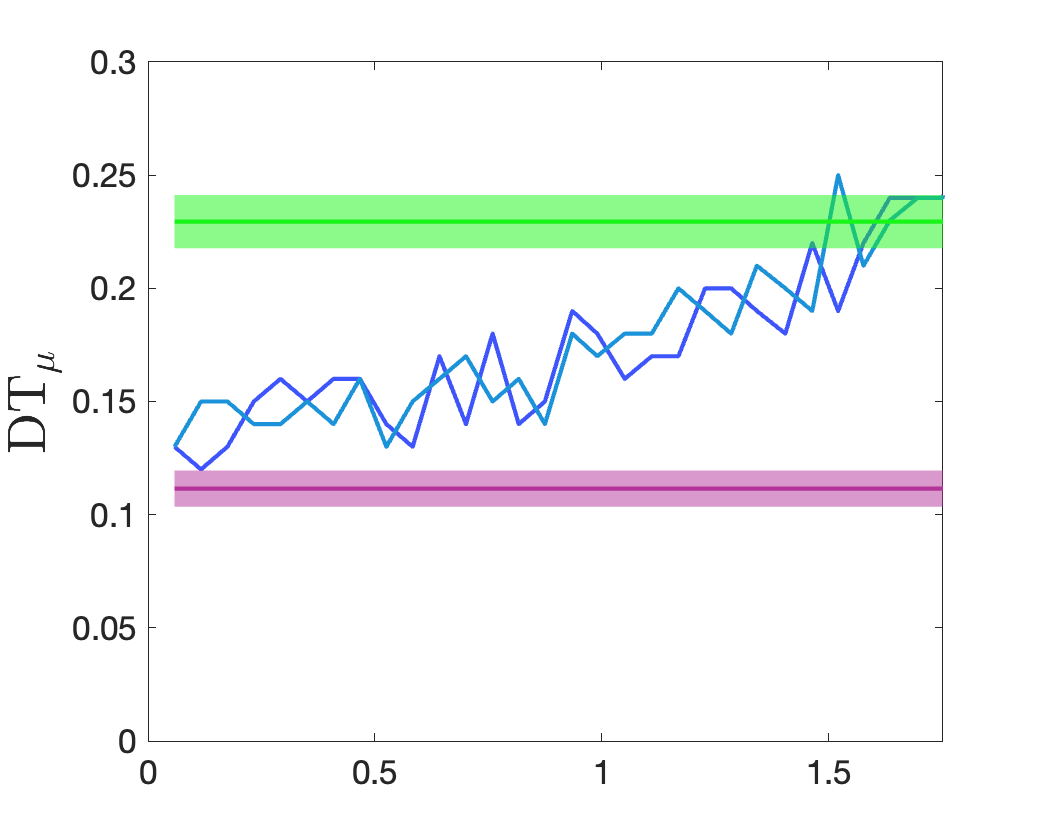}} & \parbox[c]{0cm}{
      \includegraphics[width=1.7in]{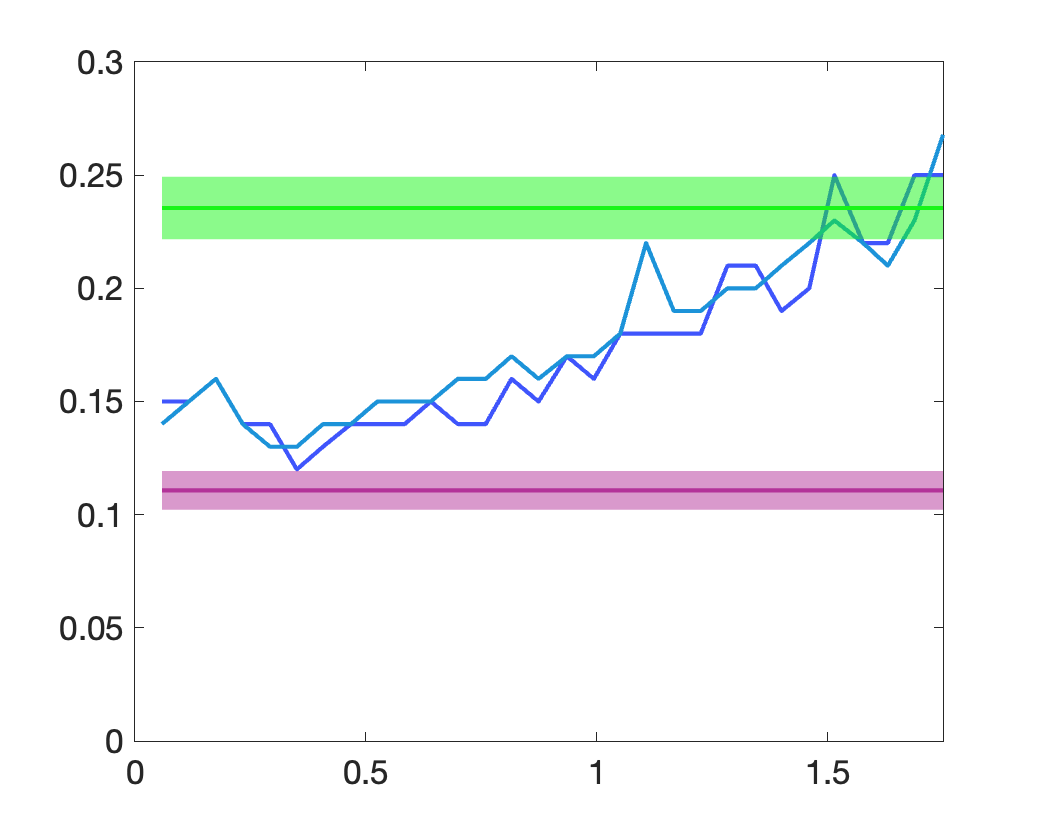}} &\parbox[c]{0em}{
      \includegraphics[width=1.7in]{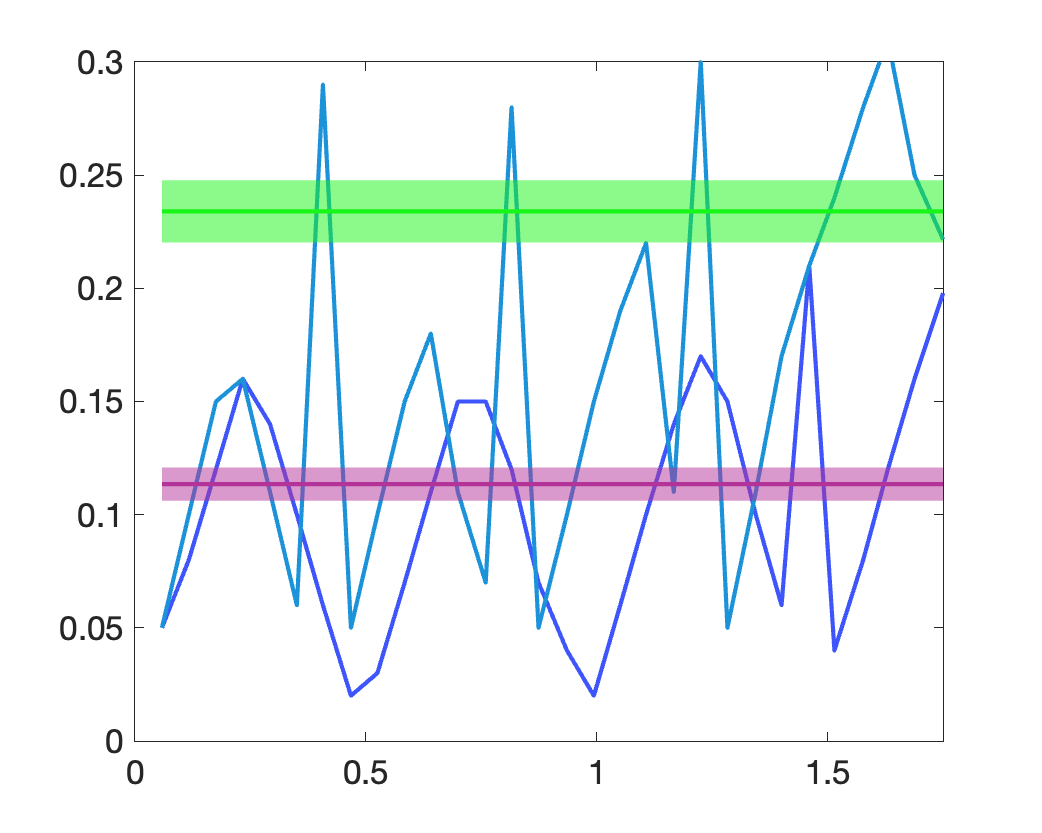}} & \parbox[c]{0cm}{
      \includegraphics[width=1.7in]{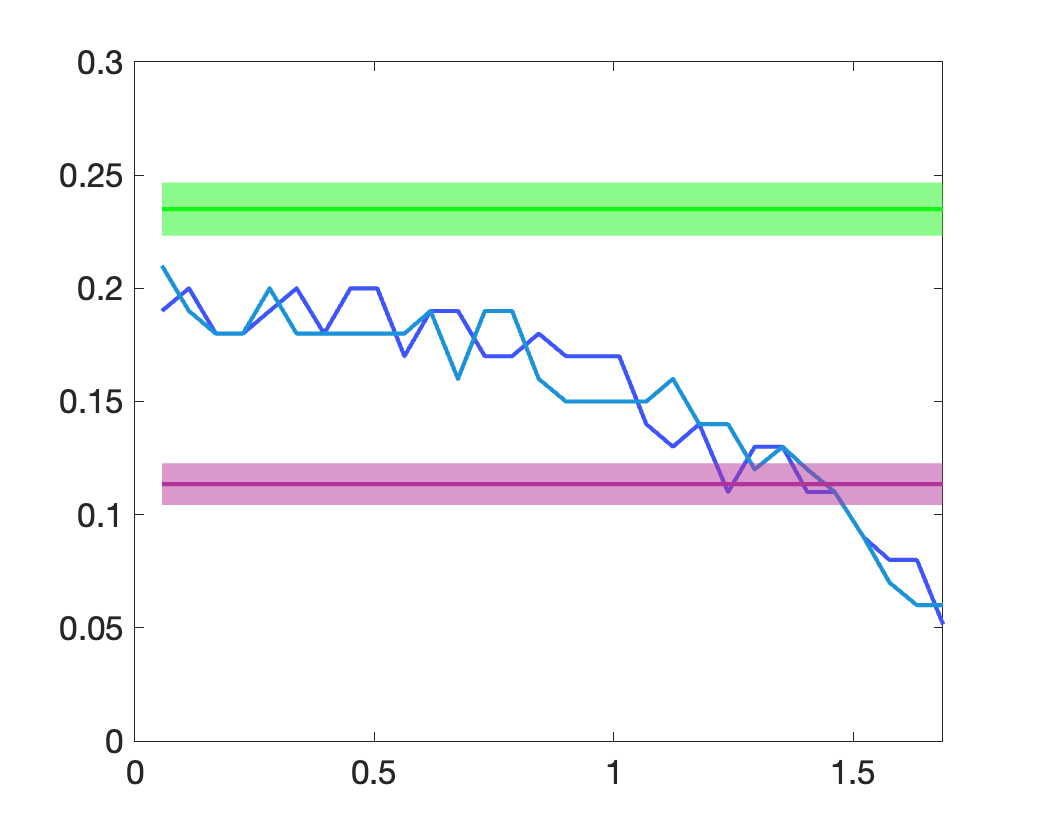}} \\
      \hline
      Sine &  \parbox[c]{0em}{
      \includegraphics[width=1.7in]{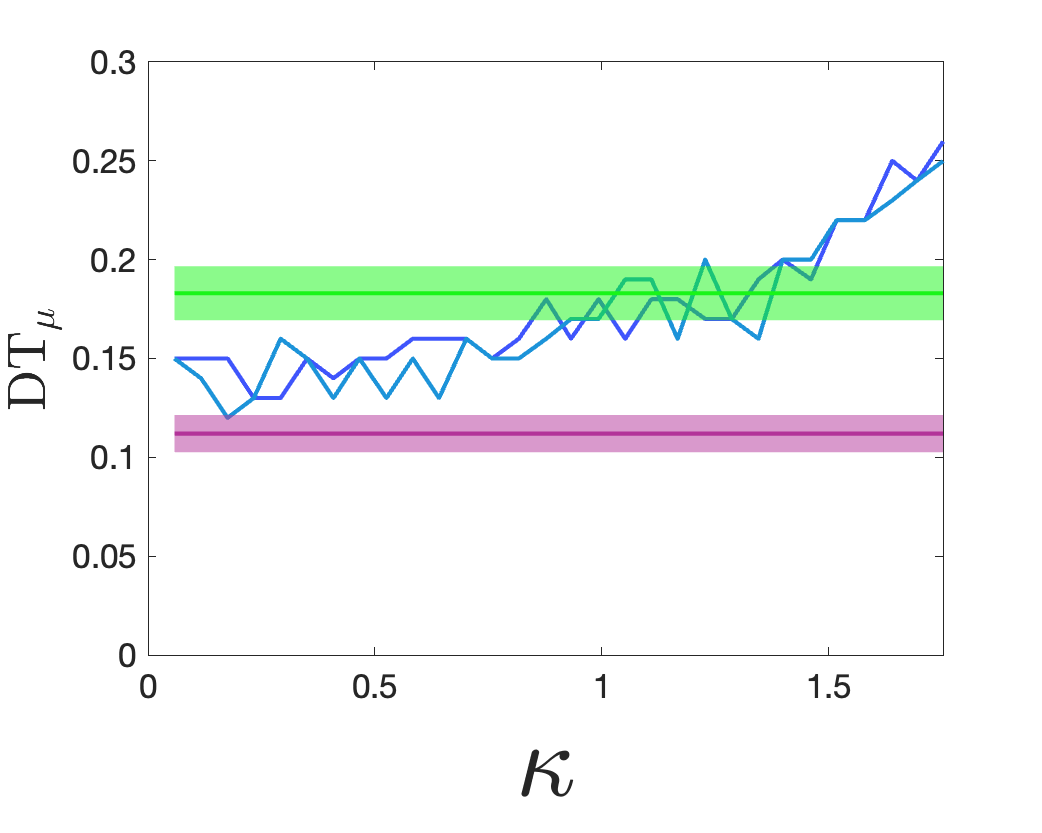}} & \parbox[c]{0em}{
      \includegraphics[width=1.7in]{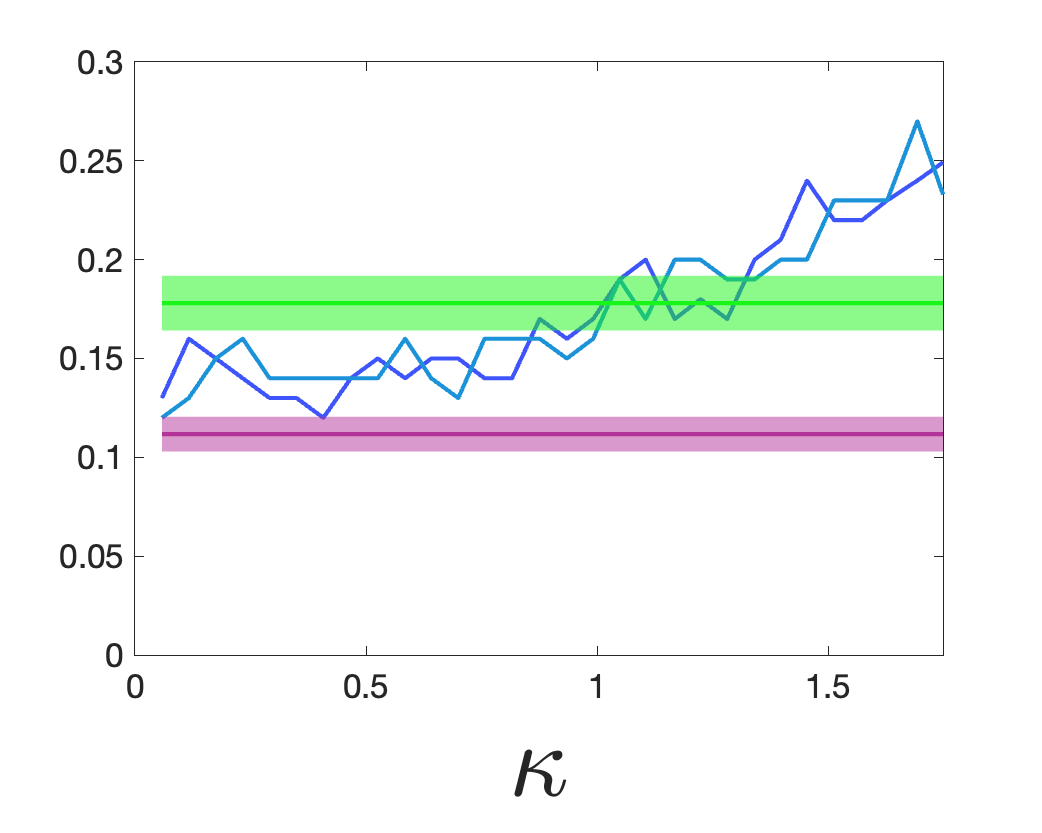}} &\parbox[c]{0em}{
      \includegraphics[width=1.7in]{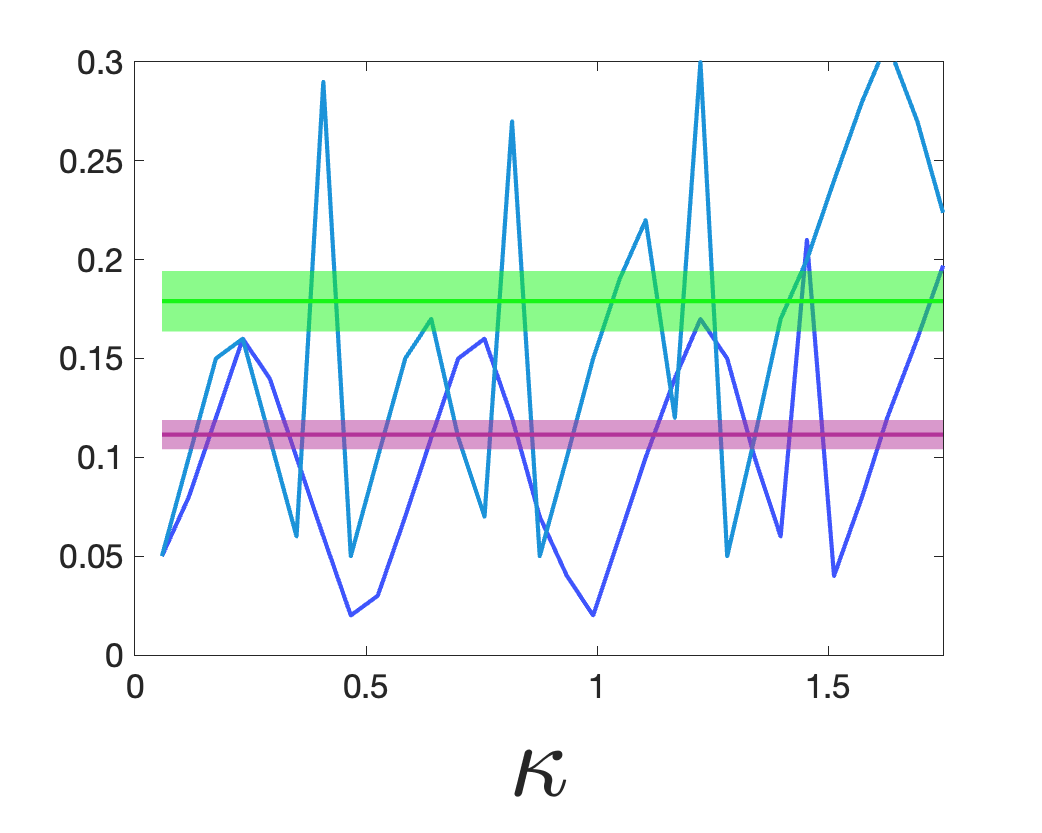}} & \parbox[c]{0em}{
      \includegraphics[width=1.7in]{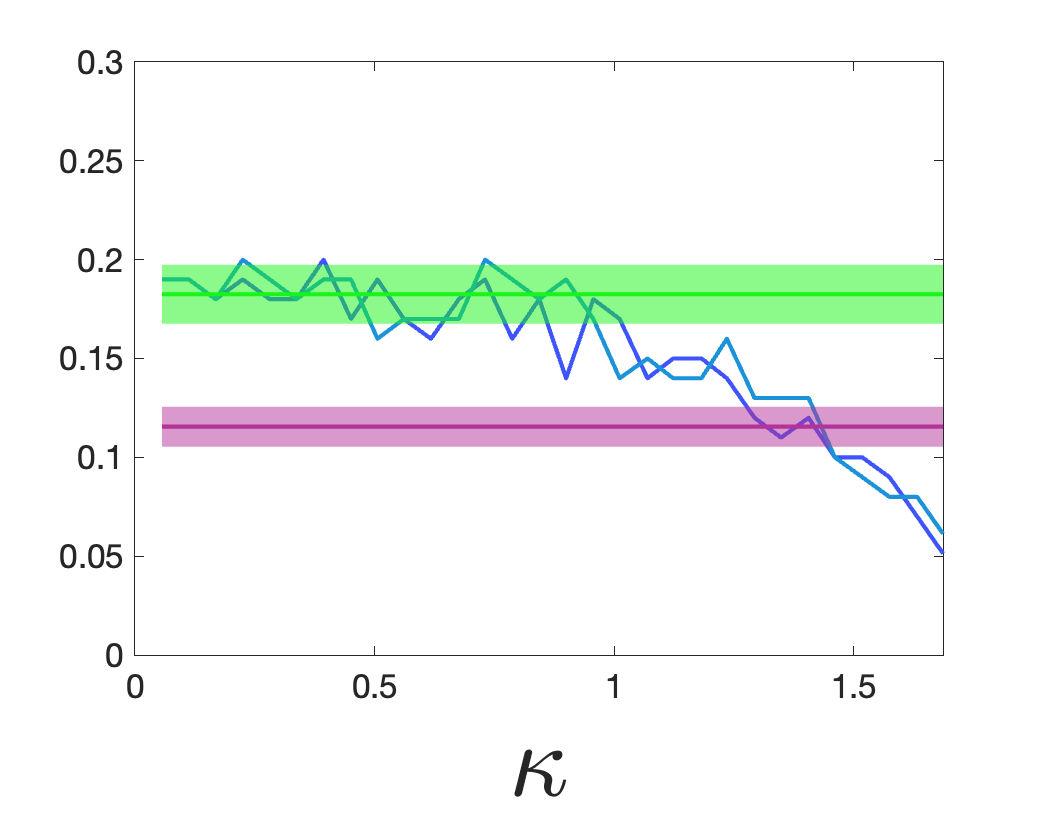}} \\
      \hline
  \end{tabular}
  \caption{Detection thresholds $DT_{\mu}$ (smaller is better) extracted from response curves (Fig. \ref{fig:departure}) for the three methods in various combinations of background and signal as a function of the cut-off parameter $\kappa$ in the application of EPR. $DT_{\sigma}$ (not shown) is similar to that of $DT_{\mu}$. 
  PC (pink) and DCC (green) provide a reference for EPR performance. PC and DCC require evaluation of full arrays, prohibiting application of (\ref{EQN_kappa}).
  $DT_\mu$ of EPR  trends with $\kappa$, notably for Poisson background $\lambda=10000$ and $\lambda=5$ shown. {{\it Note:} Due to experimental limitations the case of Poisson noise and Poisson signal is shown up to $\kappa=0.8$, though the result for higher $\kappa$ is interpreted according to these results.}}
    \label{T3}
\end{figure}

While typically comparable to PC, EPR tends to be the winner for Uniformly distributed noise data, irrespective of signal type {(Fig. \ref{T3})}.

{Although} PC may be the winner in several cases, it needs the entire array for evaluation, essentially due to its geometric meaning of a cosine between ${\bf p,q}$.  
{Figs. \ref{fig:N-kappa} and \ref{T3} show} the key gain by EPR. EPR performs similarly to PC involving modest-sized tails comprising merely 10\% of {the data ($\kappa \simeq 2$)}. In some cases, EPR even performs better than PC, namely, for uniform background noise and relatively larger $\kappa$.

\subsection{Limiting factors of EPR}
{Despite having a performance close to PC using only a small fraction of data, EPR has some limitations inherent from applying a cut-off. At high $\kappa$ the number of data points {contributing in EPR is noticeably small} creating small size array and large uncertainties in the result. {At these $\kappa$ limits,} EPR shows significant deviations from PC for Gaussian and Gaussian-like Poisson background, due to small number statistics at small array sizes. 
For Poisson and Uniform background this degradation is not likely with even better performance at smaller array size, when the background has a uniform distribution (Figs. \ref{T3} and \ref{T1}).}

\section{Conclusion and Outlook}
\label{Sec:outlook}

\begin{figure}
\centering
\includegraphics[scale=0.22]{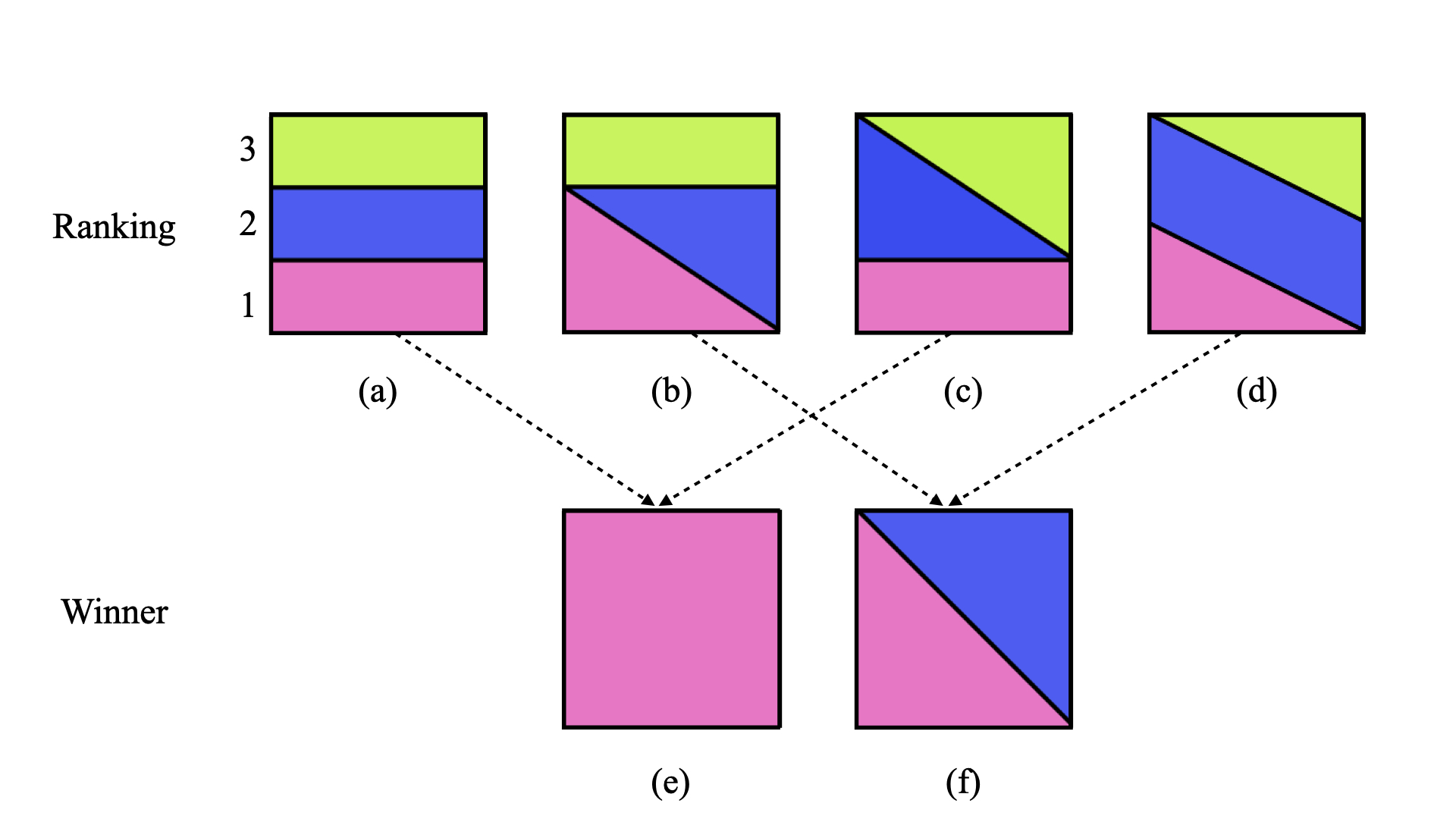}

\caption{{Sample of blocks used in {Fig. \ref{T1} in ranking} three correlation methods PC, EPR and DCC. The first row shows the blocks in the upper panel of {Fig. \ref{T1} in ranking} the above mentioned methods, and the second row blocks are for the {lower panel of the same figure representing} the winner at each specified $\kappa$. Accordingly, in block (a), the pink method has the smallest detection threshold, while the blue takes the second place and green has the largest detection threshold. Therefore, the winner is the method shown in pink i.e. block (e). Block (b) is a more complicated case in which, in ranking, blue and pink method are competing in having the smallest detection threshold, while green has the largest detection threshold. So we have a mixed winner method and it is shown as block (f). Block (c) is similar to (b), with the difference that pink has the smallest detection threshold, but either blue or green can have the second or third place. So in this case, although the ranking between three methods has a mixed part, pink is the winner i.e. block (e). The most strange case is when all three methods are competing which is shown in block (d), though there are not too many of them in our analysis. For these cases, the smallest threshold belongs to either pink or blue, while either green and blue can take the second or third place. So for this case the winner is shown as in block (f). }
}
    \label{fig:Sample}
\end{figure}

\begin{figure}[ht!] 
\hspace{-1cm}
 \begin{tabular}
      {wc{20mm}}
        \parbox[c]{0em}{
      \includegraphics[width=6in]{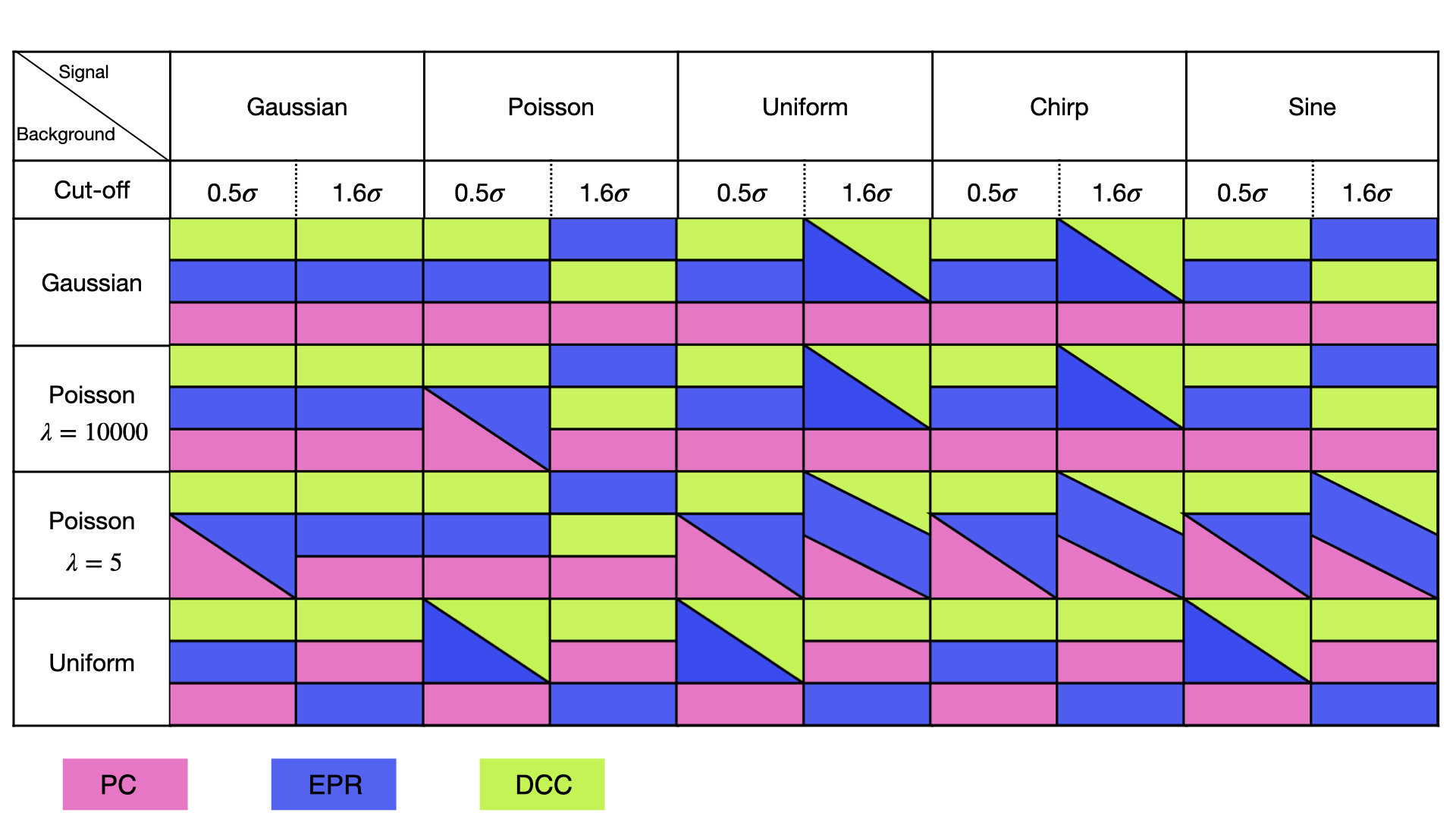}} \\
      
      \parbox[c]{0em}{
      \includegraphics[width=6in]{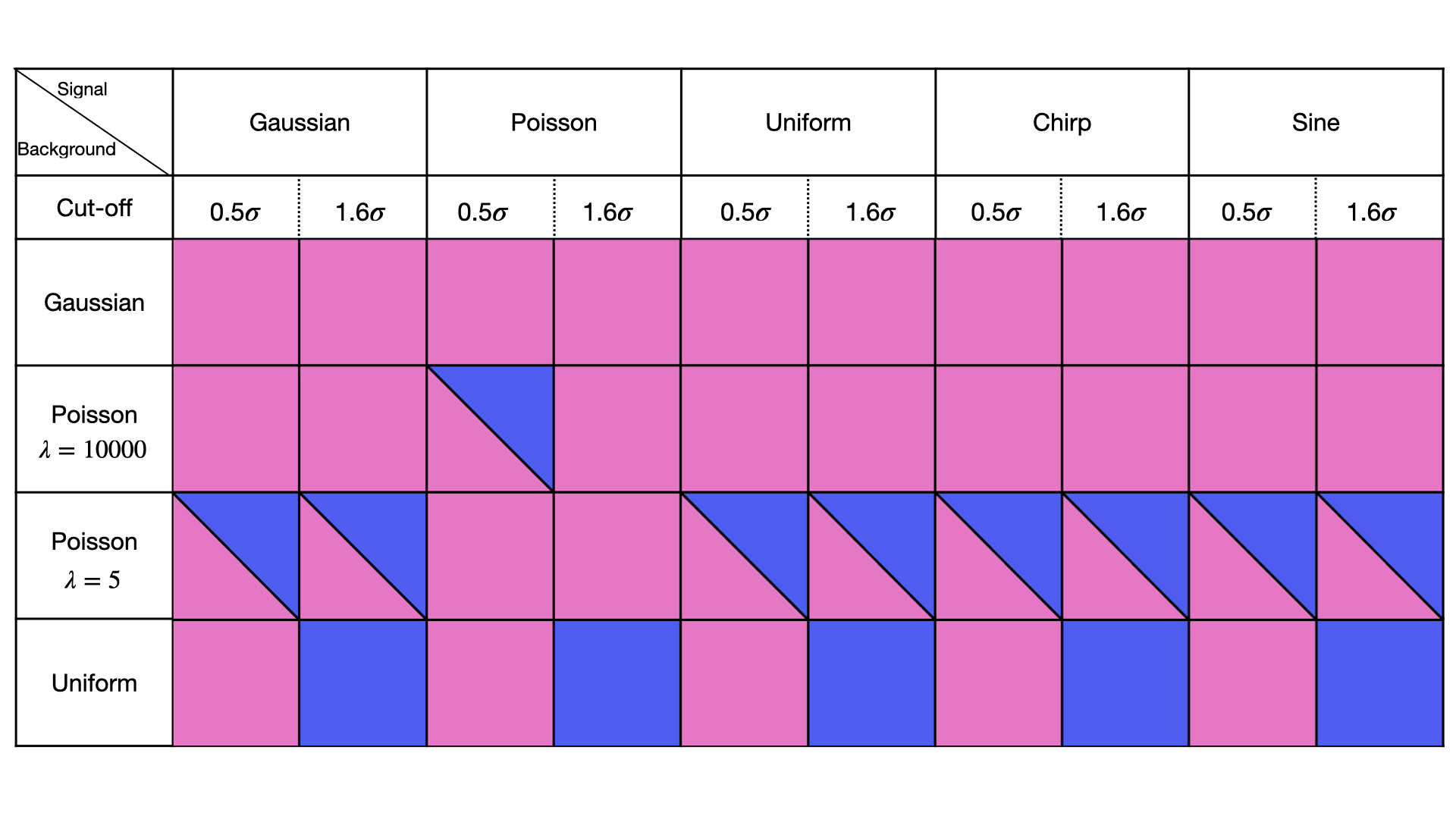}} \\
     
  \end{tabular}
\caption{(Top panel.) Ranking of DCC, PC and EPR according to detection threshold for various combinations of background and signal injections, based on Fig. \ref{fig:Sample} and associated explanation in \S \ref{Sec:outlook}. 
EPR includes the two cut-offs $\kappa=0.5, 1.6$ and we also include two values $\lambda=10000$ and $\lambda=5$ for Poisson background. 
(Bottom panel.) Shown are the winner in ranking by thresholds $DT_{\mu}$ in corresponding color coding.}
    \label{T1}
\end{figure}

\begin{figure}[ht!]
\centering
\includegraphics[scale=0.26]{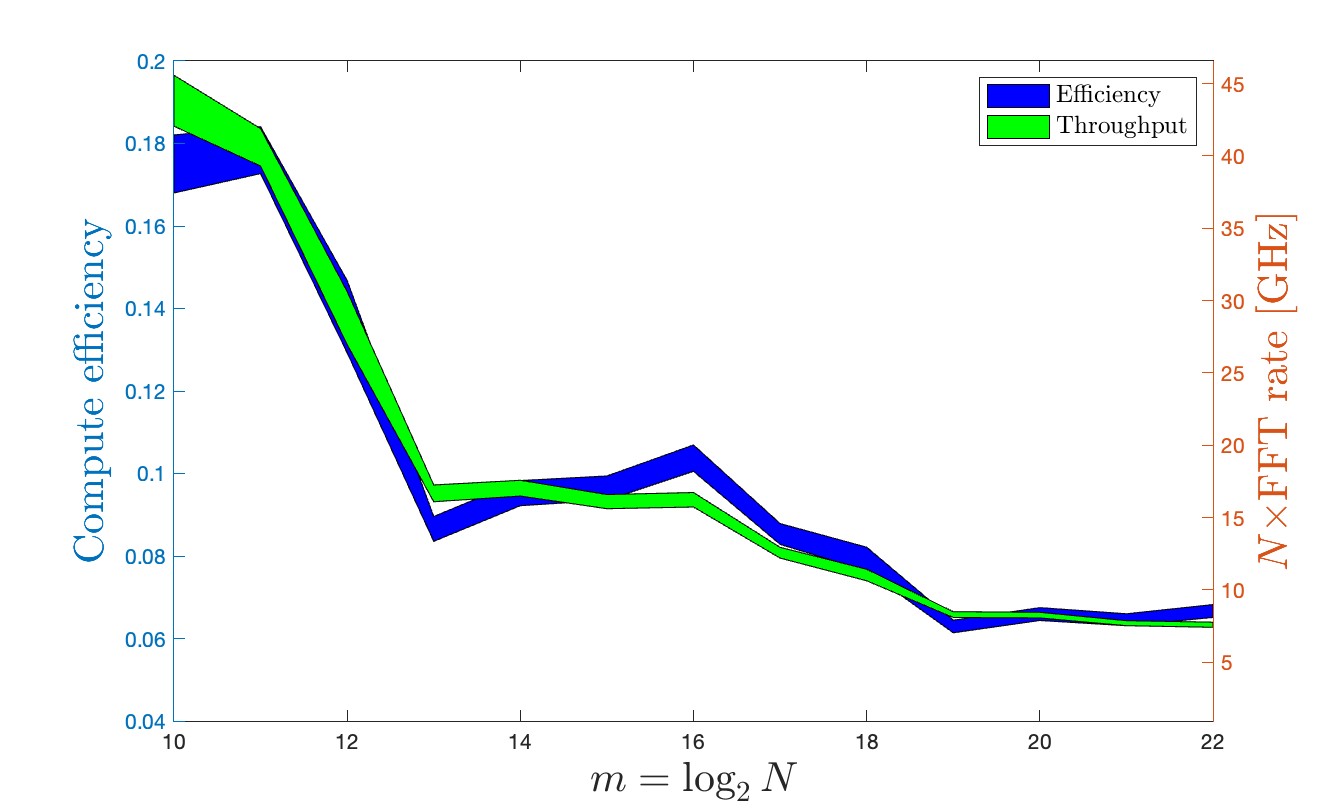}
\caption{{{Efficiency of GPU performance with respect to data array size, here shown for FFT pertinent to FFT-based signal processing. 
Efficiency critically depends on footprint in memory. Shown is a significant drop to memory bandwidth limited performance when array size exceeds Local Memory, causing calls to much slower Global Memory within the GPU.}
[Reprinted from \cite{Putten2024a}]}
}
\label{fig:GPU}
\end{figure}

\begin{figure}[ht!] 
\hspace{-2cm}
 \begin{tabular}
      {wc{20mm}|wl{45mm} wl{45mm} wl{45mm}} \hline  & \hspace{1.7cm} Gaussian  & \hspace{1.8cm} Poisson & \hspace{1.7cm} Uniform  \\ 
      
      \hline Gaussian & \parbox[c]{0em}{
      \includegraphics[width=2in]{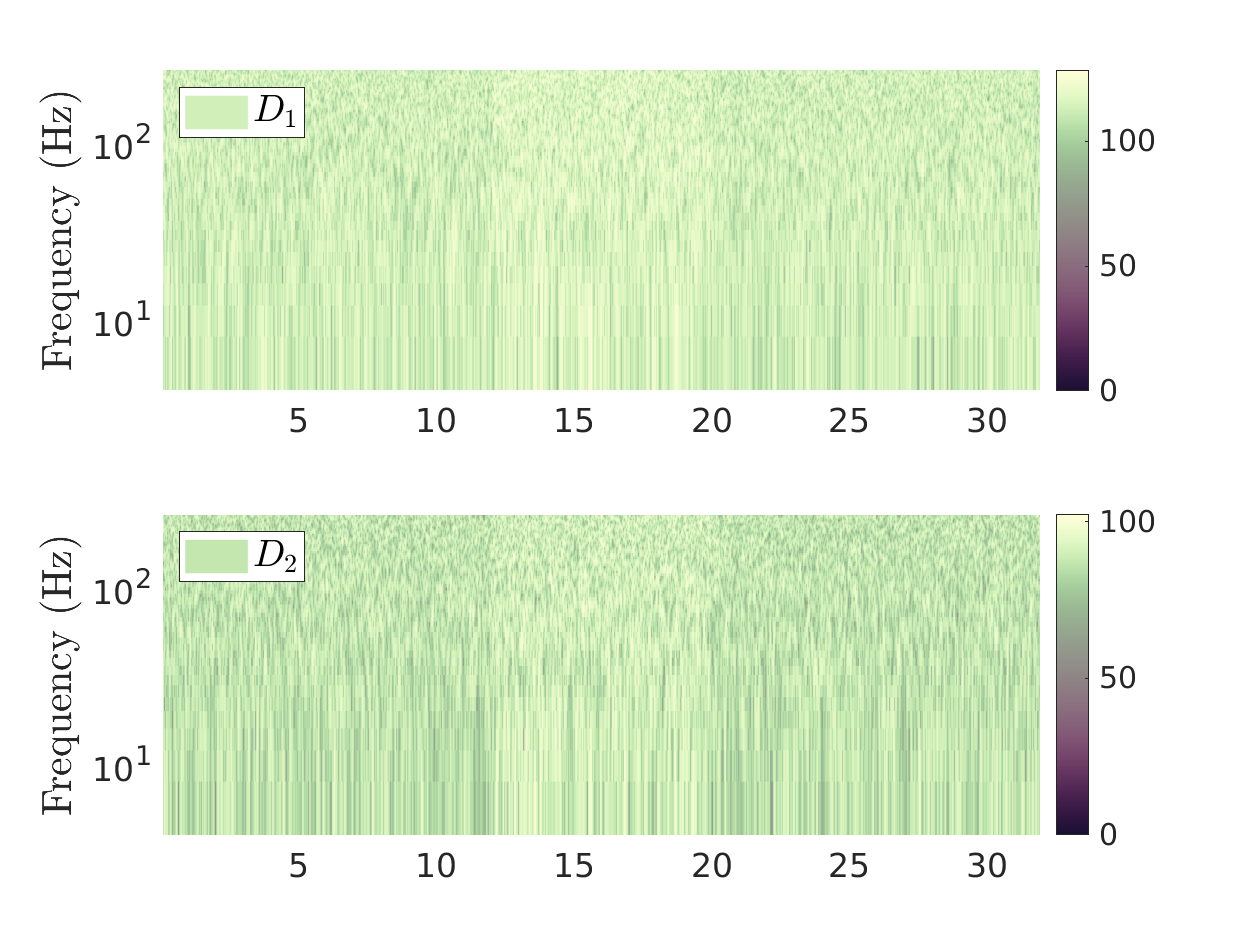}} & \parbox[c]{0em}{
      \includegraphics[width=2in]{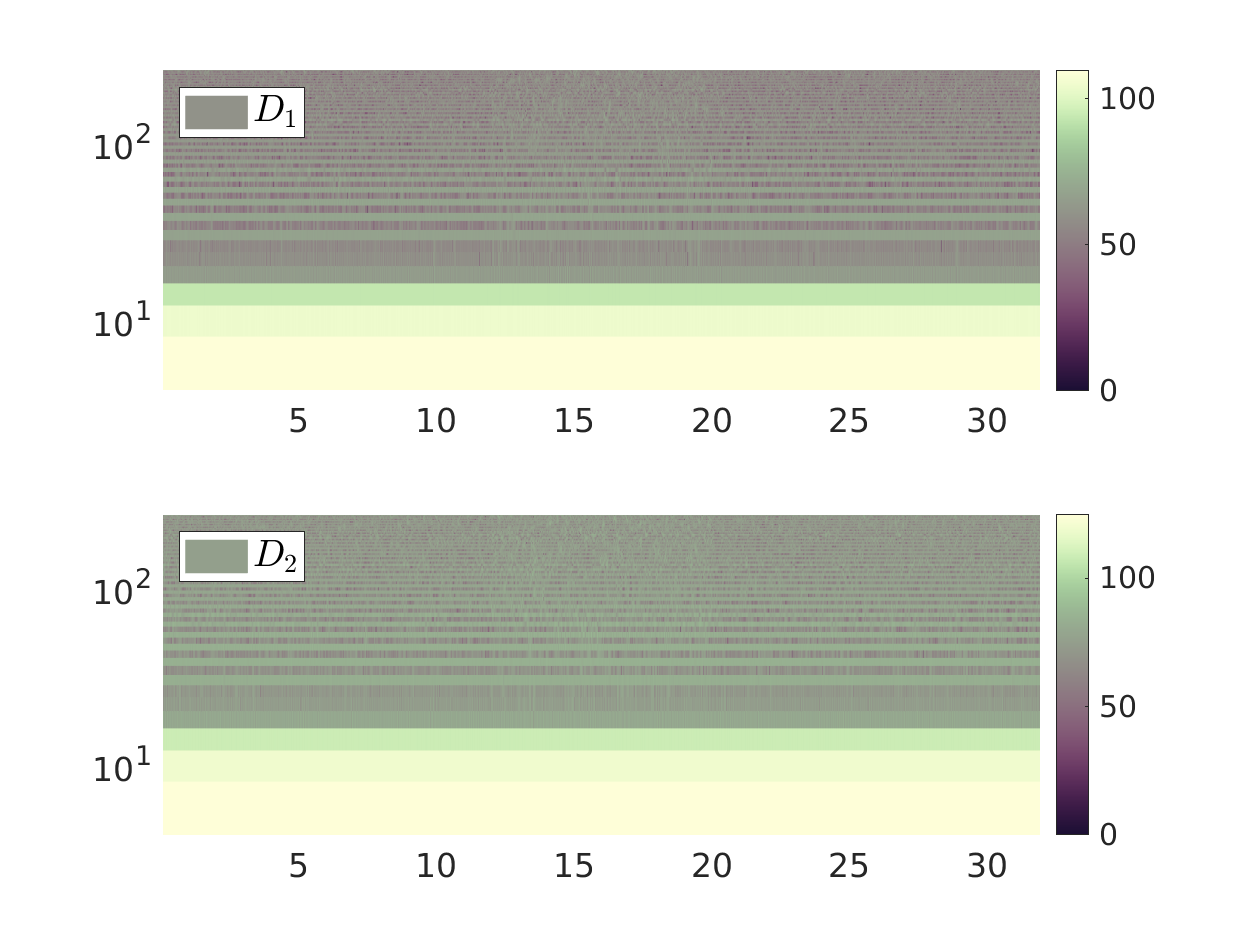}} & \parbox[c]{0em}{
      \includegraphics[width=2in]{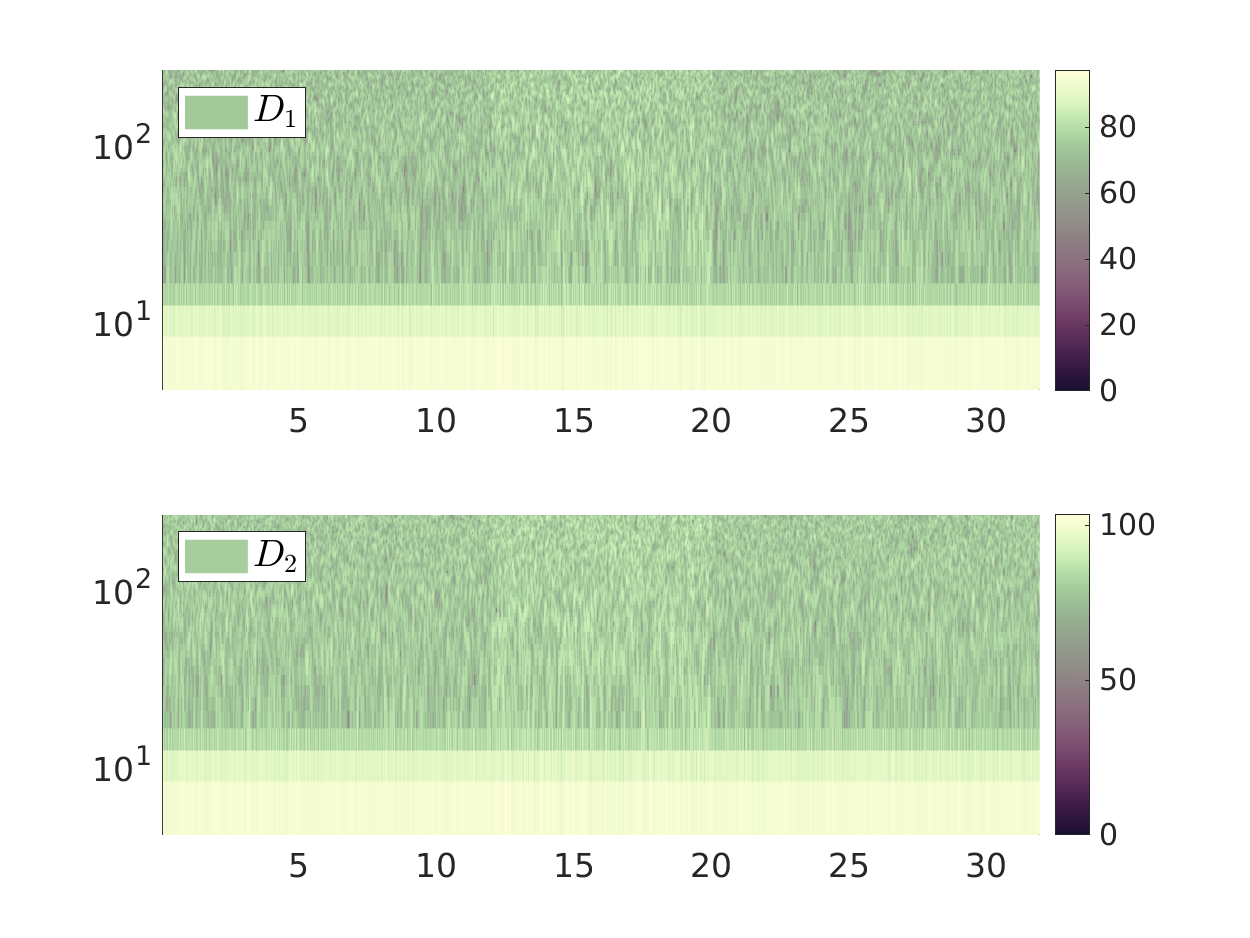}}\\
      \hline
      Poisson &\parbox[c]{0em}{
      \includegraphics[width=2in]{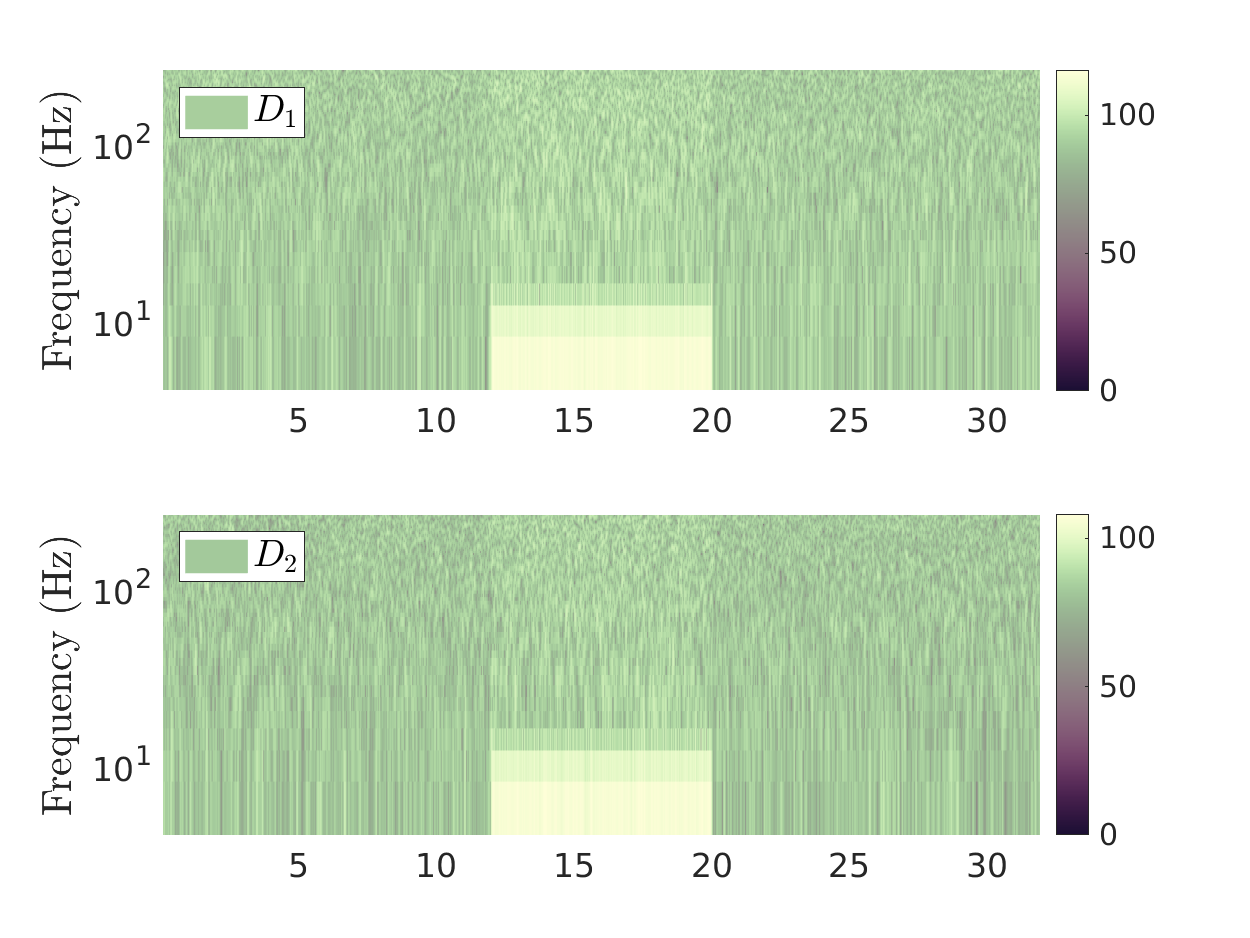}} & \parbox[c]{0em}{
      \includegraphics[width=2in]{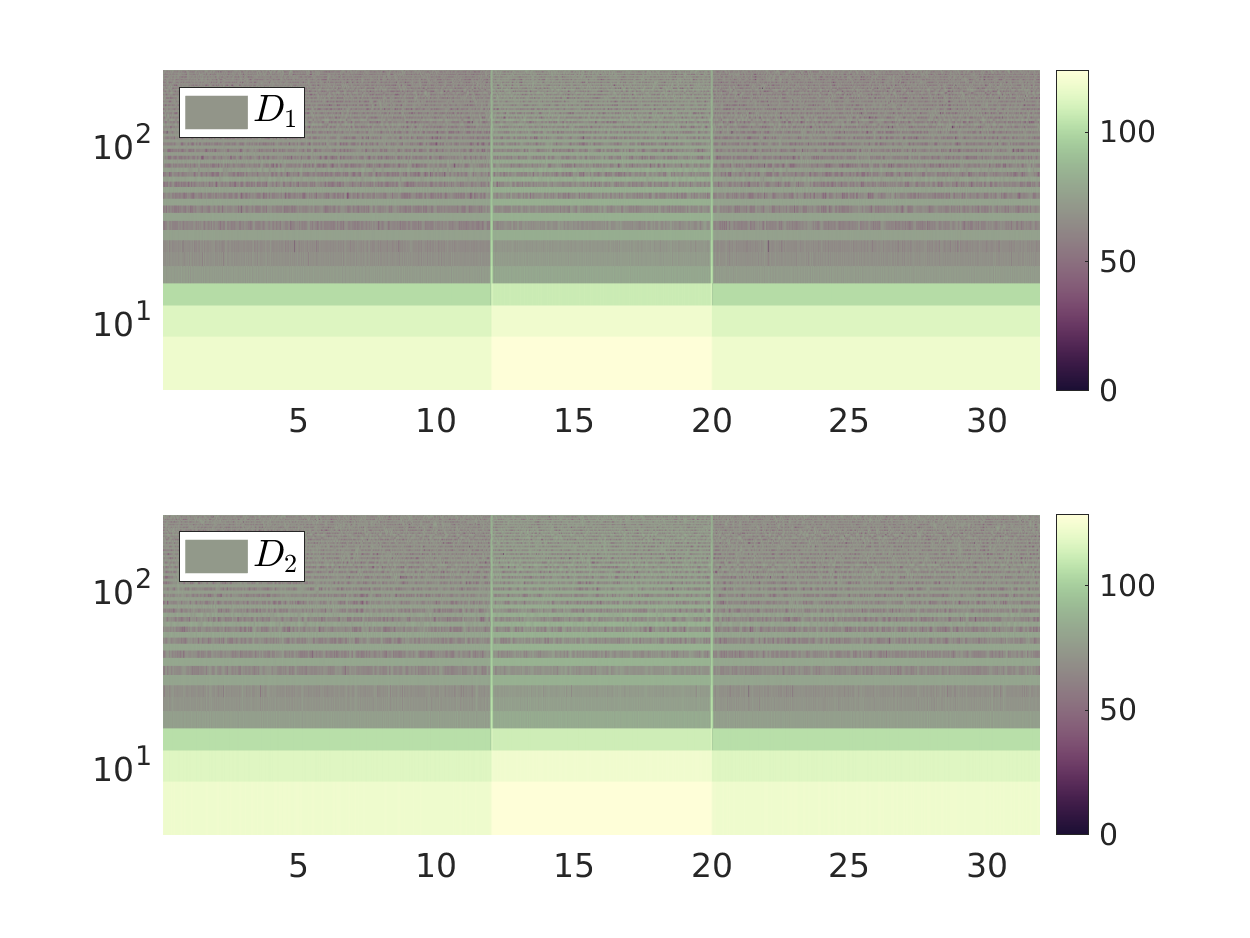}} & \parbox[c]{0em}{
      \includegraphics[width=2in]{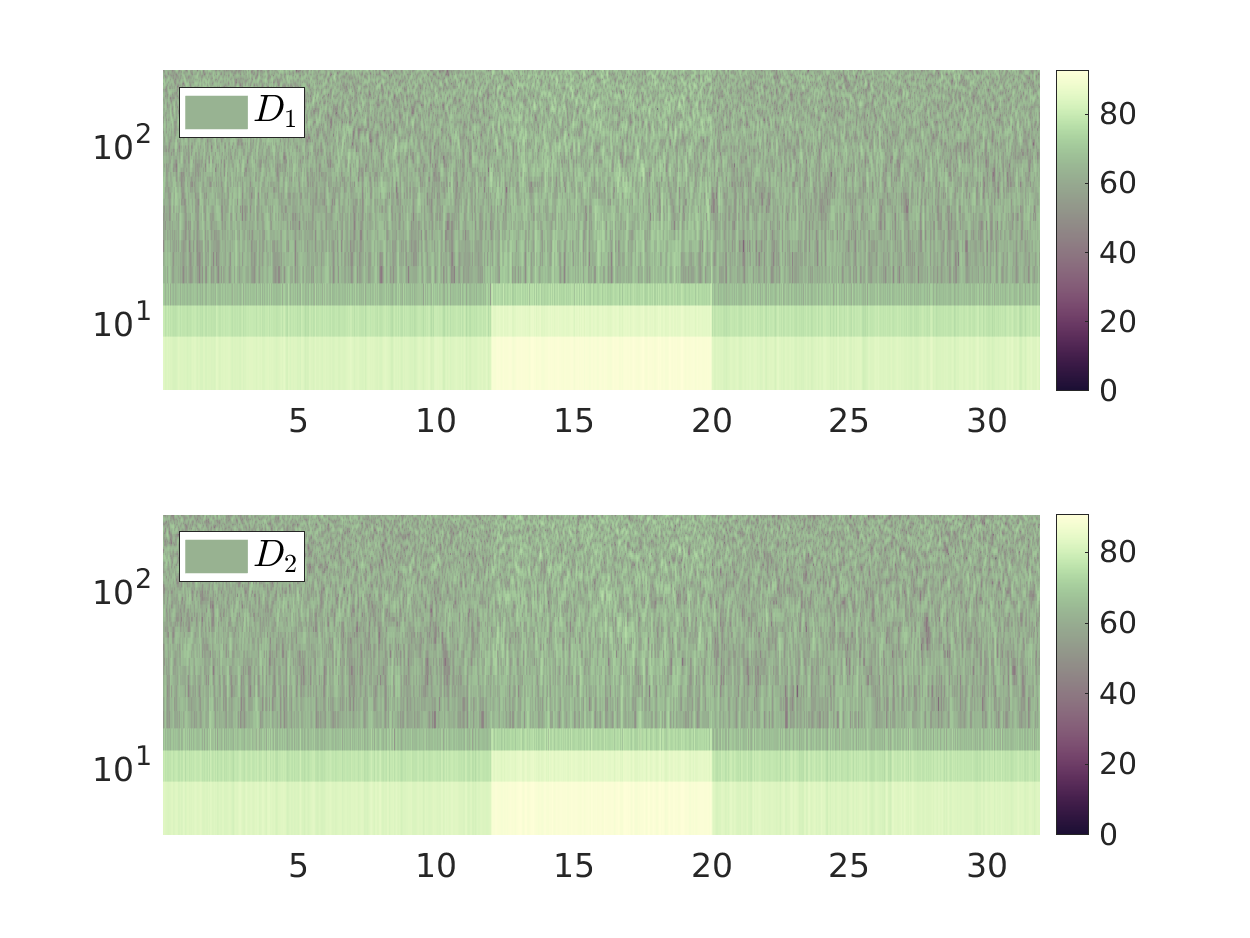}}\\
      \hline
      Uniform &  \parbox[c]{0em}{
      \includegraphics[width=2in]{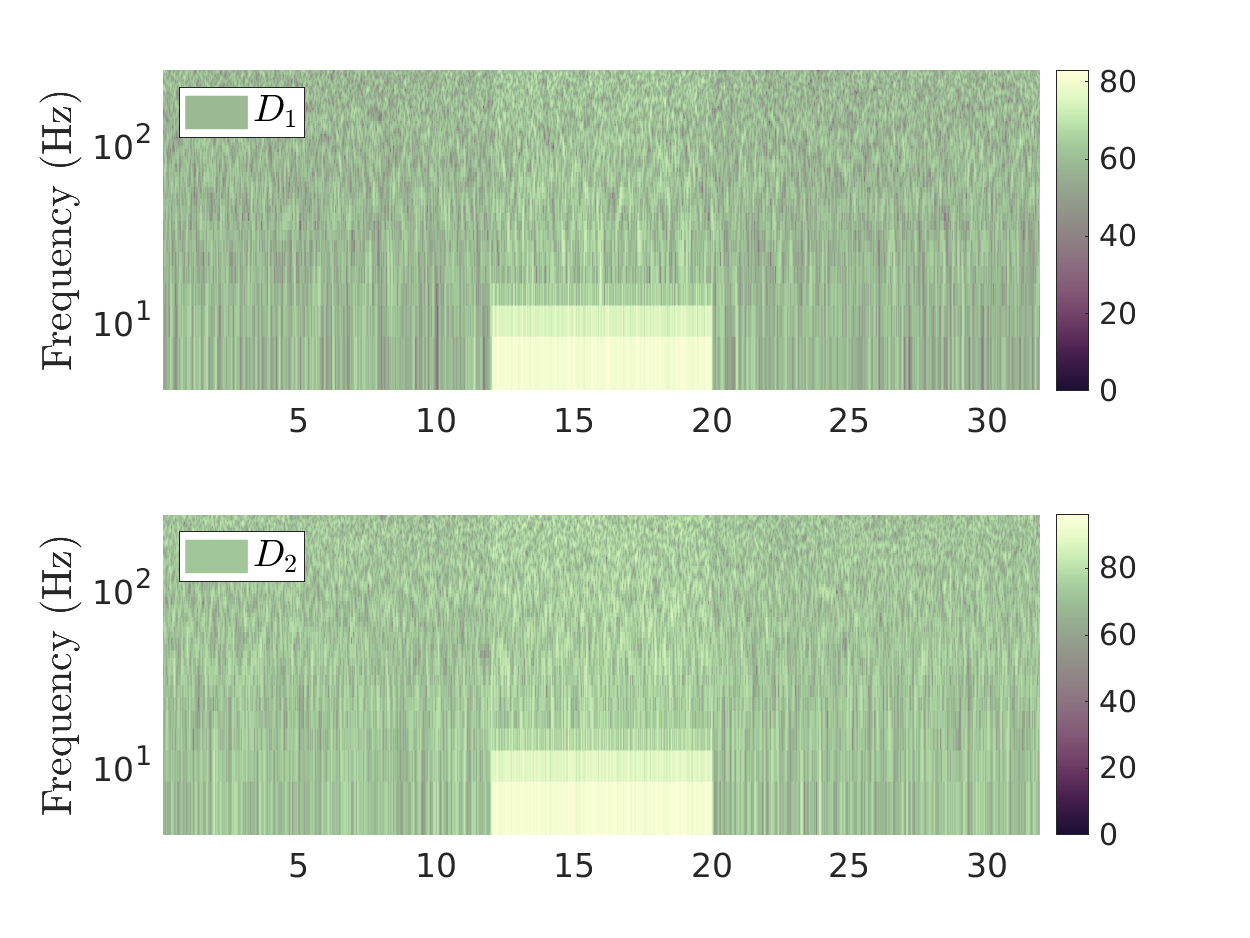}} & \parbox[c]{0em}{
      \includegraphics[width=2in]{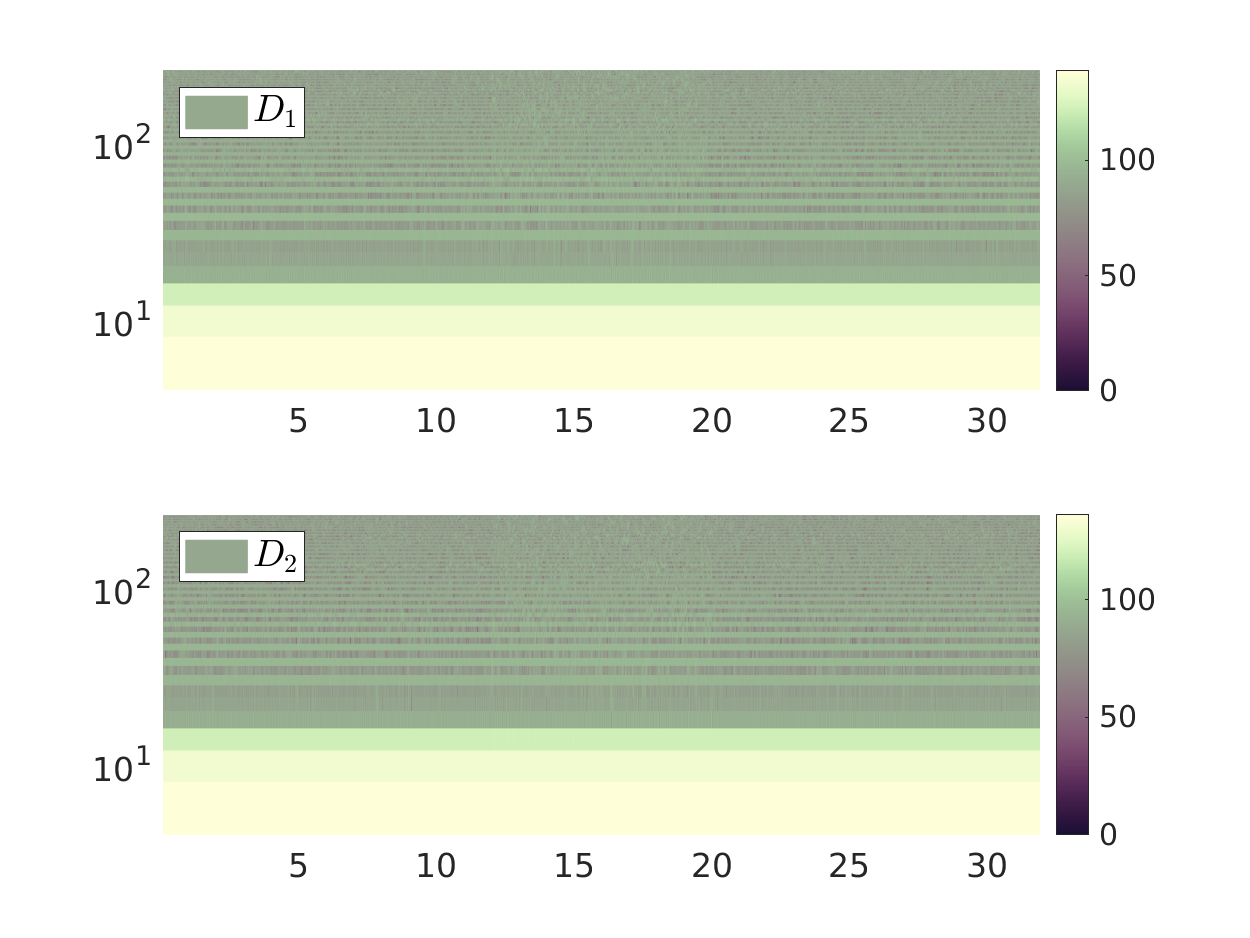}}& \parbox[c]{0em}{
      \includegraphics[width=2in]{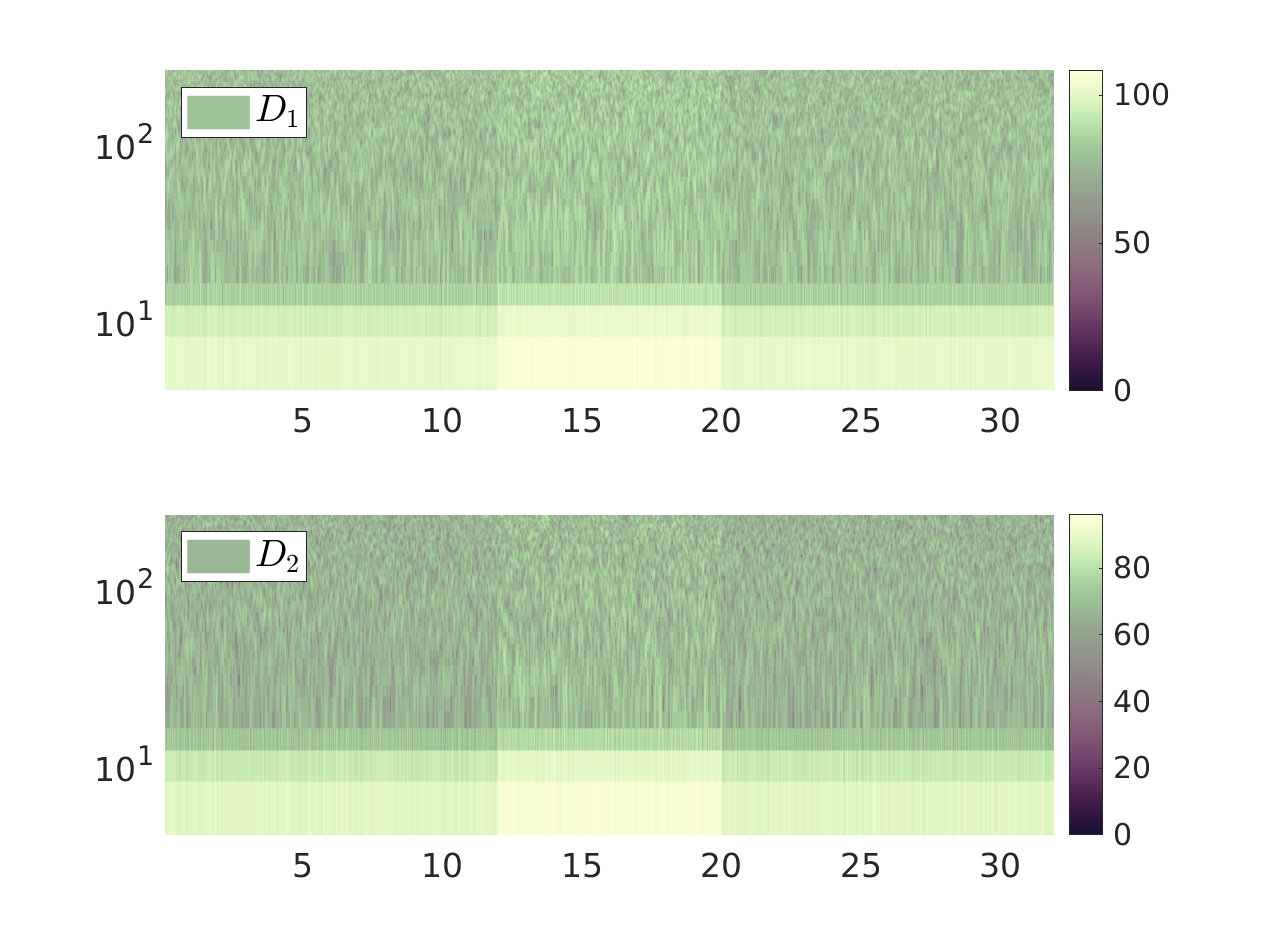}}\\
      \hline
      Chirp & \parbox[c]{0em}{
      \includegraphics[width=2in]{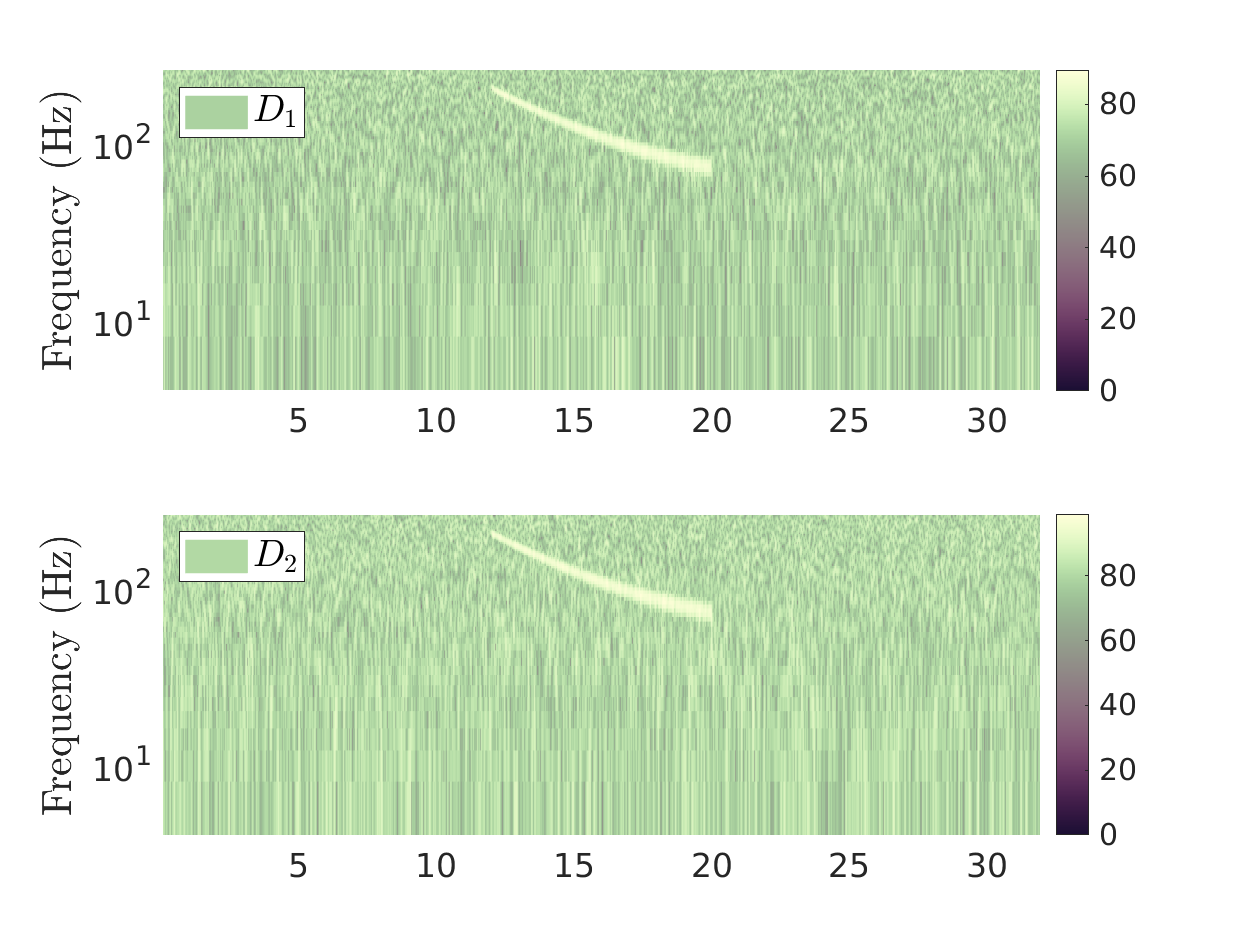}} & \parbox[c]{0cm}{
      \includegraphics[width=2in]{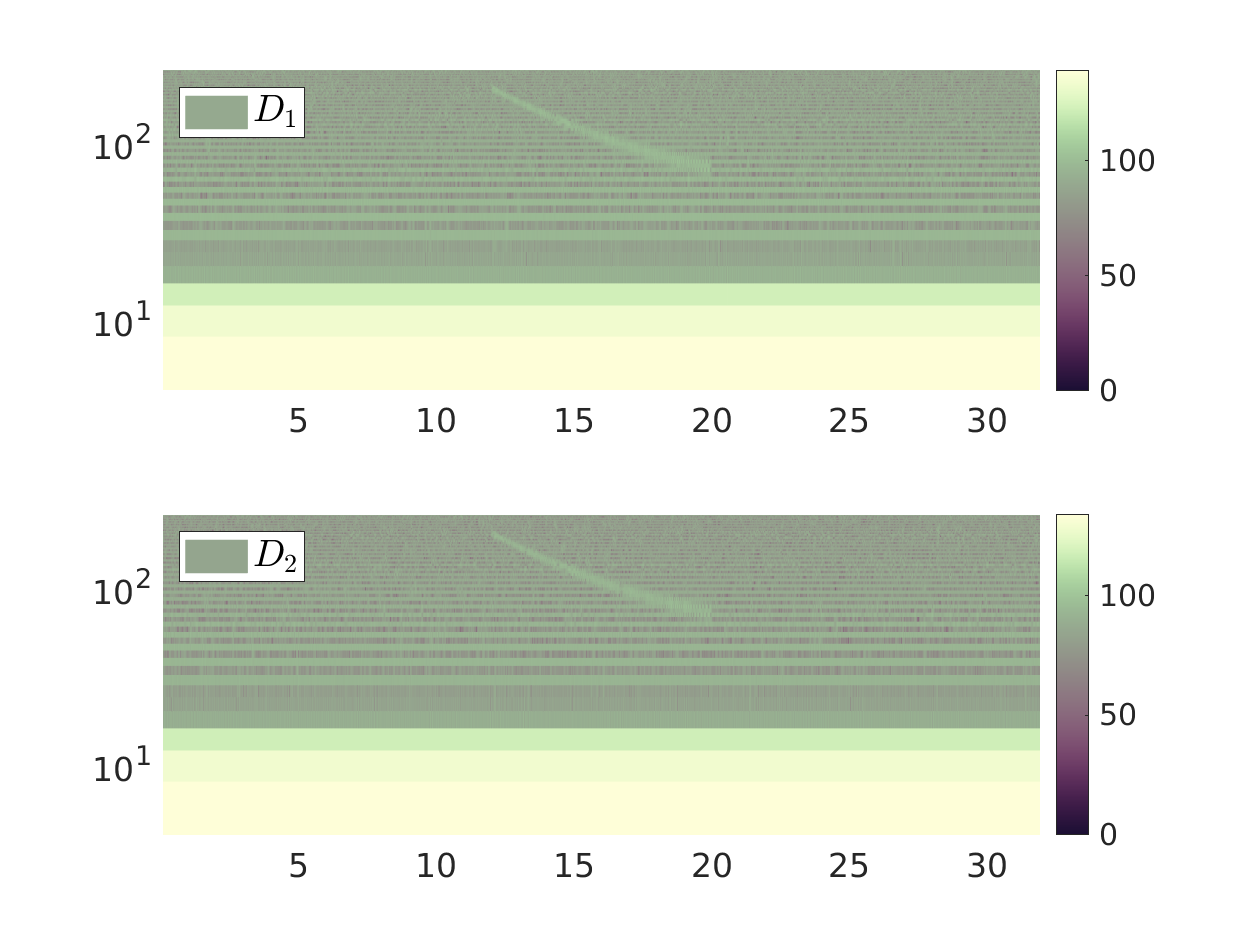}} & \parbox[c]{0cm}{
      \includegraphics[width=2in]{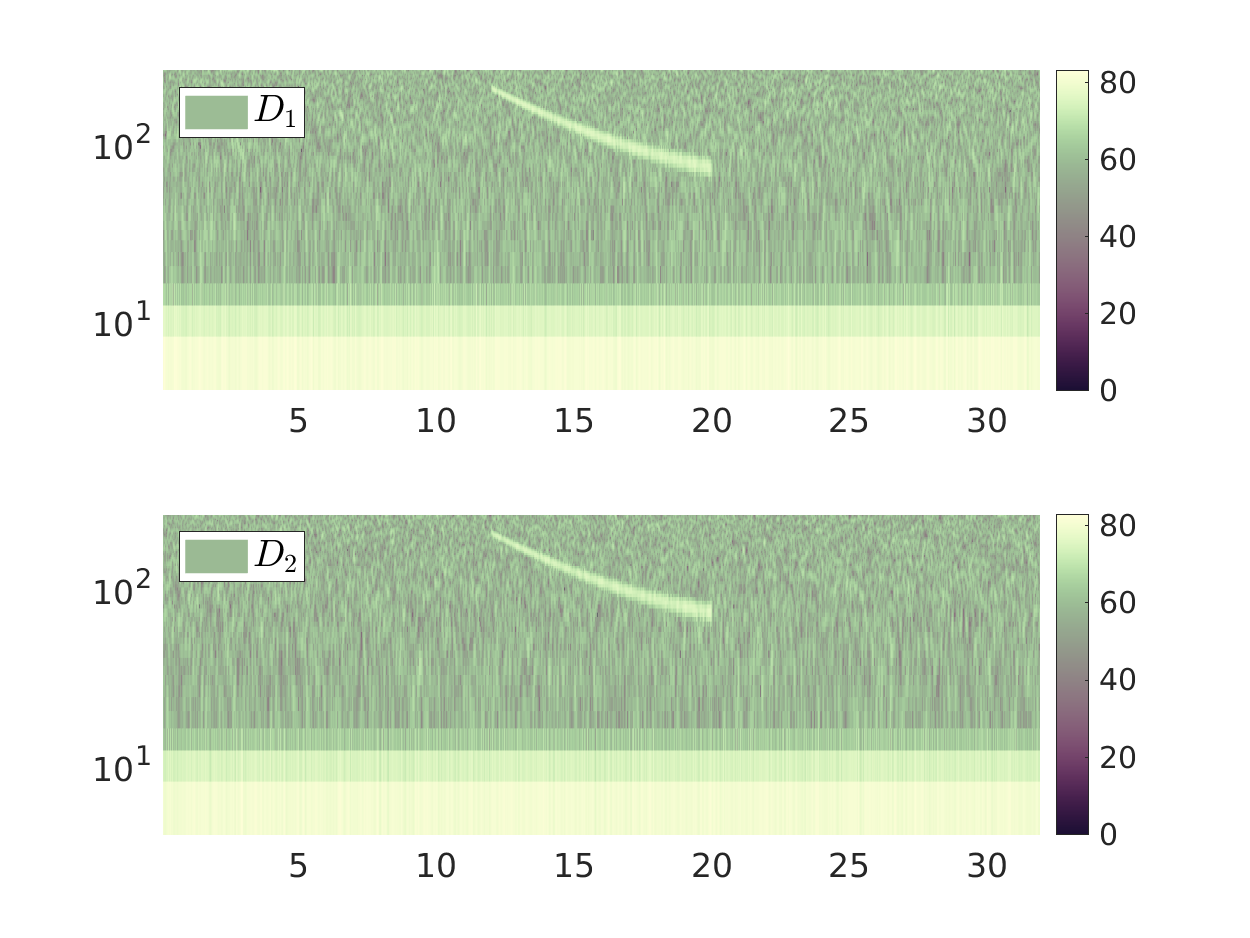}} \\
      \hline
      Sine&  \parbox[c]{0em}{
      \includegraphics[width=2in]{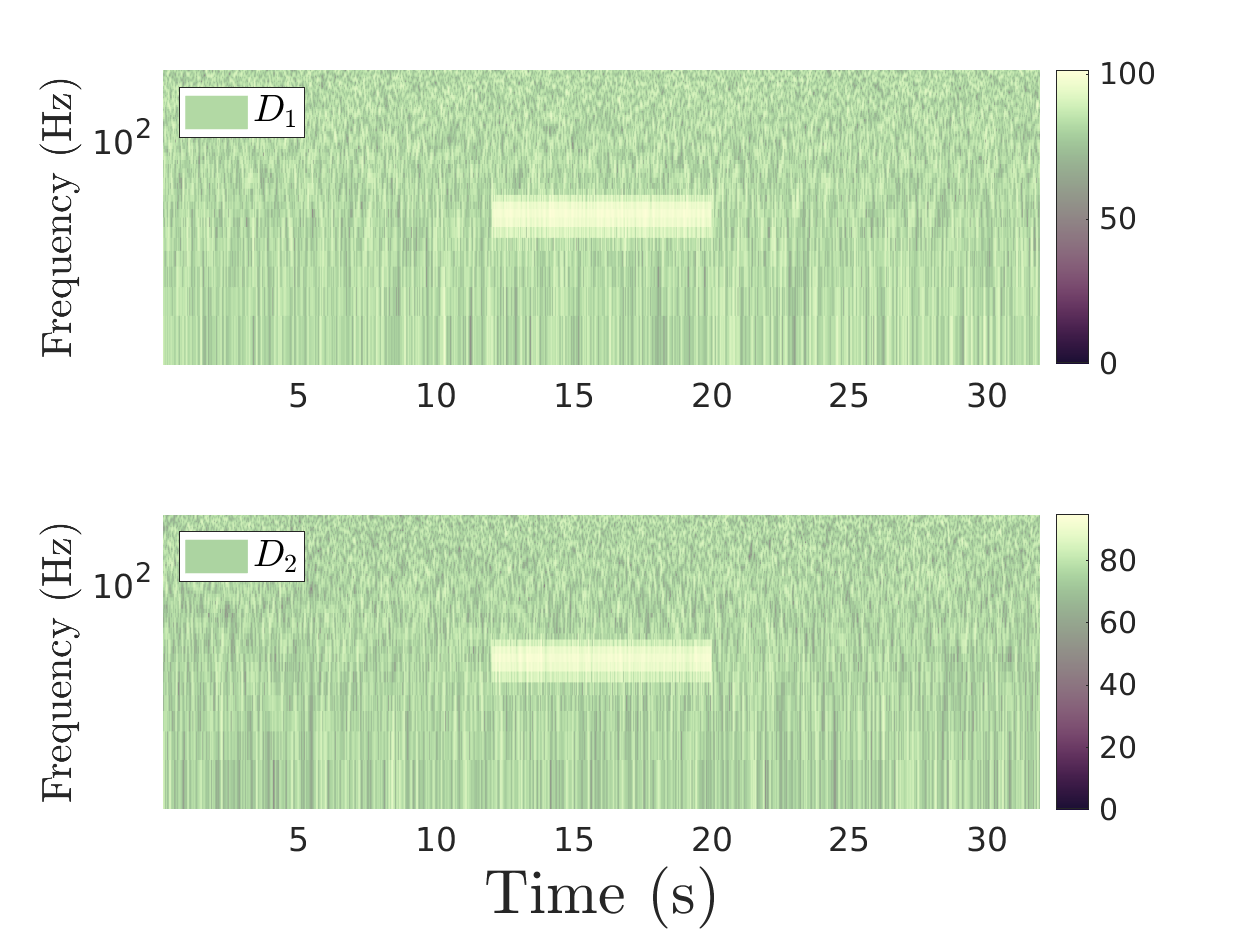}} & \parbox[c]{0em}{
      \includegraphics[width=2in]{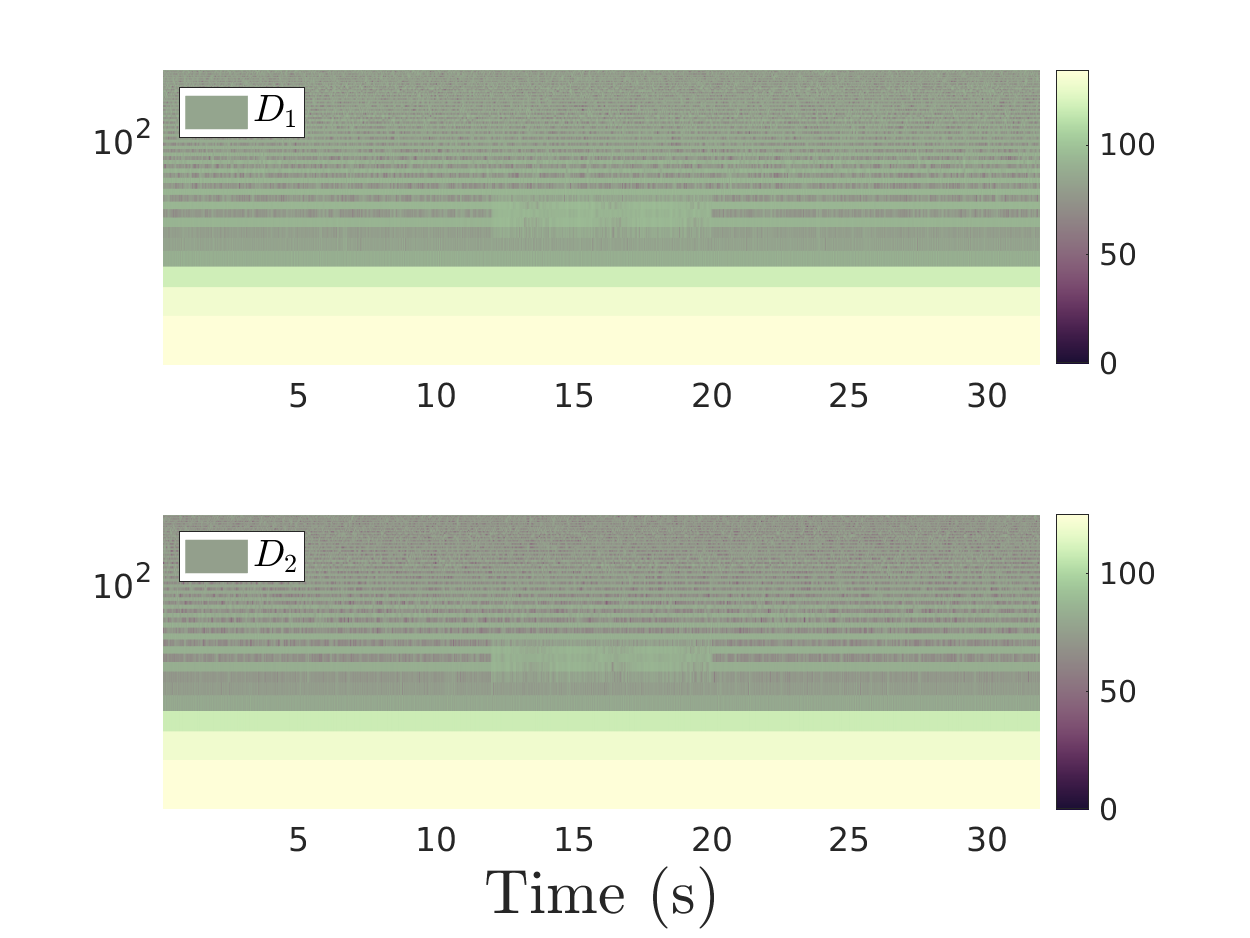}} & \parbox[c]{0em}{
      \includegraphics[width=2in]{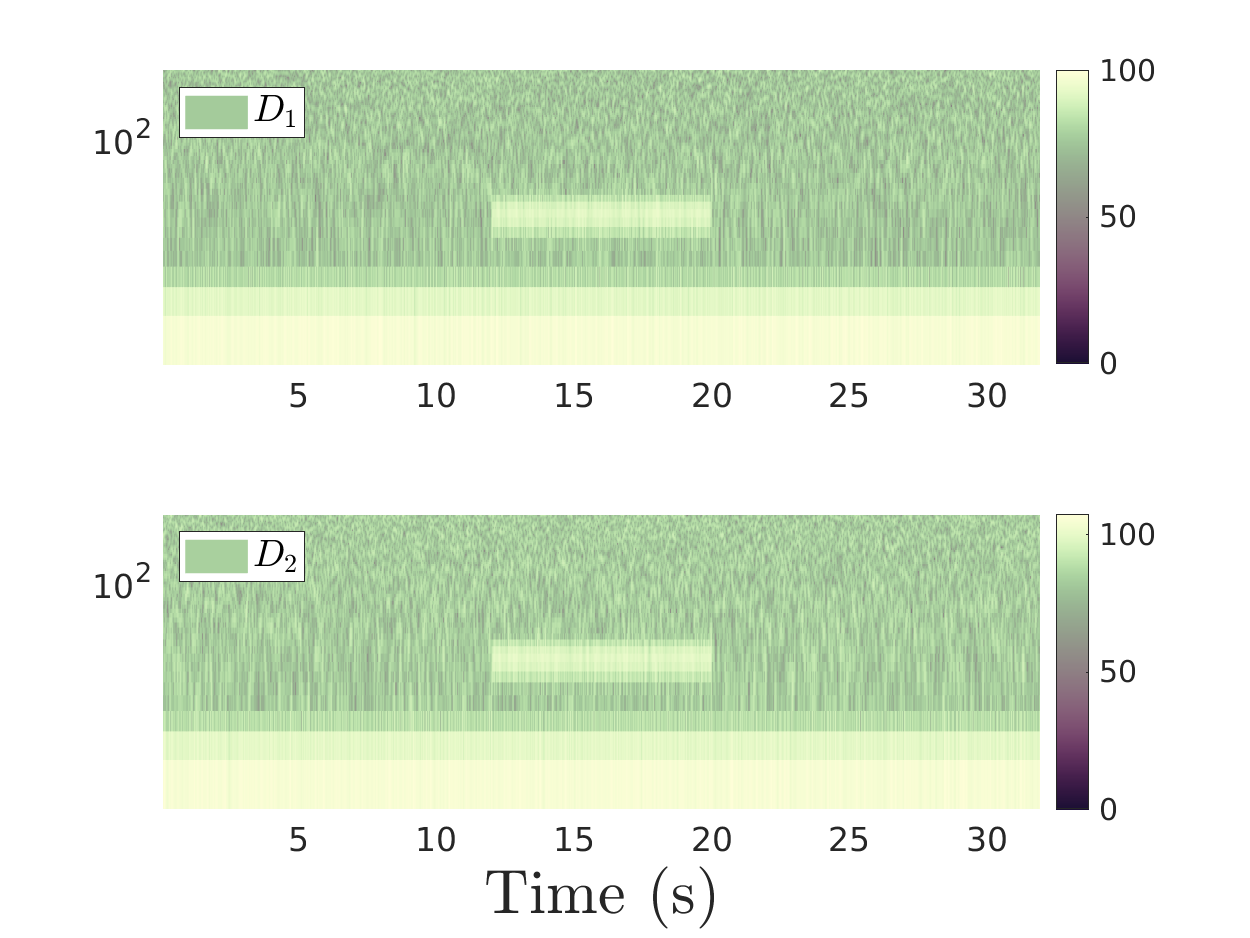}} \\
      \hline
  \end{tabular}
  \caption{Spectrograms of simulated data used in the present benchmark study of two-channel observations by correlation, covering three types of background noise - Gaussian, Poisson and Uniform - and five types of signals - Gaussian, Poisson, Uniform, Chirp and sine wave. Spectrograms are shown in pairs (one for each channel). Highlighted is the case of a relatively large signal amplitude clearly showing Chirps and Sine waves transients signals. Here we take $\lambda=10000$ for Poisson background and signal strength $\alpha=1$ for illustrative purposes.}
    \label{T2}
\end{figure}

In low signal-to-noise observations, two-channel detectors/observatories are preferred when signal detection critically depends on correlation strength across all channels (Fig. \ref{fig:Process}). In this way, two important challenges are presented: selecting the most sensitive correlation algorithm and its most efficient implementation on a choice of compute platform. They need not always be commensurate. In the era of Big Data astronomy, the latter inevitably poses novel priorities on data reduction without sacrificing sensitivity (\S \ref{Sec:Intro}), in particular in the deployment of heterogeneous computing. An example of the adverse effect of large array sizes is already apparent in {\em graphics processor units} (GPU) performance shown in Fig. \ref{fig:GPU}. In FFT-based signal processing, their performance is typically bandwidth limited, leaving modest efficiencies relative to theoretical compute performance.  The same applies (crucially) in GPU-CPU interfacing \citep{Putten2017}. On large scale platforms such as distributed computing very similar limitations tend to apply, unless data size is appropriately adjusted. 
This shows that data reduction is {pivotal} when working with massive data, consequently requiring a correlation method that {applies efficiently to a small fraction of data thereof without sacrificing discovery power}.

In this paper, we demonstrate that EPR can realize this challenge. It successfully applies to a relatively small fraction of data with {generally very little loss in sensitivity} ({Figs. \ref{T1} and \ref{fig:N-kappa}}). 

This conclusion is based on detailed benchmarks of EPR against conventional methods DCC and PC. 
While DCC and PC apply to full real-valued data array, EPR applies to Boolean-valued data typically following a cut-off (\ref{EQN_kappa}) (\S \ref{Sec:EPR_def}), which creates a significant reduction in data by focusing only on tails (Fig. \ref{fig:N-kappa}).
{Our benchmarks cover} three types of background noise (Gaussian, Poisson and Uniform) and five signal injections (\S \ref{Sec:procedure}) - a total of 15 combinations.
We rank sensitivity by detection thresholds derived from response curves as a function of signal amplitude controlled by $\alpha$ (Fig. \ref{fig:departure}). Fig. \ref{T1} summarizes our results based on the blocks shown in Fig. \ref{fig:Sample} and the results in {Fig. \ref{T3}. We} present our results by giving different colors to the three correlation methods.

{Notably, the type of background noise is seen to be critical to} the ranking of the three methods. EPR tends to be preferred when background noise is non-negative, i.e., Poisson or Uniform, while PC wins when the background is Gaussian or Gaussian-like. This is consistent with the natural conversion of non-negative to Boolean-valued data in (\ref{EQN_kappa}).  
The type of the signal appears less critical, though the ranking of EPR may change when $\kappa$ is small. The un-normalized DCC appears to be consistently sub-par to PC and EPR, demonstrating the important role of normalization in PC and EPR. 
In particular, as shown in Fig. \ref{T3}, for most of the combinations included in this study, EPR has a similar or even better detection sensitivity compared to PC by working on approximately $10\%$ of the data at $\kappa \simeq 1.6$.

{This result is even more significant when considering the computational resources and runtime requirements for these three methods. A better than 90\% reduction in the data size for EPR with modest cut-off (Fig. \ref{fig:N-kappa}), directly translates into a smaller footprint in memory and computational work load by the same factor.
In the era of Big Data astronomy, working with tails offers the potential for significant enhancement in efficiency in computational resources and portability over computational infrastructure. Ideally, efficiency is compute-limited. In practice, however, this is rarely the case due to various dependencies on architectures frequently 
encountered in heterogeneous computing.}

{Illustrative for efficiency dependence on data size is Fig. \ref{fig:GPU}, showing a dramatic drop to bandwidth limited performance when array sizes exceed Local Memory of GPUs encountered in FFT-based signal processing. The need to limit signal processing to data that matter most by potential to contain relevant signals appears similarly at the GPU-CPU interface due to their vastly different performance characteristics. In Butterfly Matched Filtering (BMF) over a dense bank of ${\cal O}\left(0.5\,{\rm M}\right)$ templates - an exascale calculation - tails are applied to GPU output to balance GPU and CPU performance \citep{Putten2023}. 
In this analysis, a cut-off $\kappa=2$ reduces output file size to less than $1\,{\rm GB}$ down from about $25\,{\rm GB}$ per LIGO frame of H1L1 data of 4096\,s. This reduction is in particular opportune when analyzing entire observational runs, comprising a large number of frames. For instance, LIGO S6 comprised over 10\,k frames \citep{Putten2017}.}

{Further optimizations may derive from integer versus floating point operations (FLOPs) in the evaluation of EPR and, respectively, PC and DCC. At array length $N$, PC and DCC require $\mathcal{O}(N^p)$ and $\mathcal{O}(N\log N)$ FlOPs, respectively, where $p=1$ for a single evaluation and $p=2$ when extended by time-slides. Applied only to tails of length $M\ll N$, EPR offers a speed-up over PC by a factor of $\mathcal{O}\left(\left(M/N\right)^p\right)$ with additional speed-up when using integer-based calculations. Implementations in C99 commonly used on GPUs offers significant gains by a factor of a few in integer or half-integer calculations over floating point operations. By its small footprint in data size and integer-based calculations combined, EPR significantly outperforms PC and may even outperform DCC unless $N$ is very large.}

The presented study {leaves open further integration with modeled or un-modeled searches by matched filtering, and respectively, Butterfly Matched Filtering applied to each channel individually} {\em before} correlations.
In Fig. \ref{fig:Process}, this is envisioned in F1, where such may serve to pre-amplify {candidate} signals prior to correlation. 
The junction at which EPR is applied may vary, e.g., close to the raw detector data or beyond at $J_i$ following $F_{i-1}$ $(i=2,3)$.
Additional {individual} channel processing can be performed by, e.g., {image rendering} such as spectrograms. 
{The significance of common features may be determined by cross-channel correlations of spectrograms using $\chi$-image analysis, {\it again} by conversion to Boolean-valued data, currently applied in searches for small un-modeled signals in gravitational-wave data \citep{Putten2023}.}


\begin{acknowledgments}
The authors thank the anonymous reviewer for the detailed reading of the manuscript and constructive comments, which greatly 
enhanced clarity of our presentation.
This work is supported by the National Research Foundation of Korea, under Grants No. NRF-2021R1A2C1092701 and NRF-RS-2024-00334550.
\end{acknowledgments}


\appendix

\section{Spectrograms}
\label{Sec:AppA}

For three different background statistics and five different signal forms, Fig. \ref{T2} presents for illustrative purposes the spectrograms. For clarity of presentation, plots are created using {two times the maximum signal strength} used in our response curve analyses.

\bibliography{mybibfile}{}
\bibliographystyle{aasjournal}



\end{document}